\documentclass{article}

\usepackage{graphicx}
\usepackage{amssymb}
\usepackage{amsthm}
\usepackage{amsmath}

\usepackage{authblk}

\title{String percolation in AA and pp collisions}

\author[1,2]{I. Bautista}
\author[3,4]{C. Pajares\thanks{Corresponding Author: pajares@fpaxp1.usc.es}}
\author[3]{J. E. Ramírez}

\affil[1]{Facultad de Ciencias F\'isico Matem\'aticas, Benem\'erita Universidad Aut\'onoma de Puebla, 1152, M\'exico.}
\affil[2]{C\'atedra CONACyT, CONACyT, 03940 Ciudad de M\'exico, M\'exico}
\affil[3]{Departamento de F\'isica de Part\'iculas, Universidad de Santiago de Compostela, E-15782 Santiago de Compostela, Espa\~na}
\affil[4]{Instituto Galego de F\'isica de Altas Enerx\'ias, Universidad de Santiago de Compostela, E-15782 Santiago de Compostela, Espa\~na}
\begin{document}
\maketitle
\begin{abstract}
A brief review of the string percolation model and its results are presented together with the comparison to experimental data. First, it is done an introduction to the quark-gluon phase diagram and the lattice results concerning the confinement and the percolation of center domains. It is studied the interaction of the strings produced in nucleus-nucleus and proton-proton collisions showing how the string percolation arises. The main consequences of the string percolation, concerning the dependence on the energy and centrality, on the multiplicities and the mean transverse momentum are obtained comparing with experimental data. It is emphasized the non-abelian character of the color field of the strings forming the cluster to reproduce the rise of the transverse momentum with multiplicity and the relative suppression of multiplicities.
It is also studied different observables like multiplicity and transverse momentum distributions, dependence with multiplicity and transverse momentum correlations, forward-backward correlations, the strength of the Bose-Einstein correlations, dependence on the multiplicity of $J/\Psi$ production and its possible suppression in pp collisions at high multiplicity, strangeness enhancement, elliptic flow, and ridge structure. The comparison with the data shows an overall agreement.
The thermodynamical properties of the extended cluster formed in the collision are discussed computing its energy and entropy density, shear viscosity over entropy density ratio, bulk viscosity, sound speed and trace anomaly as a function of temperature, showing a remarkable agreement with lattice QCD evaluations.
The string percolation can be regarded as the initial frame able to describe the collective behavior produced in AA and pp collisions.
\end{abstract}












\tableofcontents
\newpage

\section{Introduction}\label{intro}
More than four decades ago, it was raised the possibility of distributing high energy over a large volume to restore broken symmetries of the physical vacuum creating abnormal states of nuclear matter \cite{1}.
Very early, it was pointed out that the asymptotic freedom property of QCD implies the existence of a high-density matter formed by deconfined quarks and gluons \cite{2} and the exponential increase of the hadron Hagedorn spectrum was connected with the existence of a different phase \cite{3}.
The thermalized phase of quarks and gluons was called Quark Gluon Plasma (QGP) \cite{4}, and the evaluations of the required high density showed that could be reached in relativistic heavy ion collisions \cite{5,6} and, several signatures of QGP were proposed.
Quarkonium suppression \cite{7}, the excess of photons and jet quenching \cite{8,9} were some of them.
At this time it was pointed out the relevance of percolation in the study of the phase transitions of hadronic matter \cite {10,11}.

From the experimental side, there were large facilities to study the properties of large density matter starting by the AGS and ISR experiments later followed by SPS, RHIC, and LHC.
At SPS already several signatures hinted the onset of QGP formation \cite{12}.
The RHIC data show a collective elliptic flow which pointed out a very low shear viscosity over entropy density ratio, $\eta/s$, indicating strongly interacting matter.
In addition, the jet quenching was observed, indicating that this strongly interacting matter was very opaque \cite{13,14,15,16,17}.
The above mention ratio enhanced the attention to the AdS/CFT correspondence due to the result $\eta/s=1/4\pi$ \cite{18}.
The LHC experiments \cite{19,20,21} have extended the study of the elliptic flow to all the harmonics \cite{22,23} confirming the obtained strong interacting quark and gluon matter and showing that the collective behavior and the ridge structure previously observed at RHIC in Au-Au and Cu-Cu collisions \cite{24,25} also occurs in pPb \cite{26,27,28} and pp collisions at high multiplicity \cite{29}.
The collective behavior of pp and pPb interactions is a challenge to the hydrodynamics descriptions, and they raise the question whether the main experimental data can be explained by final state interactions or on the contrary, the initial state configuration should describe them.

On the other hand, the data on quarkonium confirm the validity of combined picture of a subsequent melting of the different resonances together the recombination of heavy quarks and antiquarks at high energy \cite{30,31,32}.
The departure of linear dependence on the multiplicity of the $J/\Psi$ production has been observed in pp and pPb collisions \cite{33,34} indicating multiparton interactions or multiplicity saturation \cite{35}.
Detailed studies on the jet quenching for identified particles have been done \cite{36} showing features related to the low of coherence of the gluons edited in the jet due to the high-density medium.
Finally, let us mention that at RHIC has been observed recently that the fluid produced by heavy ions is the most vortical system ever observed \cite{37}.

On the theoretical side, in addition to the hydrodynamics studies, the color glass condensate (CGC) approach \cite{38,39,40,41,42} gives a good description of most of the experimental data and is derived directly from QCD. 
In QCD, the gluon density $xG(x,Q)$ rises very fast as a function of the decreasing fractional momentum $x$ or increasing the resolution $Q$.
So the gluons showers generate more gluon showers producing an exponential increasing toward small $x$.
As the transverse size of the hadron or the nucleus rises slowly at high energy, the number of gluons and their density per unit of area and rapidity increase rapidly as $x$ decreases.
However, there will be the fusion of gluons leading to a limited transverse density of gluons at some fixed momentum resolution, $Q_s$, the gluon saturation \cite{43}.
The low $x$ gluons are closely packed, the distance between them being very small.
Hence the interaction coupling is small $a_s\ll1$.
In a given collision, the multiplicity should be proportional to the number of gluons, which at the saturation momentum $Q_s$ is \cite{44,45}
\begin{equation}
\frac{dN}{dy}\sim \frac{1}{\alpha_s (Q_s)}Q_s^2R^2.
\label{eq1}
\end{equation}
This dense system, called CGC, has a very high occupation number $1/\alpha_s$, and corresponds to a highly coherent state of strong color fields.
The high $x$ gluons can be considered as the sources of the low $x$ gluons.
The independence of the cutoff used to separate the high $x$ gluons from the low $x$ ones, gives rise to a kind of evolution equation.

In high energy physics experiments, the colliding objects move at velocities close to the speed of light.
Due to the Lorentz contraction, the collision of two nuclei can be seen as a that of two sheet of colored glass where the color field in each point of the sheets is randomly directed.
Taking these field as initial conditions, one finds that between the sheets, longitudinal color electrical and magnetic fields are formed.
The number of these color flux tubes between the two colliding nuclei is forming the called Glasma \cite{46}, which has been extensively compared with the experimental data.

Another approach to the initial state is the percolation of strings \cite{47,48,49,50,51} which is not so popular as the CGC because cannot be derived directly from QCD although it is inspired in it, and most of its results, are a direct consequence of properties of QCD.
In this approach, the multi-particle production is described in terms of older strings stretched between the partons of the projectile and target.
These string decay into $q-\bar{q}$ pairs and subsequently hadronize producing the observed hadrons. Due to the confinement, the color of these strings is confined to a small area $S=\pi r_0^2$, with $r_0=0.2$ fm in the transverse space.
The value 0.2-0.25 fm is obtained in lattice studies \cite{Campostrini1984} and also considering bilocal correlations \cite{DOSCH1988}. It corresponds to the correlation length of the QCD vacuum.
 With increasing energy and/or size and centrality of the colliding objects, the number of strings grows and the strings start to overlap forming clusters similarly than the continuum percolation theory \cite{52}.
At a given critical density, a macroscopical cluster appears crossing the collision surface, which marks the percolation phase transition. Therefore the nature of this transition is geometrical.

In string percolation, the basic ingredients are the strings, and it is necessary to know their number, rapidity extension, fragmentation and number distribution. All that requires a model and therefore string percolation
is model dependent. However most of the QCD inspired models give similar results for most of the observables in such a way that the predictions are by a large measure independent of the model used.

\begin{figure}
\centering
\includegraphics[scale=0.2]{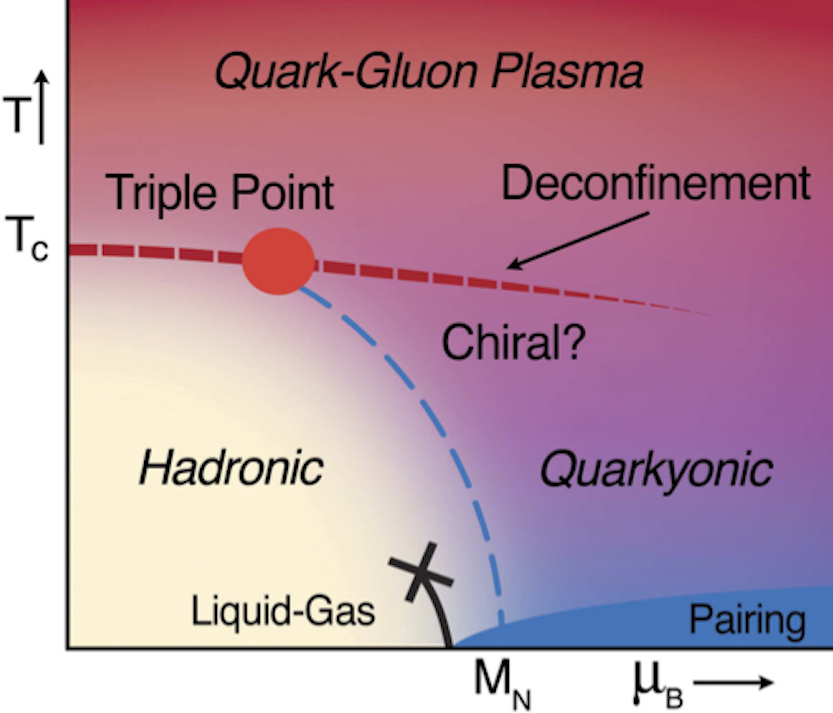}
\caption{Phase diagram of the nuclear matter. Temperature vs baryonic chemical potential \cite{53}.}

\label{fig1}
\end{figure}

The string percolation and the Glasma are related to each other \cite{54}: In the limit of high density, there is a correspondence between the physical quantities of both approaches. 
The number of color flux tubes in Glasma picture, $Q_s^2R^2$, has the same dependence on the energy and centrality of the collisions that the number of effective clusters of strings in string percolation.
In both approaches, the negative binomial distribution is obtained as the multiplicity distribution, where the parameter $k$ that controls the width of this distribution has the same energy and centrality dependence.
The role of the occupation number $1/\alpha_s$ in CGC is played by the fraction of the collision surface covered by strings.
The randomness of the color field in CGC gives rise to a reduction of the multiplicity.
Similarly, the randomness in color space of the color field of the $n$ strings of the cluster originates that the intensity of the color field of the cluster is not $n$ times the individual color field of each string but $\sqrt{n}$. This reduction implies also a reduction of the multiplicity of particle production and an increase of the transverse momentum, with the multiplicity.
Due to these similarities, the predictions of both approaches are similar for many observables. The string percolation is able to explore also the region where the high-density limit has not been reached.

The observed densities of our world have large differences which expand over many orders of magnitude, from $10^{-6} \textrm{ nucleons}/\textrm{cm}^3$ in average in the Universe to $10^{38} \textrm{ nucleons}/\textrm{cm}^3$ inside a nucleus and $10^{39} \textrm{ nucleons}/\textrm{cm}^3$ in a neutron star.
The study of the high-density limit, i. e., the study of de-confinement of quarks and gluons can be regarded as the place where high energy collision of two bodies probes the short distances and meets the thermodynamics (many body) of this short distance limit \cite{55}.
The lattices studies have shown that at low chemical potential $\mu=0$, color confinement and chiral symmetry restoration coincide and the phase transition is a crossover. Hence, in a medium of low baryon density, the mass of the constituent quark vanishes at the deconfined point $T_c$, and the screening radius of the gluon cloud vanishes.
At low $T$ and high $\mu$ there is no reason to expect similar behavior, and probably there will be an intermediate region of massive dressed quarks between the hadronic phase and the deconfined and chiral restoration phase.
However, other possibilities could exist as quarkonia and color superconductivity.
A possible diagram is shown in Fig.~\ref{fig1}.

In finite $T$ lattice QCD, the de-confinement order parameter is provided by the vacuum expectation value of the Polyakov loop $L(\vec{x})$ defined in Euclidean space
\begin{equation}
L(\vec{x})=\mathrm{Tr} \prod_{t=1}^{N_t} A_4 (\vec{x},t).
\end{equation}
Note that $L(\vec{x})$ is the ordered product of the SU(3) temporal gauge variables $A_4 (\vec{x},t)$ at a fixed spatial position, where $N_t$ is the number of lattice points in time direction and Tr denotes the trace over color indices.
The Polyakov loop corresponds to a static quark source and its vacuum expectation value is related to the free energy $F_q$ for a single quark
\begin{equation}
L(\vec{x})\sim \exp \left( -\frac{F_q}{T} \right).
\end{equation}
Below the critical temperature $T_c$ quarks are confined and $F_q$ is infinite implying $\langle L(\vec{x}) \rangle=0$.
In a de-confined medium color screening among the gluons makes $F_q$ finite, hence for $T > T_c$, $\langle L(\vec{x}) \rangle \neq 0$.
The phase transition of chiral symmetry is controlled by the chiral condensate
\begin{equation}
\sigma(T)=\langle \bar{\psi} \psi \rangle \sim M_q,
\end{equation}
which measures the constituent quark masses obtained from a Lagrangian with massless quarks.
At high temperature this mass melts, therefore
\begin{equation}
\sigma(T) \left\{ \begin{array}{lcl}
             \neq 0 &   \mathrm{if}  & T<T_\sigma, \\
              = 0 &  \mathrm{if}  & T>T_\sigma.
             \end{array}
   \right.
 \end{equation}
Here, $T_\sigma$ defines another critical temperature. The corresponding derivatives, the susceptibilities, have been studies in lattice QCD at vanishing baryon number, showing a sharp peak that defines respectively $T_c$ and $T_\sigma$.
The two temperatures, within errors, coincide.
Also, is seen a crossover, i. e., a sharp transition but without discontinuity.
The quoted value is 155$\pm$9 MeV \cite{55,56,57}. 
\begin{figure}
\centering
\includegraphics[scale=0.2]{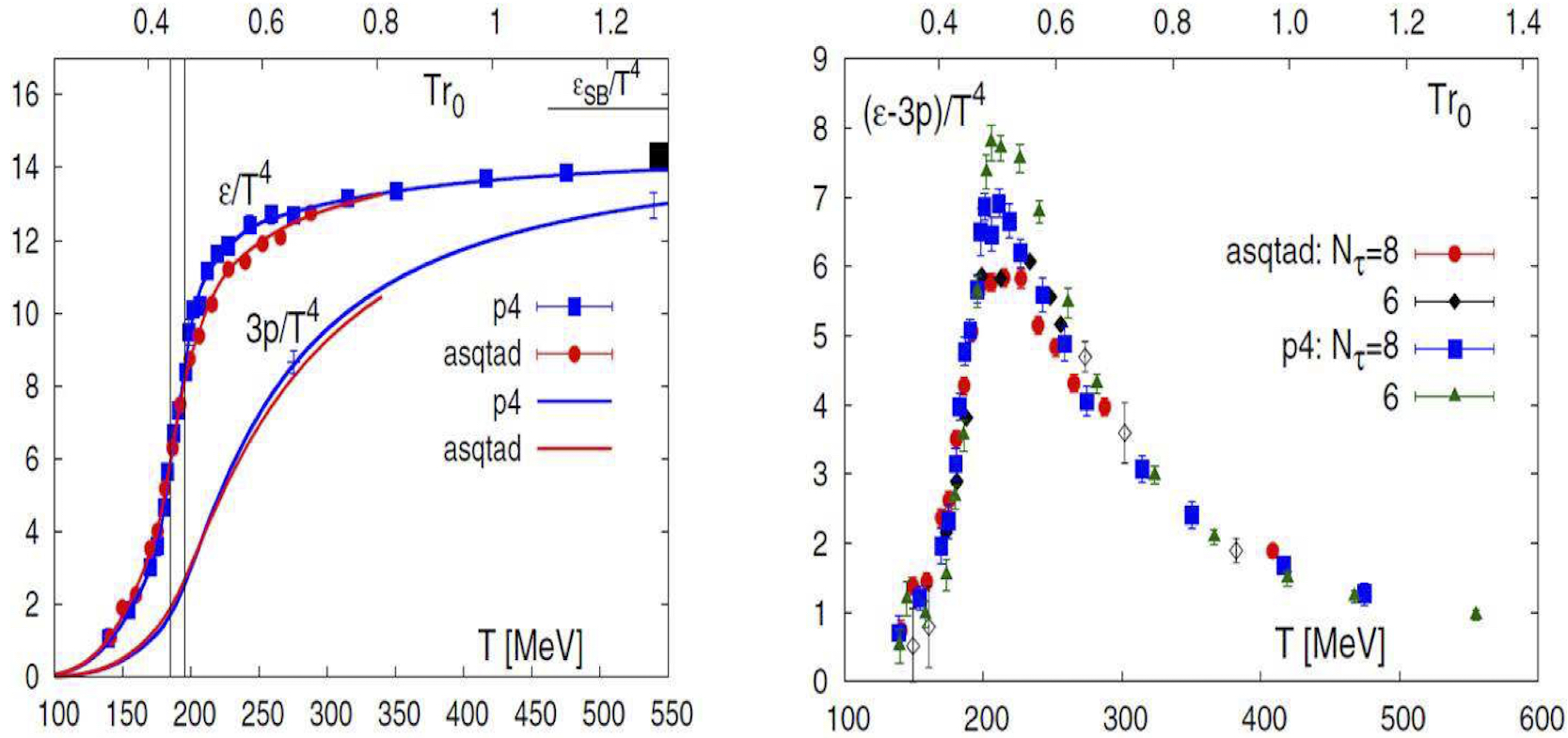}
\caption{The energy density and the pressure as the function of temperature (left). The energy density shows a sharp rise in the temperature region 170-200 MeV. The interaction measure calculated using different staggered fermion actions (right) \cite{57}.}

\label{fig2}
\end{figure}
The energy densities resulting from lattice QCD are shown in Fig.~\ref{fig2} (left), indicating that even for $T > 3T_c$ its value are far from the energy density of free gas quarks and gluons, namely
\begin{equation}
\epsilon=\frac{\pi^2}{30}\left[ g_g+\frac{7}{8} (g_q+g_{\bar{q}}) \right] T^4,
\end{equation}
where $g_g$, $g_q$ and $g_{\bar{q}}$ are the degeneracy numbers of the gluons, quarks and antiquarks.
This fact indicates that the deconfined phase is interacting strongly for a rather large range of temperatures.
This is also seen in the interaction measure
\begin{equation}
\Delta=\frac{\epsilon-3P}{T^4}.
\end{equation}
Moreover, the trace of the energy momentum tensor
\begin{equation}
T^\mu_\mu=\frac{\beta(g_s)}{2g_s}{G^a}_{\mu\nu}G^{a\mu\nu}+[1+\gamma(g_s)]m_q\bar{\psi}\psi
\end{equation}
is $\epsilon-3P$ and even for masses quarks $T^\mu_\mu\neq0$ as a consequence of the introduction of a scale in the renormalization process bearing the conformal symmetry (trace anomaly). In Fig.~\ref{fig2} (right) the results of the lattice QCD are shown.
Note that $\Delta$ decreases with $T$ very slowly, even less than $1/T^2$.

\section{Percolation model}

\begin{figure}
\centering
\includegraphics[scale=0.2]{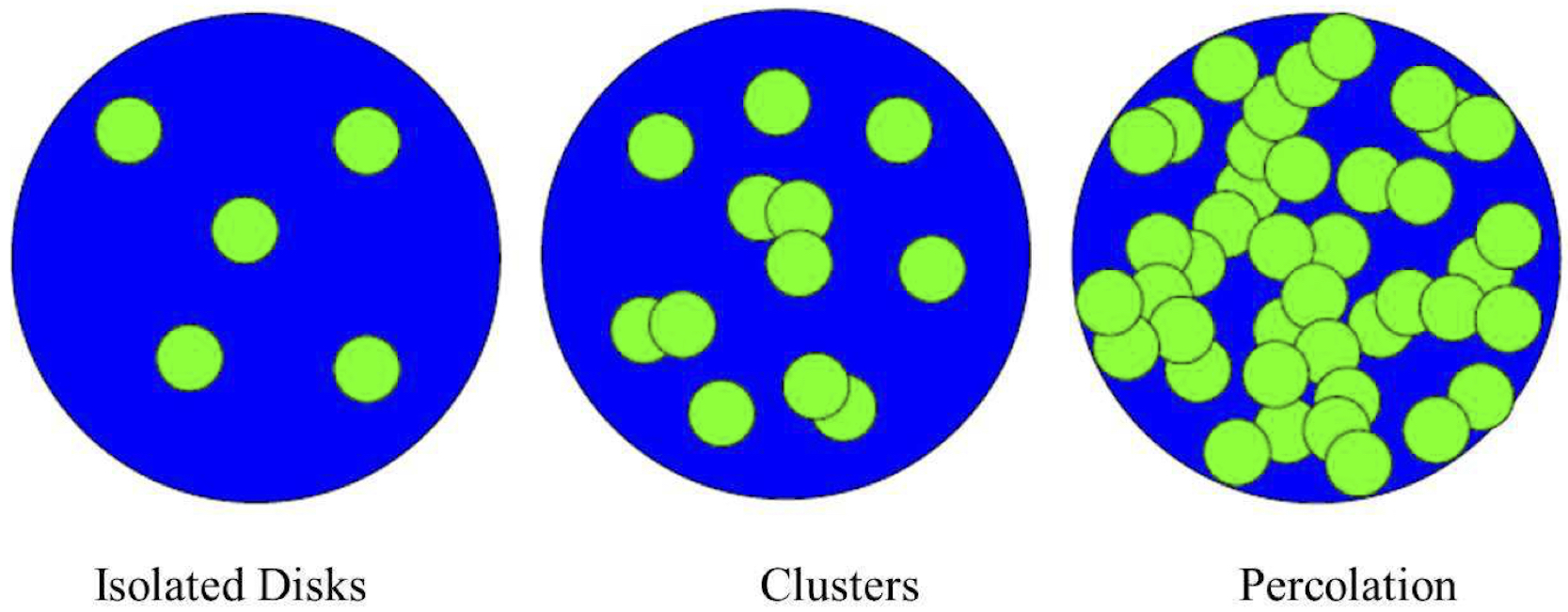}
\caption{Left panel: Disconnected discs, Middle: Cluster formation, Right panel: Over- lapping discs forming a spanning cluster \cite{55}.}

\label{fig3}
\end{figure}

Let us distribute small discs of area $\pi r_0^2$ randomly on a large surface, allowing overlap between them. As the number of discs increases, clusters of overlapping discs start to be formed.
If we regard the disc as small drops of water, how many drops are needed to form a puddle crossing the considered surface?
Given $N$ disc, the disc density is $\xi=N/S$, where S is the surface area.
The average cluster size increases with $\xi$, and at a certain critical value $\xi_c$, the cluster spans the whole surface, as is shown in Fig.~\ref{fig3}.

The critical density for the onset of continuum percolation is determined by numerical and Monte-Carlo simulations, which in the 2-dimensional case gives
\begin{equation}
\xi_c=\frac{1.13}{\pi r_0^2}.
\end{equation}
In the thermodynamical limit, $N\to \infty$ keeping $\xi$ fixed, the distribution of overlaps of the disc is Poissonian with a mean value $\rho=\xi \pi r_0^2$ \cite{perc2,perc3}:
\begin{equation}
P_n=\frac{\rho^n}{n!}\exp(-\rho).
\end{equation}
It also gives the total fraction of the plane covered by discs in $1-\exp(-\rho)$ \cite{50}.
The number 1.13 is obtained in case of discs uniformly distributed \cite{51,perc4,perc1}.
However, in cases when the discs are not uniformly distributed, this number changes.
For instance, in the cases of circular surfaces with Gaussian or Wood-Saxon profiles, the number is 1.5 and the fraction of the area covered by strings is more close to the function
\begin{equation}
\frac{1}{1+a\exp(-(\rho-\rho_c)/b)},
\end{equation}
where $\rho_c=1.5$ and the parameters a and b depend on the profile function, $b$ that controls the ratio between the width of the border of the profile $(2\pi R)$ and the total area $(\pi R2)$, and therefore is proportional to $1/R$ \cite{61}.
In the collisions of two hadrons or two nuclei, the surface where the discs are distributed is rather an ellipse or a circle, what gives rise to smaller values of the critical density \cite{62}.
For small systems where the number of discs is not large (far from the thermodynamical limit) the critical density is much smaller than above values, being 0.8 for high eccentricities \cite{62}.

\begin{figure}
\centering
\includegraphics[scale=0.2]{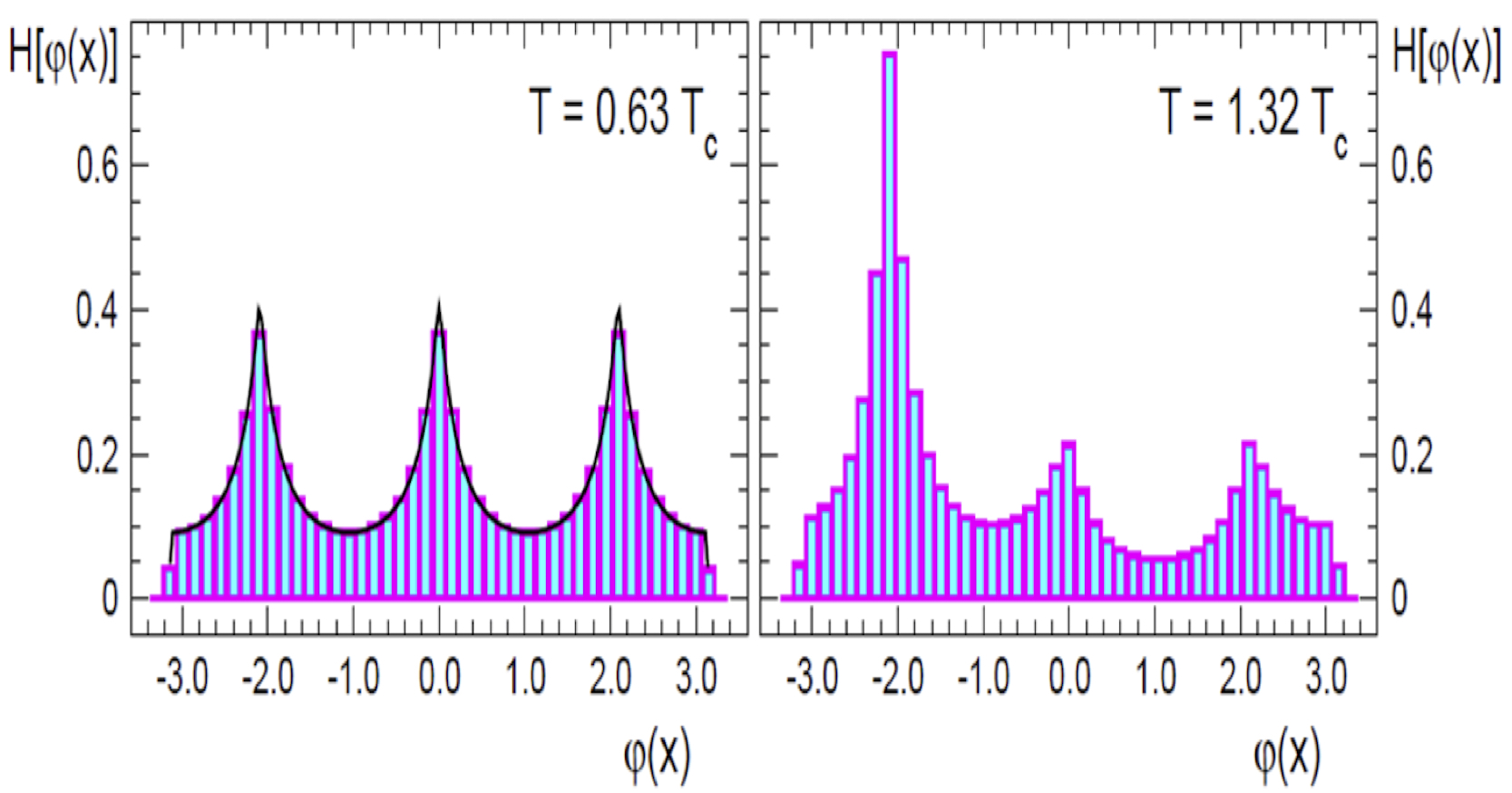}
\caption{Histograms fro the distributions of the phase $\phi(x)$ of the local loops $\langle L(\vec{x}) \rangle$. Left side show the distribution below $T_c$ and right side show the de-confined phase \cite{63,64}.}

\label{fig4}
\end{figure}

In SU(3) gauge theory, spatial clusters can be identified as those where the local Polyakov loops $\langle L(\vec{x}) \rangle$ have values close to some element of the center.
The elements of the center group $Z_3$, are a set of three phases $(0,2\pi/3,-2\pi/3)$ \cite{63}.
Below $T_c$ ($\langle L(\vec{x}) \rangle=0$), the values of $\langle L(\vec{x}) \rangle$ are grouped around there three phases, show three pronounced peaks located at the center phases.
Above $T_c$ ($\langle L(\vec{x}) \rangle\neq0$), the distribution changes: One of the peaks grows and the other two shrink. A spontaneous breaking symmetry occurs, which leads to a non- vanishing $\langle L(\vec{x}) \rangle$, as is shown in Fig.~\ref{fig4}.
Spatial clusters can be defined grouping the sites with a very similar value of $\langle L(\vec{x}) \rangle$.
The weight of the largest cluster increases sharply at $T=T_c$, indicating that the cluster percolates.
Therefore in the pure SU(3) theory the de-confinement transition is a percolation phase transition (of second order).

In high energy collisions, we expect that color strings are formed between the projectile and target partons.
These color field must have a small transverse size due to confinement.
In this way, the strings, in the transverse plane, are small discs in the surface of the collisions.
As the number of strings grows with energy and centrality degree of the collision, the strings start to overlap forming clusters which eventually percolate.
The phenomenological consequences in relation to SPS, RHIC, and LHC, pp, pA and AA data are the main subject of this brief review. A more extended version can be found in Ref. \cite{67}.

\section{String percolation}

\subsection{String models}

\begin{figure}
\centering
\includegraphics[scale=.1]{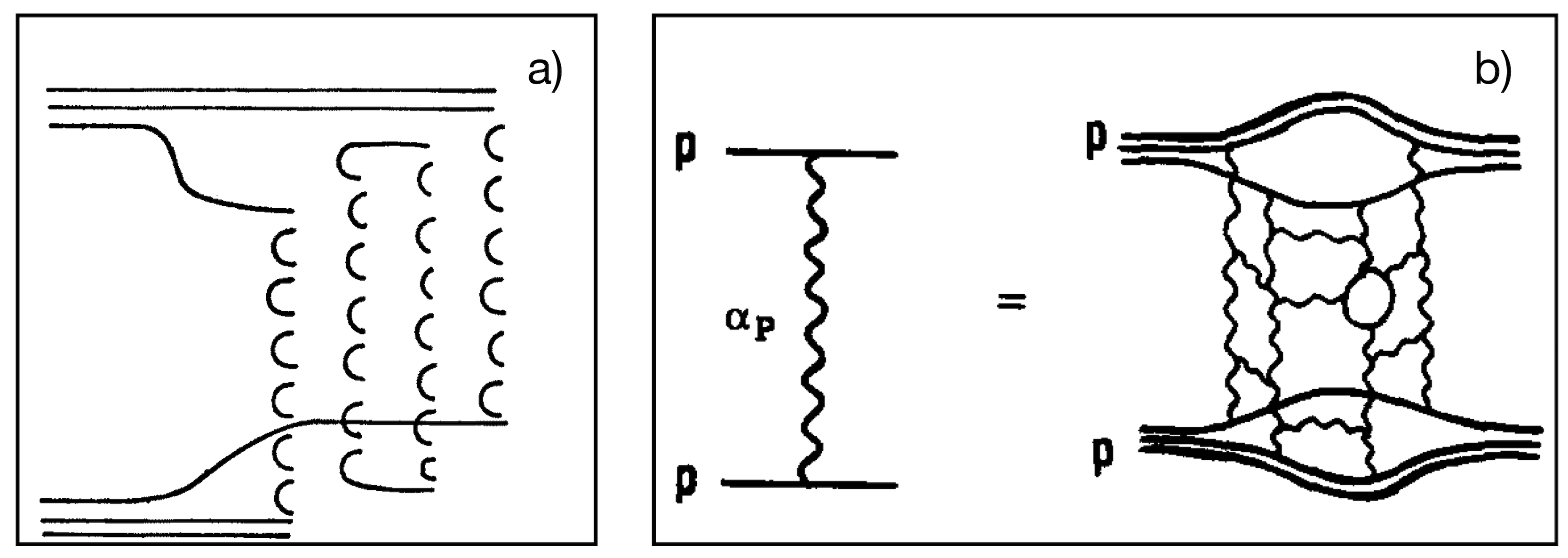}
\caption{(a)Two cut Pomeron diagram (four chain) for proton-proton collisions. (b): Single Pomeron exchange and its underlying cylindrical topology. This is a dominant contribution to proton-proton elastic scattering at high energies \cite{68}.}

\label{fig5}
\end{figure}

The basic ingredient of the string percolation are the strings. Although there are differences, most of them coincide in basic postulates as the number of strings and its dependence on energy and centrality, which is taken from the Glauber-Gribov Model.
We will concentrate in models with color exchange between projectile and target as the Dual Parton Model (DPM)  \cite{68,69,70}, Quark Gluon String Model (QGSM) \cite{71}, Venus and EPOS \cite{72}.
They are based on the $1/N_c$ QCD expansion and its terms are in correspondence with the ones of the Gribov-Reggeon calculus.
They have been extensively compared to the experimental data ISR, SPS, and Fermilab obtaining an overall agreement \cite{68}.
In DPM or QGS the multiplicity distribution dN/dy of pp collisions is given by fragmentation of 2k strings
\begin{equation}
\frac{dN^{pp}}{dy}=\frac{1}{\sigma}\sum \sigma_k [N_k^{qq-q}(s,y)+N_k^{q-qq}(s,y)+(2k-2)N_k^{q-\bar{q}}],
\label{eq12}
\end{equation}
where $N_k^{qq-q}$ and $N_k^{q-qq}$ are the inclusive spectra of hadrons produced in the strings stretched between a valence diquark of the projectile(target) and a quark of the target(projectile) and $N_k^{q-\bar{q}}$ are the inclusive spectra of the strings stretched between sea quarks and antiquarks. A schematic representation of this process is shown in Fig.~\ref{fig5} (a). 
In this figure are shown four chains, two between quark-diquark and two corresponding to quark-antiquark. Each chain corresponds to the fragmentation into $q-\bar{q}$ pairs of one string, stretched between the quark (diquark) of the projectile and the diquark (quark) of the target or from the quark and diquark from the sea. The leading term corresponds to two strings stretched between a valence quark (diquark) and a valence diquark (quark). Via unitarity the modulus square of the leading diagram corresponds to the pomeron as is shown in Fig.~\ref{fig5} (b). In this figure is shown a net of gluons which are accompanying to the quarks, which can be seen in this picture as constituent quarks. In Fig.~\ref{fig5} (b) is also shown that the pomeron corresponds to a cylinder topology.
The single particle of each string can be obtained by folding the momentum distribution of the partons at the end of the string with the fragmentation function of the string
\begin{equation}
N_k^{qq-q}(s,y)=\int_0^1 dx_1\int_0^1 dx_2\rho_k(x_1)\rho_k(x_2)\frac{dN^{qq-q}}{dy} (y-\bar{\Delta},s_s),
\label{eq13}
\end{equation}
where $\sqrt{s_s}$ is the invariant mass of the string $s_s=sx_1x_2$ and $x_1$ and $x_2$ are the light cone momentum fractions of the partons of the end of the string. 
$\bar{\Delta}$ is the rapidity shift necessary to go from the pp center of mass to the center of mass of one string
\begin{equation}
\bar{\Delta}=\frac{1}{2}\log\left( \frac{x1}{x2}  \right).
\end{equation}
The momentum distribution used for the valence quarks, valence diquarks, sea quarks and antiquarks are $x^{-1/2}$, $x^{-1}$ and $x^{-3/2}$, respectively.
In general the distribution of $2k$ partons in the proton is
\begin{equation}
\rho_k(x_1,x_{2k},x_2,x_3,x_3,\dots,x_{2k-1})=C_k^\rho x_1^{-1/2}x_2^{-1}\dots x_{2k-1}^{-1}x_{2k}^{1/2}x_{2k}^{1/2}\delta\left(1-\sum_{i=1}^{2k}x_i \right),
\end{equation}
where $C_k^\rho$ is obtained by normalizing $\rho_k$ to unity.
Due to these distributions the $N_k^{qq-q}$ and $N_k^{q-qq}$ are long in rapidity extension and centered at a point shifted with respect to the center of mass and the $N_k^{q-\bar{q}}$ string are short and centered at the center of mass.
For the fragmentation functions different ways are used, in string percolation the strings fragment according the Schwinger mechanism, such as in the Lund string.
In Eq.~\eqref{eq12} $\sigma_k$ is the cross section for producing $2k$ strings resulting from cutting $k$ pomerons.
As the pomeron has the topology of the cylinder its cutting give rise to two strings (See Fig.~\ref{fig5} (b)).
Using the AGK cutting rules \cite{74}, the cross section is calculated as follows
\begin{equation}
\sigma_k=\frac{8\pi g\exp(\Delta y)}{kz} \left[ 1-\exp(-z) \sum_{l=0}^{k-1}\frac{z^l}{l!} \right],
\end{equation}
where
\begin{equation}
z=\frac{2gC\exp(\Delta y)}{R^2+\alpha'y},
\end{equation}
and $g$ is the coupling of the pomeron to the proton, $\alpha'$ and $1+\Delta$ are the slope and the intercept of the pomeron trajectory, respectively, and $C$ is a parameter describing the inelastic diffractive states.
Summing over $k$, we obtain the total cross section
\begin{equation}
\sigma_{tot}=\exp(\Delta y) \sum_{k=0}^\infty \sum_{l=k,l>0}^\infty \left( -\frac{z}{2} \right)^{l-1} \frac{8\pi g}{l!} \left[ \delta_{k0}+(-1)^{1-k}2^{l-1} \binom{l}{k} \right].
\end{equation}
The rise of $dN/dy$ with energy is due mainly to the short strings, whose number grows with energy.
On the other hand, outside the central rapidity region, there is not contribution of short strings and the rise with energy is much slower giving rise to the approximate limiting behavior.
Assuming a Poisson distribution for cutting $k$ pomerons
\begin{equation}
P_k(n)=\frac{(kN)^n}{n!}\exp(-kN),
\end{equation}
where $N$ is the mean multiplicity production when cutting one pomeron, therefore the multiplicity distribution is 
\begin{equation}
P_k(n)=\sum_k\omega_kP_k(n),
\end{equation}
where $\omega_k=\sigma_k/\sigma$.
Usually, $\langle n \rangle P(n) $ is plotted as a function of $n\langle n \rangle$.
When the result is independent of energy, one has the known Kobe-Nielsen-Olsen scaling (KNO), which is violated at SPS Fermilab, RHIC and LHC.
The reason for that in DPM is due to the contribution of the short strings that they contribute mostly at high multiplicities pushing upwards the high multiplicity tail of the distributions.
The increase with $s$ of the short strings contributions is due to the increase of the invariant mass of the short strings, formed between quarks and antiquarks of the sea, and to the $s$-dependence of the weights.
DPM can be generalized to hA and AA collisions in the following way \cite{75}:
Consider a collision with $N_A$ participants nucleons of A, $N_B$ participant nucleons of B and a total number of $N_c$ collisions.
In this configuration are produced $2N_c$ strings, of these $2N_A$ are stretched between valence quarks and valence diquarks ($q_v^A-qq_v^B$) and ($qq_v^A-q_v^B$).
The remaining $N_B-N_A$ valence quarks and diquarks of B have no valence partner of A and have to form $2N_B-2N_A$ strings with sea quarks and antiquarks of A ($q_s^A-qq_v^B$) and ($\bar{q}_s^A-q_v^B$). The remaining $2N_c-2N_B$ strings are formed between sea quarks and antiquarks of A and B ($q_s-\bar{q}_s$)
\begin{eqnarray}
\frac{dN^{AB}}{dy}&= & \frac{1}{\sigma_{AB}} \sum_{N_A,N_B,N_c}  \sigma^{AB}_{N_A,N_B,N_c} \theta(N_B-N_A)\Big[N_A\left( N^{qq_v^A-q_v^B}(y) +N^{q_v^A-qq_v^B}(y)\right)+\nonumber \\
 & & (N_B-N_A) \left( N^{\bar{q}_s^A-q_s^B}(y) +N^{q_s^A-qq_v^B}(y)\right)+\nonumber\\
  & & (N_c-N_B)\left(N^{q_s^A-\bar{q}_s^B}(y)+N^{q_s^A-q_s^B}(y) \right)\Big]+\mathrm{sym}(N_A\leftrightarrow  N_B),
\label{eq21}
\end{eqnarray}
where $\sigma^{AB}_{N_A,N_B,N_c}$ is the cross section for $N_c$ inelastic nucleon-nucleon collisions involving $N_A$ and $N_B$ nucleons of A and B, respectively. This cross section have been studied extensively \cite{76,77}.
The inclusive spectra, as in the pp case, are given by a convolution of momentum distribution and fragmentation functions. In the case of A=B, we have approximately
\begin{eqnarray}
\frac{dN^{AA}}{dy} & \approx & \langle N_A \rangle (2N^{qq-q_v}(y)+(2\langle k \rangle -2)N^{q_s-\bar{q}_s}(y))+\nonumber \\
 & & (\langle N_c \rangle-\langle N_A \rangle)2\langle k \rangle N^{q_s-\bar{q}_s}(y),
 \label{eq22}
\end{eqnarray}
where we have introduced the possibility of having $k$ multiple scattering in the individual nucleon-nucleon collisions, which was neglected in Eq.~\eqref{eq21}. 
Notice, that there is not any reason to assume that the term proportional to $N_c$ is due to hard collisions. 
There are many soft collisions included in this term. 
In the central rapidity region we have $2Nk$ strings which for heavy nuclei collisions and high energy is very large number, even larger than 1500. 
Due to that, we expect interactions between them and they will not fragment in an independent way.

In the case of pA collisions, the Eqs.~\eqref{eq21} and \eqref{eq22} transform into
\begin{eqnarray}
\frac{dN^{pA}}{dy} &=& \frac{1}{\sigma_{pA}}\sum_{N_A}\sigma_{N_A}^{pA}\Big[ \left( N^{qq^p-q^A_v}(y)+N^{q^p_v-qq^A}(y) \right)\nonumber\\
 & &+ (N_A-1) \left(N^{\bar{q}_s^p-q^A_s}(y)+N^{q_s^p-qq^A}(y) \right)\Big],
\end{eqnarray}
where $N_A$ matches with the number of collisions.

\subsection{String fusion and percolation}

\begin{figure}
\centering
\includegraphics[scale=.25]{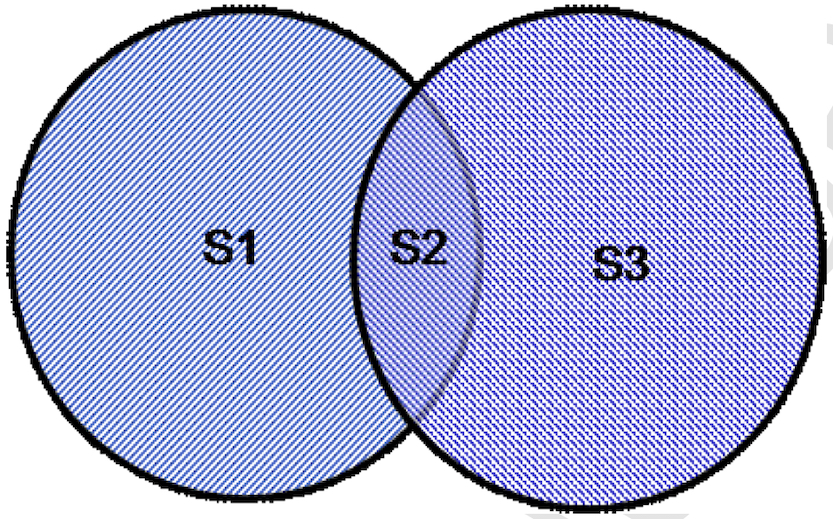}
\caption{Projections of two overlapping strings onto the transverse plane.}

\label{fig6}
\end{figure}

As we have said before, at large energy we expect that the strings overlap in the transverse plane. 
The transverse space occupied by a cluster of overlapping strings splits into a number of areas with different number of strings overlap, including areas where no overlapping takes place. 
In each area color field coming from the overlapping strings add together. 
As a result, the cluster is split in domains with different color strength. 
One may assume that emission of $q\bar{q}$ pairs in the domains proceeds independently, governed by the strength of the color field (string tension) of the corresponding domain. 
Evidently, these new formed strings domains have not only different color strength.
 Other assumption that one may do is that emission of $q\bar{q}$ pairs in the domains proceeds independently, governed by the strength of the color field (string tension) of the corresponding domain. Evidently, these new formed string domains have not only different color field but also different transverse area.
As an example, let us consider a cluster of two partially overlapping string as it is shown in Fig.~\ref{fig6}.
 In this scenario, a simple string with transverse area $S_1$ emits partons with transverse momentum distribution
\begin{equation}
\frac{d\sigma}{dyd^2p}=C\exp\left( -\frac{m_T^2(p_T)}{t_1} \right),
\label{eq23}
\end{equation}
where $t_1$ is the tension and $m_T^2=m+p_T^2$, being $p_T$ and $m$ the transverse momentum and the mass of the emitted parton.
The tension, according to the Schwinger mechanism, is proportional to the field and thus to the color charge of the ends of the string \cite{78,79,80,81,82}, which we denote by $Q_0$.
The mean transverse momentum squared is $\langle p_T^2\rangle_1=t_1$ and proportional to $Q_0$.
We denote the mean multiplicity of produced particles by string per unit of rapidity as $\mu_1$ which is also proportional to the color charge.
In Fig.~\ref{fig6}, we have the overlapping of two strings which partially overlap in the area $S^{(2)}$ (region 2 in the figure), so that $S^{(1)}=S_1-S^{(2)}$ is the area in each string not overlapping with the other, where $S_1$ is the transverse area of a single string.
The color density of a simple string is $q=Q_0/S_1$.
Then, the color in each of the non overlapping areas will be
\begin{equation}
Q_1=qS^{(1)}=Q_0S^{(1)}/S_1,
\end{equation}
and in the overlapping area each string will have color
\begin{equation}
\bar{Q}_2=qS^{(2)}=Q_0S^{(2)}/S_1,
\end{equation}
The total color in the overlap area will be a vector sum of the two overlapping colors $qS_2$.
In this summation the total color charge should be conserved \cite{50,51}. 
Thus $Q_2^2=(\vec{Q}_{ov}+\vec{Q}'_{ov})^2$, where $\vec{Q}_{ov}$and $\vec{Q}'_{ov}$ are the two vector colors in the overlap area.
Since the colors in the two strings may generally be oriented in arbitrary directions respective to one another, the average of $\vec{Q}_{ov}\vec{Q}'_{ov}$ is zero, then $Q_2^2=\vec{Q}^2_{ov}+\vec{Q}'^2_{ov}$, which leads to
\begin{equation}
Q_2=\sqrt{2} q S^{(2)}=\sqrt{2} Q_0 S^{(2)}/S_1.
\end{equation}
Notice that due to the vector nature, the color in the overlap is less than the sum of the two overlapping colors.
This effect has important consequences concerning the saturation of multiplicities and the rise of the mean transverse momentum with multiplicity which we will study in the next section.
Thus, assuming independent emission from the three regions of Fig.~\ref{fig6}, we obtain for the multiplicity weighted by the multiplicity for a single string ($\mu_1$)
\begin{equation}
\mu/\mu_1=2(S^{(1)}/S_1)+\sqrt{2}(S^{(2)}/S_1),
\end{equation}
and for the mean transverse momentum squares (we divide the total transverse momentum squared by the multiplicity)
\begin{equation}
\frac{\langle p_T^2 \rangle}{\langle p_T^2 \rangle_1}=\frac{2(S^{(1)}/S_1)+\sqrt{2}\sqrt{2}(S^{(2)}/S_1)}{2(S^{(1)}/S_1)+\sqrt{2}(S^{(2)}/S_1)}=\frac{2}{2(S^{(1)}/S_1)+\sqrt{2}(S^{(2)}/S_1)},
\end{equation}
where we have used the property $S^{(1)}+S^{(2)}=S_1$. Generalizing to any number $N$ of overlapping strings, we have
\begin{eqnarray}
\frac{\mu}{\mu_1}&=&\sum_i \sqrt{n_i}(S^{(i)}/S_1),\\
\frac{\langle p_T^2 \rangle}{\langle p_T^2 \rangle_1} &=&\frac{\sum_i (S^{(i)}/S_1)}{\sum_i \sqrt{n_i}(S^{(i)}/S_1)}=\frac{N}{\sum_i \sqrt{n_i}(S^{(i)}/S_1)},\label{eq30}
\end{eqnarray}
where the sum runs over all individual overlaps of $n_i$ strings having areas $S^{(i)}$.
We have used the identity $\sum_iS^{(i)}=NS_1$. These equations are not easy to apply because we have to identify all individuals overlaps of any number of strings with their areas.
However one can avoid these difficulties realizing that one can combines all terms with a given number of overlapping strings $n_i=n$ into a single term, which sums all such overlaps into a total area of exactly $n$ overlapping strings $S_n^{Tot}$.
Then, one can write
\begin{eqnarray}
\frac{\mu}{\mu_1}&=&\sum_{n=1}^N \sqrt{n}(S_n^{Tot}/S_1),\\
\frac{\langle p_T^2 \rangle}{\langle p_T^2 \rangle_1} &=&\frac{N}{\sum_{n=1}^N (S_n^{Tot}/S_1)}.
\end{eqnarray}
The total area can be easily computed in the thermodynamic limit. One finds that the distribution of overlap strings over the total surface $S$ in the variable $n$ is Poissonian with mean $\rho=NS_1/S$, which corresponds to the filling factor in the percolation context. Therefore, the fraction of the total area covered by strings will be $1-\exp(-\rho)$. Note that the multiplicity in Eq.~\eqref{eq30} is damped by a factor
\begin{equation}
F(\rho) =\frac{\mu}{N\mu_1}=\frac{\sqrt{n}}{\rho}=\sqrt{\frac{1-\exp(-\rho)}{\rho}}.
\label{eq37}
\end{equation}
Finally, we can write for the mean values
\begin{eqnarray}
\mu & = & NF(\rho)\mu_1, \label{eq38}\\
\langle p_T^2 \rangle & = & \langle p_T^2 \rangle_1/F(\rho) \label{eq39}.
\end{eqnarray}
In the rest of this review, these last equations will be used extensively.

\subsection{Quenching of the low $p_T$ partons}\label{quenching}
The $p_T$ distribution of the partons from the decay of a cluster of strings is given by Eq.~\eqref{eq23}, where the tension is now scaled by $1/F(\rho)$, then it is computed as $t=t_1/F(\rho)$.
The parton will be emitted in different azimuthal direction and have to travel paths of different longitudes before they of out and are observed. 
Parton going through the overlap meet stronger field than those going only through the field of a simple string. 
In this way the partons loose their energy passing through the field and the observed distribution will depend on their azimuthal angle even if initially they were emitted isotropically.
Radiative energy loss has been extensively studied in QCD for a parton passing through a quark gluon plasma medium \cite{83,84}.
In our case the situation is different and is more similar to a charge particle moving in an external electromagnetic field. 
The corresponding force causes a loss of energy, which is given by \cite{85}
\begin{equation}
\frac{dp(x)}{dx}=-0.12e^2(eEp(x))^{2/3},
\end{equation}
where $E$ is the external electric field. This equation leads to the quenching formula
\begin{equation}
p_0(p)=p\left(1+\gamma p^{-1/3}t_1^{2/3}\right)^3,
\label{eq42}
\end{equation}
where $eE/\pi=t_1$ can be identified as the longitude of the path travelled by the parton.
The quenching coefficient is known in QED (is very small) but in our case has to be adjusted to the experimental data and it turns out to be very small of the order of $10^{-2}$.
Retaining the first term of the equation we obtain
\begin{equation}
p_0-p=3\gamma p^{2/3}t_1^{2/3},
\end{equation}
which will be used to compute the harmonic of the $p_T$ distribution. 
Notice that due to the smallness of the quenching parameter the effect of the quenching is only felt in the azimuthal distribution, not in the $p_T$ distribution, once the azimuthal angel is integrated. 
The use of QED formulas for the QCD case may raise some doubts.
However, it has been shown that at least in $N = 4$ SUSY Yang Mills case, the expression is essentially the same \cite{86}.

\subsection{Multiplicity distributions}
The multiplicity distributions in the DPM of QGSM in pp and AA collisions are given by Eqs.~\eqref{eq12}-\eqref{eq13} and Eqs.~\eqref{eq21}-\eqref{eq22}, respectively.
However, as the energy or centrality of the collision increases one expects interaction among strings. 
As discussed before, due to the randomness of the color field in color space non-abelian field) the resulting color field in a cluster of $n$ overlapping strings is only $\sqrt{n}$ times the strength of the color field of a single string, giving rise to a suppression of the multiplicity of particles produced by the cluster. 
The same reason lies at the origin of the enhancement of the mean $p_T$.
According to Eq.~\eqref{eq37}, the multiplicity distribution in pp collisions in the central rapidity region is given by
\begin{equation}
\frac{dN^{pp}}{dy}=F(\rho_p)N^s_p\mu_1,
\end{equation}
where $N^s_p$ is the number of string in the central rapidity region.
In AA collisions the number of string stretched between the sea quark and antiquarks in the central rapidity region is proportional to $N_A^{4/3}-N_A$, which is the total number of nucleon-nucleon collisions.
Hence, the total number of strings in a central heavy ion collisions is very large. 
However, each string must have a minimum of energy to be produced and decay subsequently into particles. 
On the other hand, the total energy available.
In the collision grows as $A$, whereas the number of strings as $A^{4/3}$ in the central collisions.
Therefore at not very high energy (for instance RHIC energies), the energy is not sufficient to produce such huge number of strings. In order to take into account this energy conservation effect one may reduce the number of sea quark and antiquarks changing \cite{87}
\begin{equation}
N^{4/3}_A\to N^{1+\alpha(\sqrt{s})}_A,
\end{equation}
where
\begin{equation}
\alpha(\sqrt{s})=\frac{1}{3}\left( 1-\frac{1}{1+\ln(\sqrt{s/s_0}+1)} \right).
\end{equation}
Here, the parameter $s_0$ marks the energy squared above which energy conservation effects become small and $\alpha\to3$. One thus can write
\begin{equation}
\frac{dN^{AA}}{dy} \sim N_A(N^{\alpha(\sqrt{s})}_A-1)\frac{dN^{pp}}{dy}.
\label{eq47}
\end{equation}
Taking into account the interaction of strings we can write a closed formula for the multiplicity distribution in AA in terms of the multiplicity distribution of pp, namely \cite{87}
\begin{equation}
\left. \frac{1}{N_A}\frac{dN}{dy} \right|_{y=0} = \left. \frac{dN^{pp}}{dy}  \right|_{y=0} \left( 1+\frac{F(\rho_{N_A})}{F(\rho_p)}(N^{\alpha(\sqrt{s})}_A-1) \right),
\end{equation}
where
\begin{eqnarray}
\rho_{N_A} & =& \rho_p N^{\alpha(\sqrt{s})+1}_A\frac{S_1}{S_{N_A}},\\
\rho_p &= & N_p^s \frac{S_1}{S_{p}},
\end{eqnarray}
and $S_{N_A}$ is the transverse area of the collision formed when there are $N_A$ participant nucleons of the projectile and $N_A$ participant nucleons of the target.
Note that $S_{N_A}$ depends on $N_A$ and A.

\begin{figure}
\centering
\includegraphics[scale=0.2]{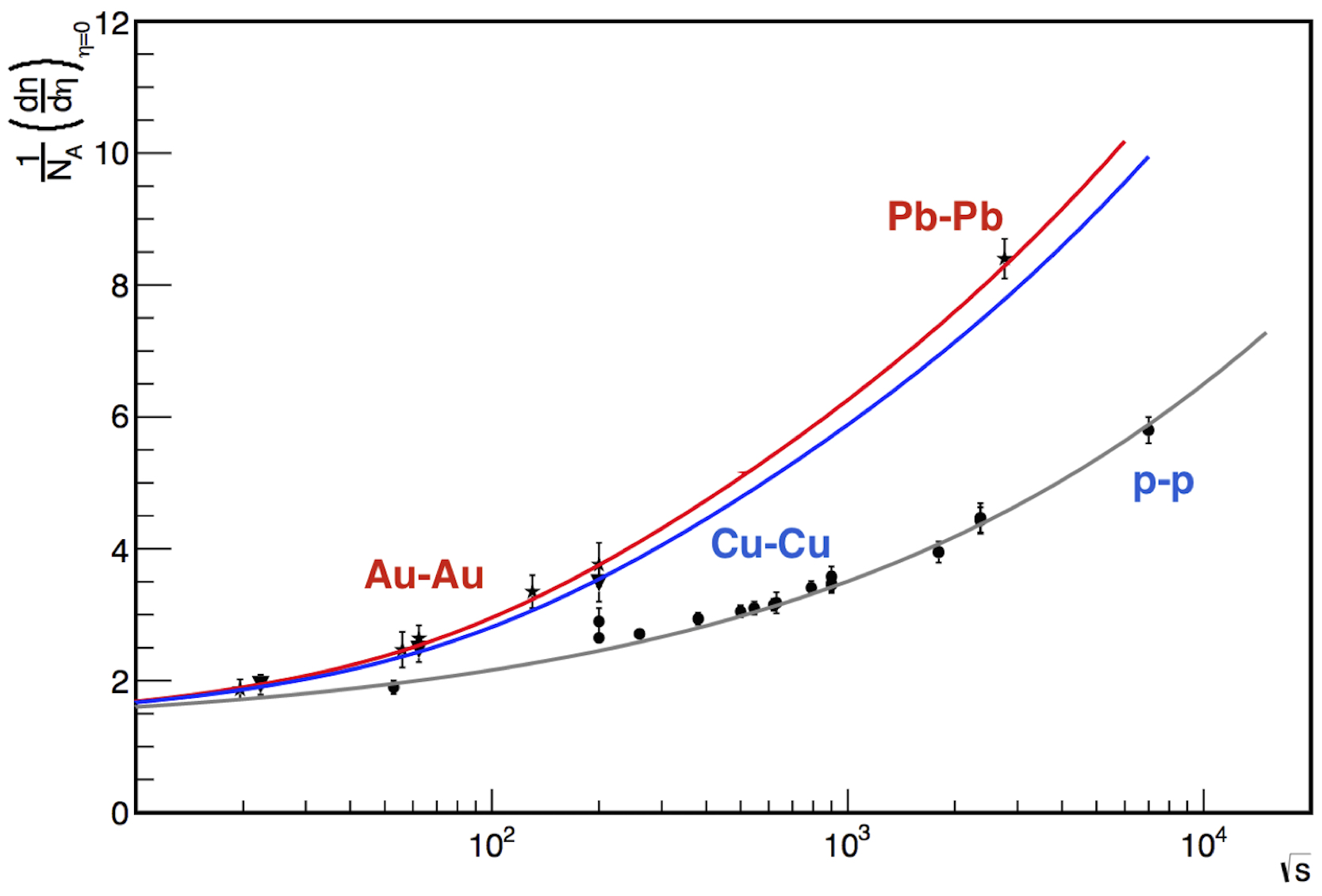}
\caption{Comparison of the evolution of the mid-rapidity multiplicity with energy from the CSPM and data for pp and A-A collisions. Lines are from the model for pp (gray), Cu-Cu (blue) and red lines for Au-Au/Pb-Pb \cite{87,88,89,90,91}.}

\label{fig7}
\end{figure}

\begin{figure}
\centering
\includegraphics[scale=0.25]{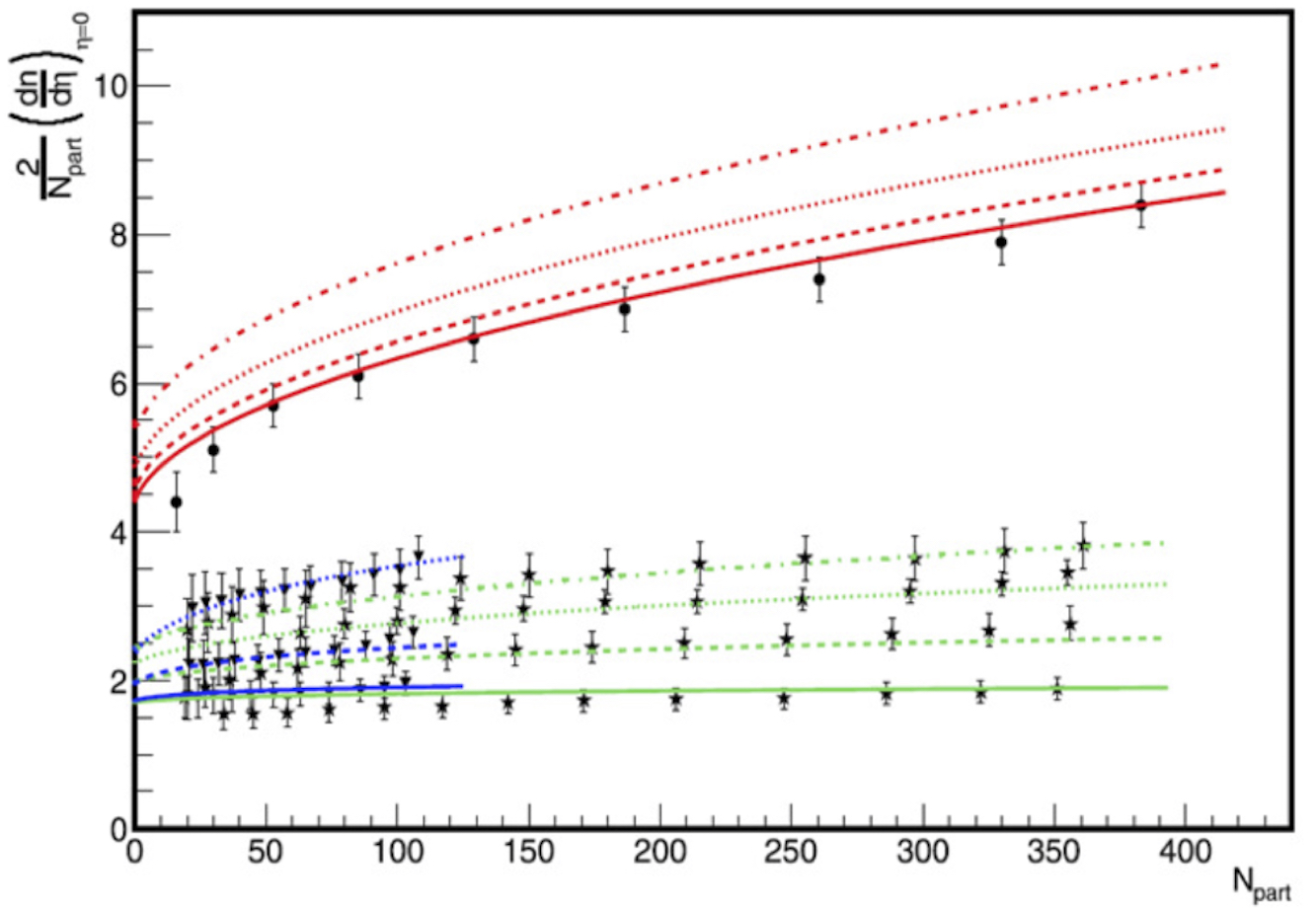}
\caption{Multiplicity dependence on centrality ($N_{part}$). Cu-Cu (triangles), Au-Au (stars) and Pb-Pb (circles).Curves represent the model calculations. Blue line for Cu-Cu, green line for Au-Au and red for Pb-Pb.}

\label{fig8}
\end{figure}

Moreover, the dependence of the multiplicity on $\sqrt{s}$ is full specified, once the average number of strings $N$ in a pp collision is known. At low energies it is 2, growing as
\begin{equation}
N_p^s=2+4\left(\frac{r_0}{R_p} \right) ^2 \left( \frac{\sqrt{s}}{m_p} \right)^{2\lambda}.
\end{equation}
Notice that here a single parameter $\lambda$ describes the rise of the multiplicity with energy for both pp and AA, even though in central AA collision the multiplicity increase faster than in pp collisions due to the energy dependent factor $\alpha$, arising from energy conservation. A fit to pp collisions data in the range $53<\sqrt{s}<7000$ GeV and to AA collisions (Au-Au, Cu-Cu and Pb-Pb) at different centralities for $19.6<\sqrt{s}<2760$ GeV has been done. The values obtained for the two parameters are $\sqrt{s_0}=245$GeV and $\lambda=0.201$. 
Figure~\ref{fig7} shows a comparison of the energy dependence results with data for pp \cite{87,88,89} and central Cu-Cu \cite{90} and for Au-Au and Pb-Pb \cite{91}.
The results of the dependence of the the multiplicity per participant nucleon on the number of participants is shown in Fig.~\ref{fig8} together with the experimental data Cu-Cu, Au-Au and Pb-Pb at different energies.

\begin{figure}
\centering
\includegraphics[scale=0.18]{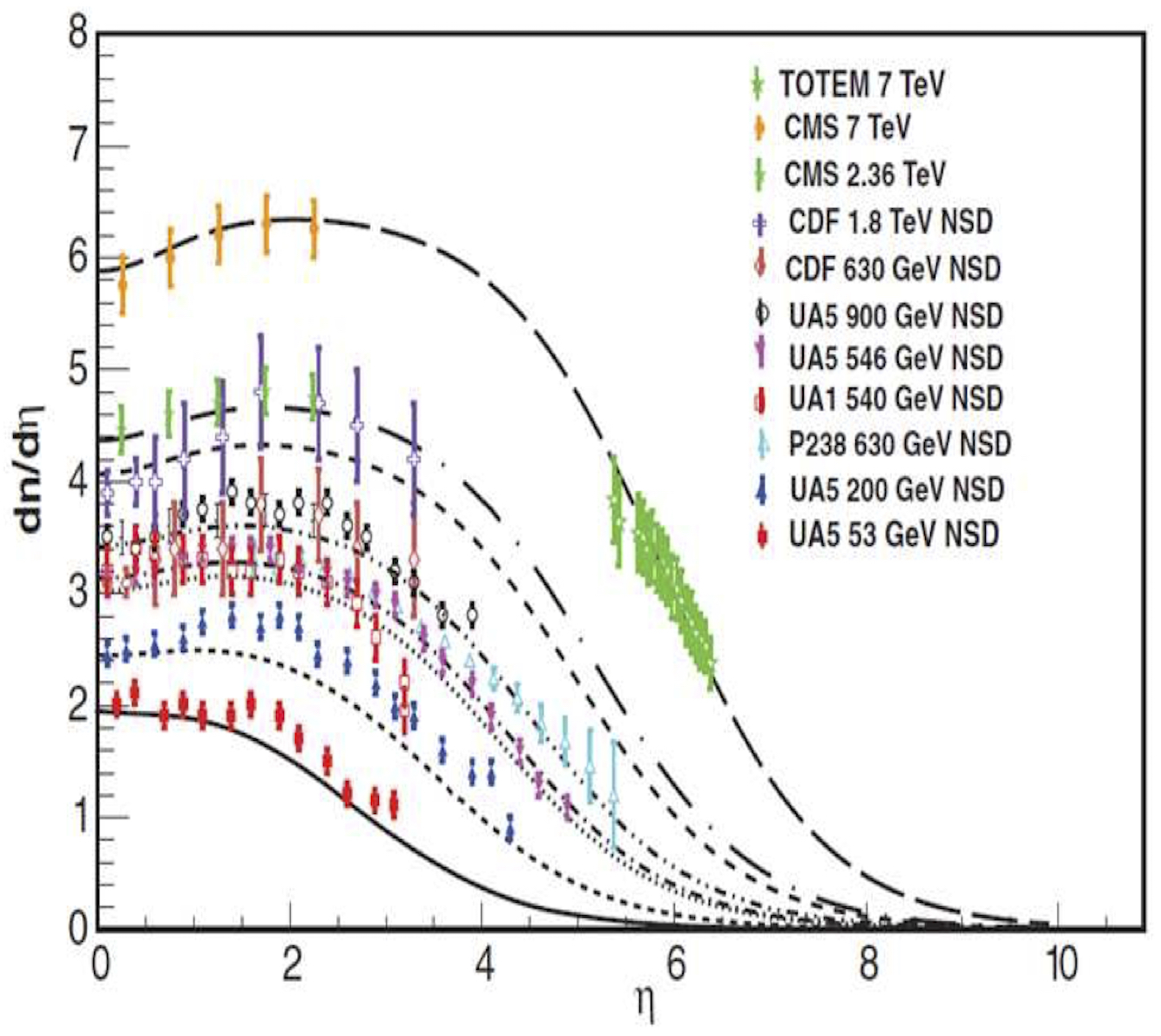}
\caption{Comparison of results from the evolution of $d_{n_{ch}}/d\eta$ with dependence on pseudorapidity for pp collisions at different energies (lines).}

\label{fig9}
\end{figure}

\begin{figure}
\centering
\includegraphics[scale=0.25]{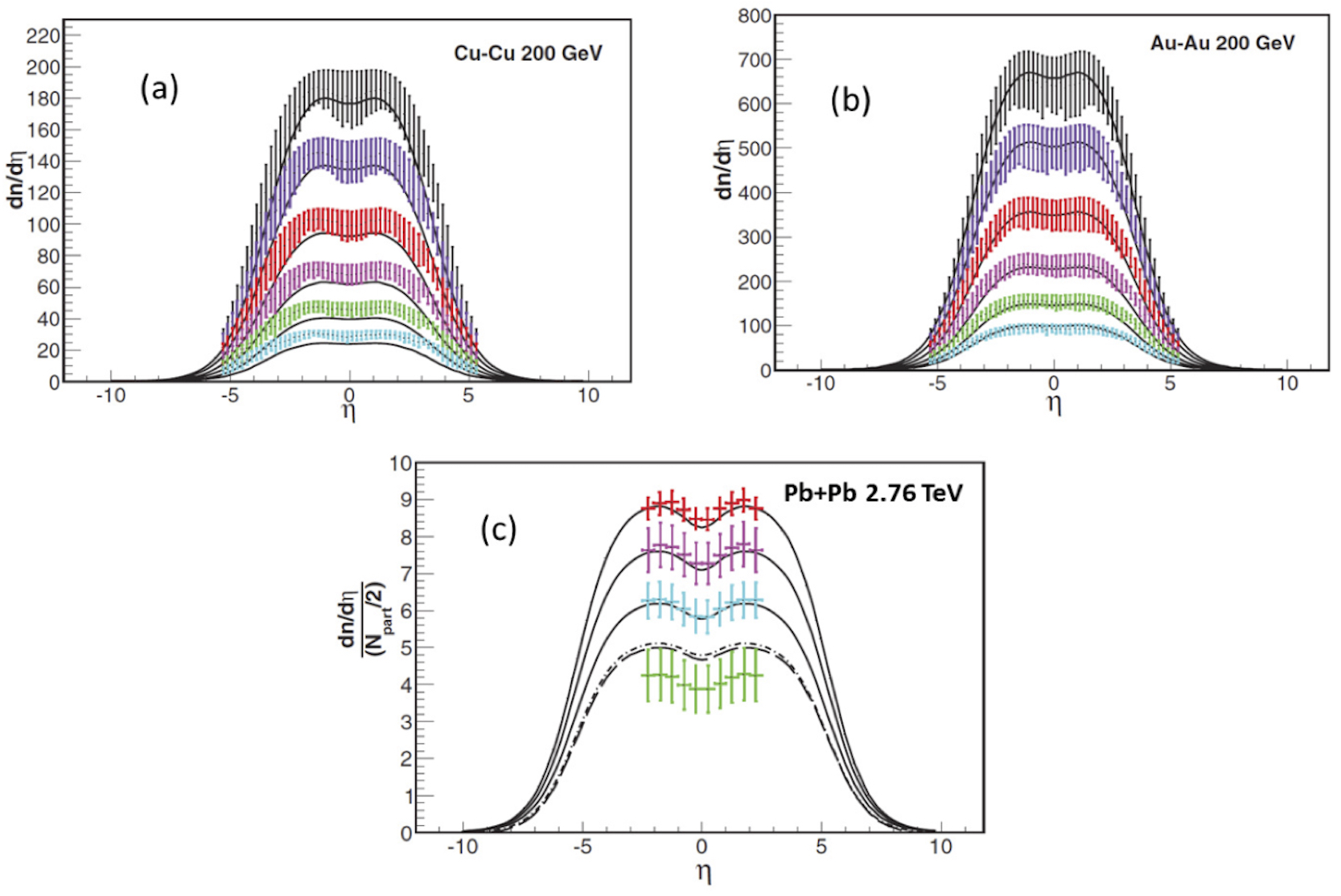}
\caption{Comparison of results from the evolution of $dn/d\eta$ with dependence on pseudorapidity for (a).Cu-Cu at 200 GeV, (b) Au-Au 200 GeV. Plot (c) shows $\frac{dn_{ch}}{\eta} \frac{1}{N_{part}/2}$ for Pb-Pb collisions at 2.76 TeV.}

\label{fig10}
\end{figure}

\begin{figure}
\centering
\includegraphics[scale=.2]{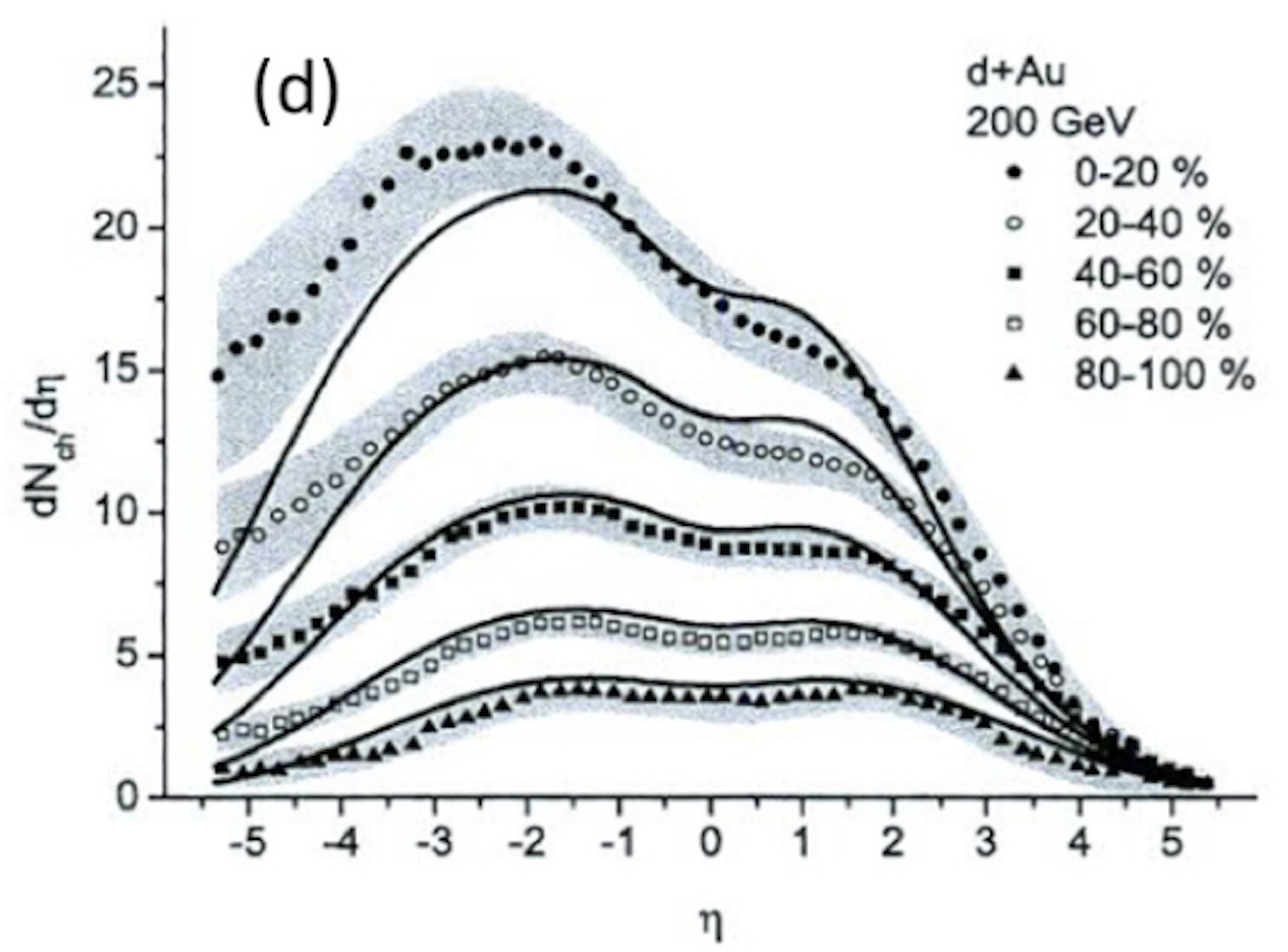}
\caption{Comparison of results from the evolution of $dn/d\eta$ with dependence on pseudorapidity at different centralities for d+Au collisions at 200 GeV.}

\label{fig11}
\end{figure}

The evolution outside the central rapidity region has been studied extensively extending Eq.~\eqref{eq47} to all rapidities  \cite{92,93,94,95,96,97}. 
The limiting fragmentation property is not satisfied exactly. 
In Fig.~\ref{fig9}, we show the results together with the experimental data for pp collisions at all rapidities at different energies \cite{96,97} and in Fig. \ref{fig10} the results for Cu-Cu, Au-Au \cite{98} and Pb-Pb \cite{99} together the experimental data. In Fig.~\ref{fig11}, we compare the results \cite{97,98,99,100} for d-Au collisions together the experimental data. 
A good description of all experimental data is obtained.

The behavior of $dN/dy$ in pp and AA collisions is very similar to the Glasma picture of CGC expressed in Eq.~\eqref{eq1}.
In fact as the saturation momentum squared $Q_s^2$ behaves like $N_A^{1/3}$, and $R_A^2$ as $N_A^{2/3}$, the multiplicity per participant is almost independent of $N_A$ and only a weak dependence arises from the logarithmic dependence of the running coupling constant $\alpha_s(Q_s)\sim 1/\log(N_A)$. 
In  percolation the multiplicity per participant is also almost independent of $N_A$ and the only dependence arises from the factor $1-\exp(-\rho)$, which also grow weakly with $N_A$.
 Concerning the energy dependence, this is given by $Q_s^2$, which behaves like $s^\lambda$, so that, the same behavior than in percolation. 
There is an extra energy dependence in CGC due to $1/\alpha_s$, which corresponds in percolation again to the factor $1-\exp(-\rho)$. 
Since both are measure of the fraction occupied area by color fields, it is not surprising the correspondence between $1/\alpha_s$, the occupying number of gluons, and $1- \exp(-\rho_{N_A} )$. 

\subsection{Multiplicity and transverse momentum distributions}\label{mult}

\begin{figure}
\centering
\includegraphics[scale=.2]{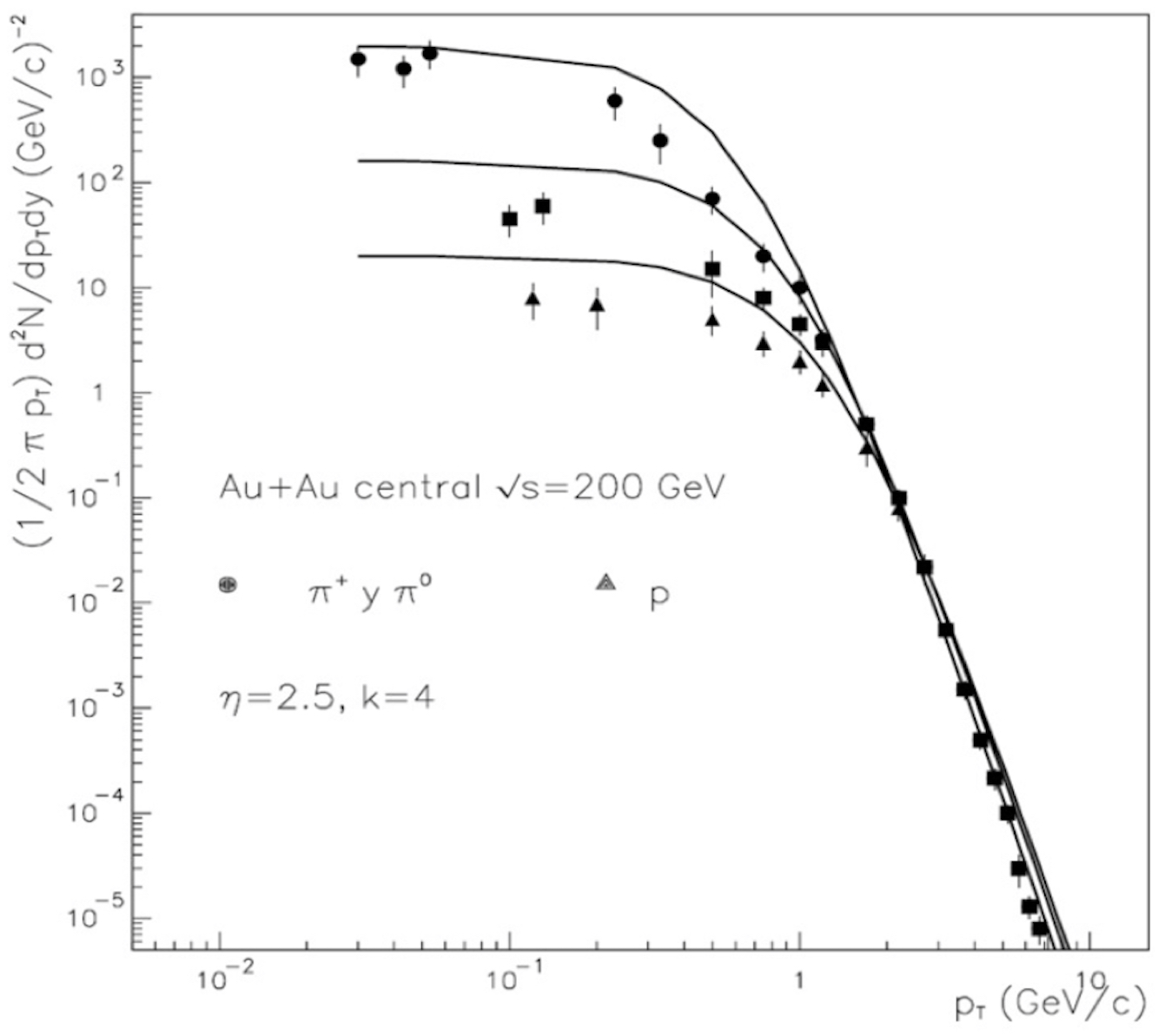}
\caption{Experimental PHOBOS data on low pt distributions for pions, kaons and protons along with our results for central Au-Au collisions at $\sqrt{s}=$200 GeV.}

\label{fig12}
\end{figure}

Let us start considering a set of overlapping strings which depend both on the number of strings and the overlapping area, which combine to give an average multiplicity $N$. 
We may characterize the different overlaps just by the average multiplicity that combines both the number of collisions and the area. With a lot of overlapping strings $N$ will change practically continuously.
We can introduce a probability $W(N)$ to have overlaps with size $N$ in a collision and write the total multiplicity distribution as
\begin{equation}
P(n)=\int dN W(N)P(N,n),
\label{eq52}
\end{equation}
where P(N,n) is the multiplicity distribution of the overlap of a given $N$, which we take Poissonian with the average multiplicity $N$
\begin{equation}
P(N,n)=\frac{N^n}{n!}\exp(-N).
\end{equation}
The normalization conditions $\sum_nP(n)=1$, and $\sum_n nP(n)=N$ lead to the following relations
\begin{subequations}
\begin{eqnarray}
\int dN W(N)=1,\\
\langle n \rangle=\langle N \rangle=\int dN NW(N). \label{eq54}
\end{eqnarray}
\end{subequations}
For the weight function we assume the gamma distribution \cite{101,102,103}
\begin{equation}
W(N)=G(N,k_N,\tau_N)= \frac{\tau_N}{\Gamma(k_N)}(\tau_NN)^{k_N-1}\exp(-\tau_NN),
\label{eq55}
\end{equation}
where $\tau_N=k_N/\langle N \rangle$.
There are several reasons for this choice. 
The growth of the centrality can be seen as a transformation of the cluster size distribution. 
Start with a set of single strings with a few clusters formed of a few overlapping strings. 
As the centrality increases, there appear more strings and more clusters composed of more strings.
This change can be considered as substitution of strings in a cluster by the new formed clusters, defined by a new $N$ corresponding to a higher color field in the cluster. 
This transformation, similar to the block transformation of Wilson type, can be seen as a transformation of the cluster size probability of the type
\begin{equation}
P(x)\to x\frac{P(x)}{\langle x \rangle} \to \dots \to x^k\frac{P(x)}{\langle x^k \rangle} \dots 
\label{eq56}
\end{equation}
Transformations of this type were studied long time ago by Jona Lasinio in connection with the renormalization group in probabilistic theory \cite{105}, showing that the only probability distribution function $P(x)$ which is stable  under such transformation are the generalized gamma functions, among them the simplest one is the gamma function which has one parameter less.
We point out that transformation of type \eqref{eq56} have been used previously, to study the probability associated with some special event which are shadowed by themselves and not for the total of events \cite{106,107,108,109,110}. 
We will come back to this point, studying the underlying events when one high $p_T$ particle is triggered. Notice that $W(N)$ satisfy KNO scaling, namely the product $NW(N)$ is only a function of $N/\langle N \rangle$, if the parameter $k$ of Eq.~\eqref{eq56} is energy independent. 
This property is a consequence of the invariance of the gamma functions under the transformation of type Eq.~\eqref{eq55} \cite{110}.
Let us discuss now the transverse momentum distribution (TDM) $f (p_T )$. As in the case of multiplicity distribution we consider a cluster which decay in the same way that a single string with the only difference of its ``size" $x$
\begin{equation}
f(p_T,x)=\exp(-xp_T^2).
\end{equation}
Actually, $x$ denotes the inverse of the color field in the cluster which depends not only on the size but also on the degree of overlapping strings inside the cluster.
Assuming that $x$ varies continuously one can write the total TMD, similarly to the multiplicity distribution case as
\begin{equation}
f(p_T)=\int dx W_p(x) f(x,p_T).
\label{eq58}
\end{equation}
We must realize the normalization condition
\begin{equation}
\int dp_T^2f(p_T)=\langle n \rangle,
\end{equation}
which gives the relation
\begin{equation}
\langle n \rangle=\int dx x \alpha^2W(\alpha x).
\end{equation}
Comparing the latter with Eq.~\eqref{eq54}, we can make the identification $W_p(x) = (\alpha x)^2W(\alpha x)$ If we take the gamma distribution in Eq.~\eqref{eq55} for $W(N)$, then $W_p(x)$ turns out up to a factor to be also the gamma distribution with different $k$ and $r$ given by
\begin{equation}
W_p(x)=\frac{r}{r_p}G(x,k+2,r_p),
\label{eq62}
\end{equation}
with $r_p = \alpha r$. 
So, at the end, both the multiplicity distribution and the TMD are given by a convolution of the cluster distribution and its TMD with the size probability $W(x)$ which in both cases can be taken the gamma distributions although with different parameters.
Introducing Eq.~\eqref{eq55} into Eqs. \eqref{eq52} and \eqref{eq58}, we obtain
\begin{equation}
\frac{1}{(1+p_T^2/r)^k}=\int_0^\infty dx \exp (-p_T^2x)\left( \frac{r}{\Gamma(k)} \right) (rx)^{k-1}\exp(-rx),
\label{eq63}
\end{equation}
and
\begin{equation}
\frac{\Gamma (n+k')}{\Gamma(n+1)\Gamma(k')}\frac{r^{k'}}{(1+r')^{k'}(n+k')^n}=\int_0^\infty dN\frac{e^NN^n}{n!}\frac{r'}{\Gamma(k')(r'N)^{k'-1}}\exp(-r'N).
\label{eq64}
\end{equation}
The mean value and the dispersion of the distributions \eqref{eq63} and \eqref{eq64} are
\begin{align}
\langle x \rangle=\frac{k}{r}, \hspace{0.3cm} \frac{\langle x^2 \rangle-\langle x \rangle^2}{\langle x \rangle^2}=\frac{1}{k},\\
\langle n \rangle=\langle N \rangle \frac{k'}{r'},  \hspace{0.3cm}  \frac{\langle N^2 \rangle-\langle N \rangle^2}{\langle N \rangle^2}=\frac{1}{k'},\\
 \frac{\langle n^2 \rangle-\langle n \rangle^2}{\langle n \rangle^2}=\frac{1}{k'}+\frac{1}{\langle N \rangle}.
\end{align}
The distribution \eqref{eq64} is a negative binomial distribution. 
Eqs. \eqref{eq63} and \eqref{eq64} are superposition of clusters and $1/k$ and $1/k'$ control the transverse momentum fluctuations and the fluctuations on the number of string in the cluster.
At small density there are no strings overlapped, and $k$ and $k'$ go to infinity. 
When the density increases, the strings start to overlap forming clusters, and therefore the $k'$ decreases. 
Their minimum is reached when the fluctuations in the number of strings per cluster reach their maximum. 
Above this point, increasing the string density, these fluctuations decrease and the $k'$ increases.
Now, if we take into account that the mean multiplicity and transverse momentum given by Eqs.~\eqref{eq38} and \eqref{eq39}, the Eqs. \eqref{eq63} and \eqref{eq64} become
\begin{equation}
f(p_T,y)=\frac{dN}{dp_T^2dy}=\frac{dN}{dy}\frac{k-1}{k}\frac{F(\rho)}{\langle p_T^2 \rangle_1}\frac{1}{\left(1+\frac{F(\rho)p_T^2}{k\langle p_T^2 \rangle_1} \right)^k},
\label{eq68}
\end{equation}
and
\begin{equation}
P(n)=\frac{\Gamma(n+k')}{\Gamma(n+1)\Gamma(k')}\frac{\left(\frac{k'}{\langle n \rangle_1 F(\rho)} \right)^{k'}}{\left( 1+\frac{k'}{\langle n \rangle_1 F(\rho)} \right)^{n+k'}}.
\label{eq67}
\end{equation}
We observe that
\begin{eqnarray}
\langle n \rangle = F(\rho)N_s\langle n \rangle, & & \langle p_T^2 \rangle = \frac{k}{k-2} \frac{\langle p_T^2 \rangle_1}{F(\rho)}.
\end{eqnarray}
Eqs.~\eqref{eq68} and \eqref{eq67} give the distributions for any projectile, target, energy and degree of centrality and are universal functions which depend of only two parameters, $\langle p_T^2 \rangle_1$ and $\langle n \rangle_1$, the average transverse momentum and multiplicity of particles produced by one string.
In case of identified secondary particles, it should be used the corresponding quantities for each identified particles, $\langle p_T^2 \rangle_{1i}$ and $\langle n \rangle_{1i}$. Sometimes instance $p_T$ is used $m_T$. At $\rho\to \infty$ and $k\to\infty$ the TMD becomes $\exp(-F(\rho)p_T^2/\langle p_T^2\rangle_1)$ very similar to the behavior at $\rho\to0$. From Eq.~\eqref{eq68} we have
\begin{equation}
\frac{d\ln f}{d\ln p_T}=\frac{-2F(\rho)}{\left( 1+\frac{F(\rho)p_T^2}{k\langle p_T^2 \rangle_{1i}} \right)} \frac{p_T^2}{\langle p_T^2 \rangle_{1i}}.
\label{eq69}
\end{equation}
At $p_T^2\to0$, the latter reduces to $-2F(\rho)p_T^2/\langle p_T^2 \rangle_{1i}$ and vanishes at $p_T^2=0$.
On the other hand, as $\langle p_T^2 \rangle_{1\pi}<\langle p_T^2 \rangle_{1k}<\langle p_T^2 \rangle_{1\mathrm{p}}$, the absolute value of Eq.~\eqref{eq69} is larger for pions than for kaons than for protons, this is the well known hierarchy that often it is advocated in favor of a hydrodynamic picture hadronic interactions.
However, we describe very well the data as it is seen in Fig.~\ref{fig12}, where we show our results together the PHOBOS data \cite{106} for central Au-Au collisions.

\begin{figure}
\centering
\includegraphics[scale=0.2]{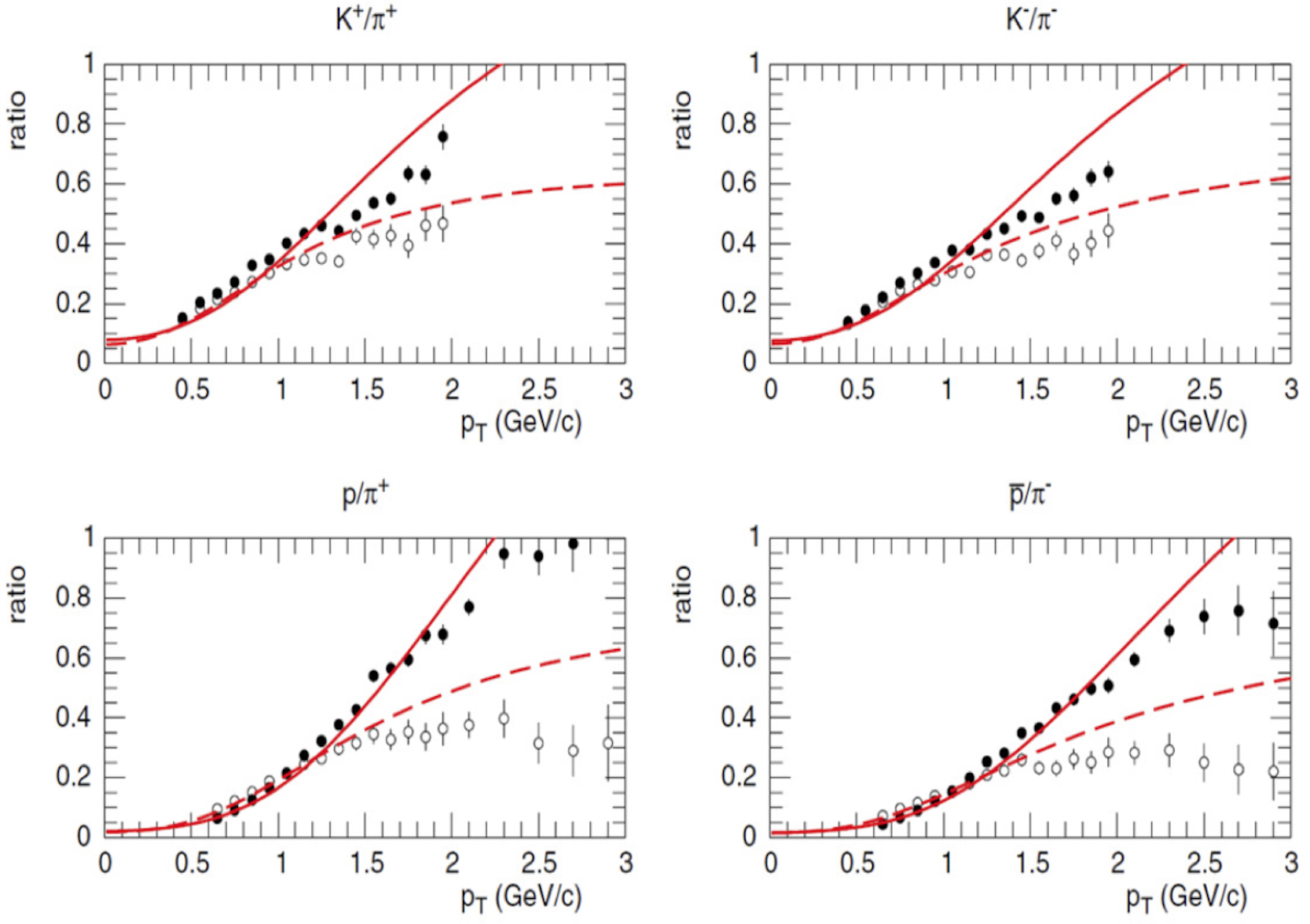}
\caption{Ratios for different distributions $k/\pi$, $\mathrm{p}/\pi$ in Au-Au collisions at $\sqrt{s}=$200 GeV at two different centralities : 0-5$\%$ (solid circles), 60-70$\%$ (open circles) in comparison with the data.}

\label{fig13}
\end{figure}

Let us now discuss the interplay of low and high $p_T$. One defines the ratio between central and peripheral collisions as
\begin{equation}
R_{CP}(p_T)=\frac{f'(p_T,y=0)/N'_{coll}}{f(p_T,y=0)/N_{coll}}.
\label{eq70}
\end{equation}
The normalization on the number of collisions in the latter, essentially eliminates $N_s$ from $dN/dY$, this is true at mid rapidity.
From Eq.~\eqref{eq58} and \eqref{eq68} we obtain
\begin{equation}
R_{CP}(p_T)=\frac{((k'-1)/k')}{((k-1)/k)} \left( \frac{F(\rho')}{F(\rho)} \right)^2 \frac{\left(1+\frac{F(\rho)p_T^2}{k\langle p_T^2\rangle_{1i}} \right)^k}{\left(1+\frac{F(\rho')p_T^2}{k'\langle p_T^2\rangle_{1i}} \right)^{k'}}.
\end{equation}
Here $k$ and $k'$ are values of the parameter $k$ for TMD for peripheral and central collisions. In the limit $p_T^2\to 0$, as $F(\rho')<F(\rho)$ we have
\begin{equation}
R_{CP}\simeq \left(\frac{F(\rho')}{F(\rho)} \right)^2<1,
\end{equation}
which is independent of $k$ and $k'$. As $\rho'/\rho$ increases the ratio $R_{CP}$ decreases, in agreement with experimental data. 
As $p_T$ increases, we have
\begin{equation}
R_{CP}(p_T)\sim \frac{1+\frac{F(\rho)p_T^2}{k\langle p_T^2\rangle_{1i}}}{1+\frac{F(\rho')p_T^2}{k'\langle p_T^2\rangle_{1i}}}
\end{equation}
and $R_{CP}$ increases. At large $p_T$
\begin{equation}
R_{CP}(p_T)\sim \frac{F(\rho)k'}{F(\rho')k}p_T^{2(k-k')}.
\end{equation}
At low density in the region where decreases with the string density $k' < k$ and $R_{CP} (p_T ) > 1$. It is the Cronin effect.
As $\rho'/\rho$ increases the ration $R_{CP}$ increases. With the growth of the energy of the collision, the energy density increases reaching the region where $k$ increases. 
Now at $\rho'>\rho$ and $k' > k$, there will be a suppression of $p_T$.
In the forward rapidity region, the normalization of Eq.~\eqref{eq70}, does not cancel $N_s$ from $dN/dy$, since in this region $N_s$ is proportional to $N_A$ instead of $N_{coll}$.
Now, an additional factor $(N'_A/N'_{coll})/(N_A/N_{coll})$ appears in $R_{CP} (p_T )$.
As $N'_{coll}-N_A$ for central collisions is larger, than for peripheral collisions, we have $R_{CP} (p_T , y = 3) < R_{CP} (p_T , y = 0)$, thus a further suppression occurs in agreement with experimental data \cite{115}.
The results for the TMD for $\pi^+$, $k^+$ and p in Au-Au at $\sqrt{s}=$ 200 GeV are in good agreement with the Phenix data \cite{138p}.
In Fig.~\ref{fig13}, we show the ratios kaons/pion, 
and proton/pion as a function of $p_T$ at the two extreme centralities. 
The obtained values of $k$ as a function of the string density increase as it was expected.

The experimental data on pp in the range $\sqrt{s}=$23, 200, 630 GeV and 1.38, 7 TeV can also be described by the distribution of Eq.~\eqref{eq68} \cite{140p}.
In this case the values of $k$ decreases with energy as expected. 
At higher energy and high multiplicity $k$ should increase \cite{67}.

Even though the parametrization \eqref{eq68} describes well the data up to 5-10GeV/c, most of the considerations concerning the string fragmentation are only  valid for low and intermediate $p_T$.
In order to include the high $p_T$ part of the spectrum more refined study is necessary.

\begin{figure}
\centering
\includegraphics[scale=.2]{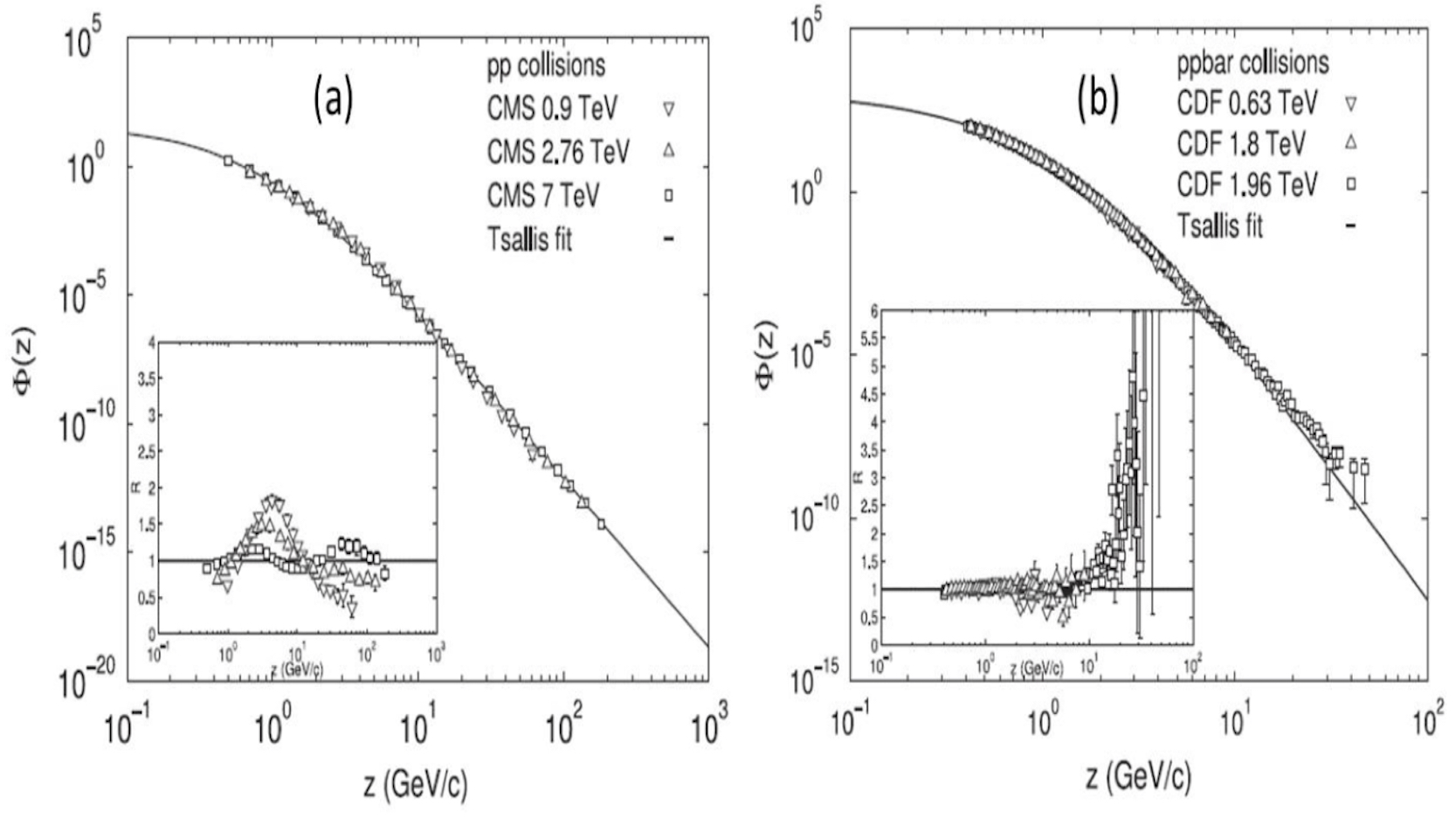}
\caption{Scaling behavior of the charged hadron pt spectra presented in z (a) in pp collisions and (b) $\mathrm{p}\bar{\mathrm{p}}$ collisions with different energy scales. The inset is the distribution of the ratio between the experimental data and the fitted results \cite{122}.}

\label{fig15}
\end{figure}

The differences between the baryon and meson spectrum is not only due to the mass differences, which results $\langle p_T^2 \rangle_{1M}<\langle p_T^2 \rangle_{1B}$.
This effect only causes a shift in the maximum of the nuclear modified factor $R$, but keeps the height at the maximum, contrary to the data. 
In the fragmentation of a cluster formed of the overlapping of several strings the flavor properties follow from the corresponding properties of the
flavor of the valence partons of the end of the
 individual strings and hence the resulting flavor of the cluster $F$ is the flavor composition of the individual strings $f$ as well as the color composition.
The clusters have higher color and different flavor ends. 
The fragmentation of a cluster will be by means of the creation of a pair $F\bar{F}$
, where $F$ and $\bar{F}$ denotes the sets of flavor quarks and antiquarks of the end of the cluster.
After the decay, the two new $F\bar{F}$ strings will be treated in the same way decaying into more $F\bar{F}$ strings until they come to objects with mass comparable to hadron masses, which we identified with the observable hadrons by combining the produced flavor with statistical weights. 
In this way the production of baryons and antibaryons will be enhanced
with the number of strings of the cluster. The additional quarks (antiquarks) required to form a baryon (antibaryon) are provided by the quarks (antiquarks) of the overlapping strings that form the cluster
. 
In some sense the coalescence picture of particle production is incorporated in a natural way.
An effective way of taking into account these flavor considerations can be seen in Ref.~\cite{ex1}.
Very often, it is used an exponential instead of a gaussian for the decay of one string. 
Indeed, the tension of a cluster fluctuates around its mean value because the chromo-electric field is not constant. 
Such fluctuations lead to a Gaussian distribution of the string tension \cite{138p,140p,118,116}
\begin{equation}
\frac{dn}{dp_T}\sim \int_0^\infty dx \exp \left( -\frac{x^2}{2 \langle x^2 \rangle}  \right) \exp(-\lambda p_T^2/x^2),
\end{equation}
which give rise to the thermal distribution
\begin{equation}
\frac{dn}{dp_T}\sim \exp\left(-p_T\sqrt{\frac{2\pi}{\langle x^2 \rangle}} \right),
\end{equation}
where $\langle x^2 \rangle=\pi \langle p_T^2 \rangle_1/F(\rho)$. The temperature is expressed as \cite{118,119}
\begin{equation}
T(\rho)=\sqrt{\frac{\langle p_T^2 \rangle}{2F(\rho)}}.
\end{equation}
Now the total TMD is changed and instance of the gamma distribution in Eq~\eqref{eq62} a Tsallis type distribution is obtained, namely
\begin{equation}
f(p_T)=\left(1+\frac{\sqrt{f(\rho)}p_T}{k\langle p_T \rangle_1} \right)^k.
\end{equation}
There are several scaling properties found in TMD related to string percolation. 
The experimental data for pp collisions exhibit a universal behavior in a suitable variable $z=p/B$ \cite{121,122}.
Indeed, the TMD of pp and $\mathrm{p}\bar{\mathrm{p}}$ at all energies are at the same curve as it is shown in Fig.~\ref{fig15}. The parameter $B$ it is found proportional to $1/F(\rho)$ and therefore increasing with energy. A similar scaling is found in Au-Au collisions \cite{120} at different centralities.

The experimental data on the mean $p_T$ as a function of the multiplicity show that in pp, pA and AA collisions all of them grow, being larger in pp than in pA and in AA collisions.
In PbPb the rise of mean $p_T$ with multiplicity is flattened above a certain low multiplicity. 
The same occurs at pPb although in this case is flattened at higher multiplicity. 
This behavior is understood as a consequence of Eq~\eqref{eq68}. In fact, the factor $1/F(\rho)$ is responsible of the rise of $\langle p_T\rangle$ with multiplicity because grows with multiplicity.
The flattening of PbPb and pPb is due to the dependence of $k$ on the string density. 
In PbPb for most of the multiplicities the corresponding string densities are above the percolation threshold.
In this region $k$ grows with $\rho$, and according to Eq.~\eqref{eq68}, the rise is lowered. In the case of pp collisions, on the contrary, the corresponding string densities lie below the critical density.
In this region $k$ is a decreasing function of $\rho$ hence there is not flattening. We expect that at higher energy, larger than 14 TeV, the critical density will be reached, even in pp collisions.

\subsection{Transverse momentum fluctuations}
\begin{figure}
\centering
\includegraphics[scale=0.2]{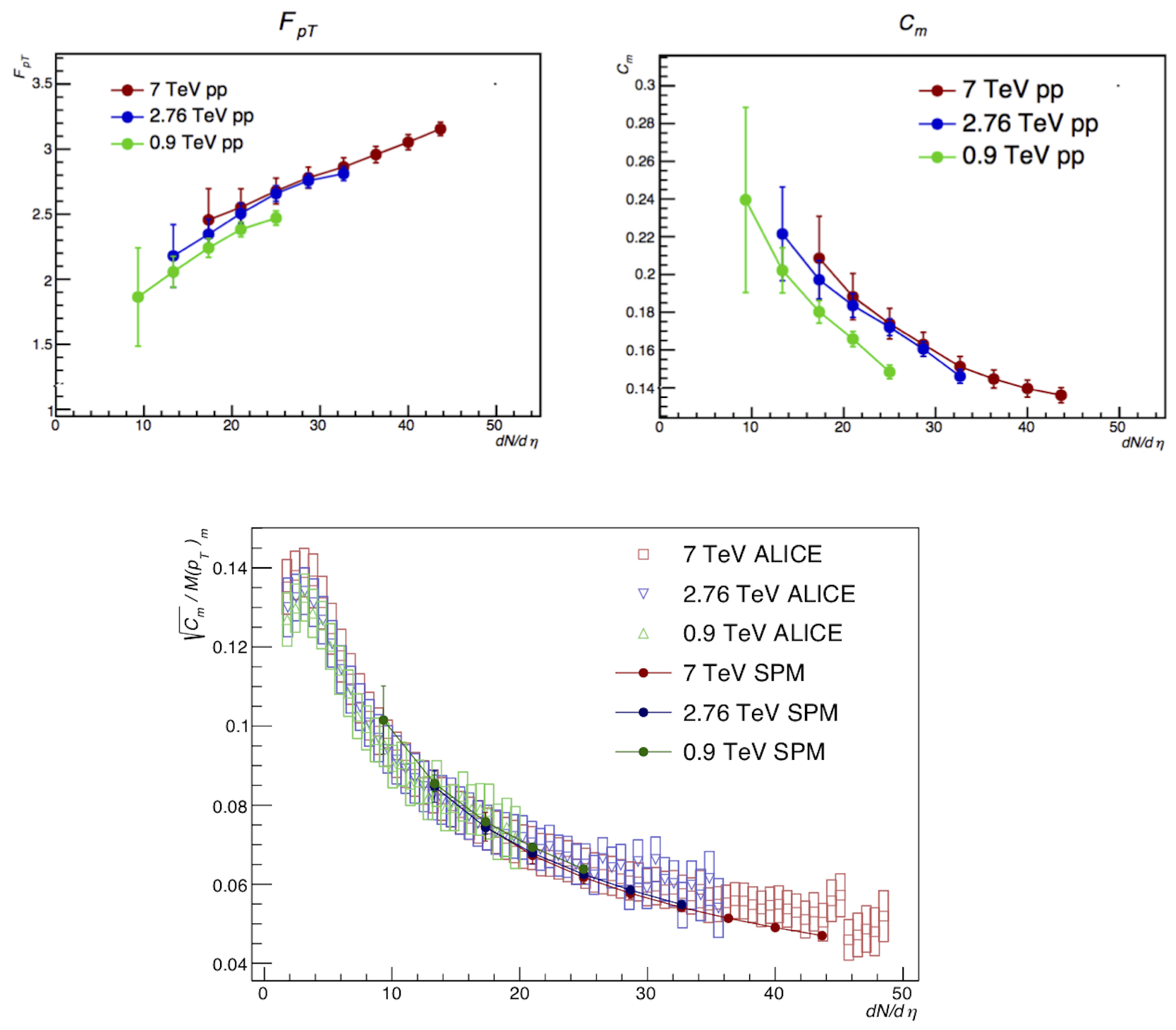}
\caption{Scaling behavior of the charged hadron $p_T$ spectra presented in z (a) in pp collisions and (b) p$\bar{\mathrm{p}}$ collisions with different energy scales. The inset is the distribution of the ratio between the experimental data and the fitted results.}

\label{fig16}
\end{figure}

The event by event fluctuations of thermodynamical quantities as the temperature were proposed as a probe for the deconfined phase. Due to that, the study of the fluctuations on the mean $p_T$ is very interesting. These fluctuations are measured using the observables:
\begin{equation}
F_{p_T}=\frac{w_{data}-w_{random}}{w_{random}}, \hspace{0.3cm} w=\frac{\sqrt{\langle p_T^2 \rangle-\langle p_T\rangle^2}}{\langle p_T \rangle},
\end{equation}
and the correlation between the transverse momemtum
\begin{equation}
\langle \Delta p_{Ti},\Delta p_{Tj} \rangle=C_m\simeq 2F_{p_T}(\langle p_T^2 \rangle-\langle p_T\rangle^2)/\langle p_T \rangle \langle N\rangle,
\label{eq79}
\end{equation}
where $\Delta p_{Ti}=(p_{Ti}-\langle p_{Ti} \rangle)$. And
\begin{equation}
M(p_T)=\frac{1}{\sum_{k=1}^{n_{ev},m}N_{acc,k}}\sum_{k=1}^{n_{ev},m} \sum_{i=1}^{N_{acc},k}p_{Ti},
\end{equation}
and $C_m/M(p_T)$.
The last observable is used because suppresses the statistical fluctuations in string percolation. 
The observable $F_{p_T}$, using Eq.~\eqref{eq79} is given by \cite{125}
\begin{equation}
F_{p_T}=\sqrt{\frac{N_s\langle p_T \rangle^2_1\mu_1 -2N_s^{3/4}(\frac{S_n}{S_1})^{1/4}\langle p_T \rangle_1\langle p_T \rangle\mu_1+N_s(\frac{S_n}{S_1})^{1/2}\langle p_T \rangle^2\mu_1    }{(N_s\frac{S_1}{S_n})^{1/2}\langle p_T \rangle^2_1-2(N_s\frac{S_1}{S_n})^{1/4}\langle p_T \rangle_1\langle p_T \rangle+\langle p_T \rangle^2}}-1.
\end{equation}
In Fig.~\ref{fig16}, we show the result for $C_m$ in Pb-Pb collisions together the ALICE data.
\begin{figure}
\centering
\includegraphics[scale=.1]{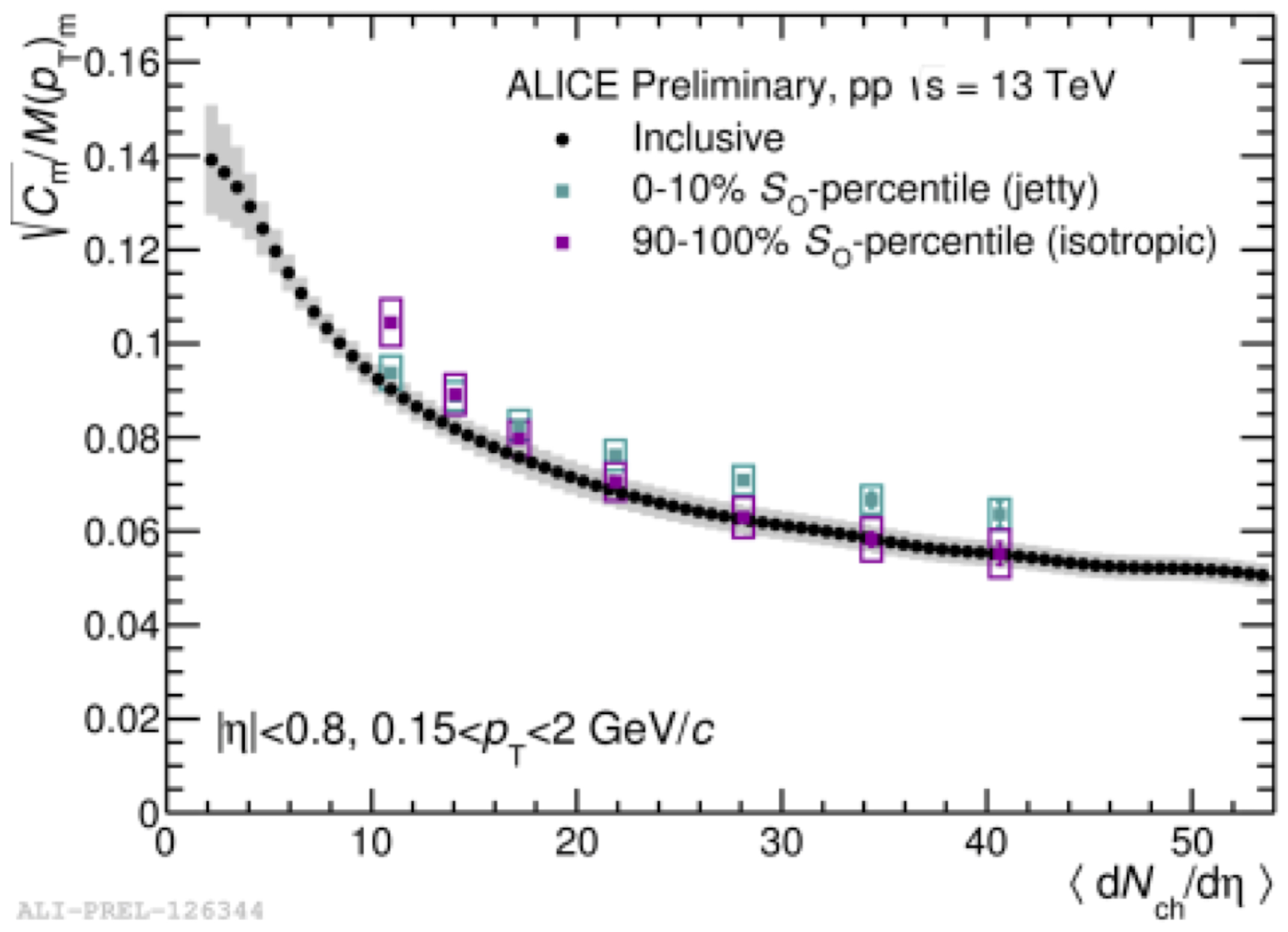}
\caption{Preliminary results for $\sqrt{C_m}/M(p_T)$ as a function of the multiplicity for pp at different energies as well as Pb-Pb collisions at 2.76 TeV together with the experimental data.}

\label{fig17}
\end{figure}
In Figs.~\ref{fig16} and \ref{fig17}, we show the results for $F_{p_T}$, $C_m$, $\sqrt{C_m}/\langle p_T \rangle$ for pp collisions at 0.9, 2.76 and 7 TeV together the CMS data.
In Fig.~\ref{fig17}, we show the results for $\sqrt{C_m}/M(p_T)_m$ as a function of the multiplicity for pp at different energies with the experimental data. 
It is observed a change in the slope at high multiplicities that is reproduced in the string percolation but not for the usual Monte Carlo code models. 
In string percolation the change of slope arises naturally due to the formation of a large cluster above a critical density (corresponding to a high multiplicity) and therefore suppression of the fluctuations.

\subsection{Forward-backward correlations}
The width of the KNO scaling shape is related to the fluctuations on the number of strings or the number of clusters (independent color sources). This width is also related to the forward-backward ($F$-$B$) correlations. These correlations can be described by a linear approximation
\begin{equation}
\langle n_B \rangle=a+bn_F,
\end{equation}
where $n_F$ is the number of particles observed in the forward (backward) rapidity window and the slope $b$ measures the correlation forward-backward
\begin{equation}
b=\frac{\langle n_F n_B \rangle-\langle n_F \rangle\langle n_B \rangle}{\langle n_F^2 \rangle-\langle n_F \rangle^2}.
\end{equation}
Usually, the F and B rapidity intervals are taken separated by a central rapidity window $|y|<y_c$ in such a way that the short range correlations are eliminated ($y_c=0.5$) because its range is less than unit of rapidity. 
In any multiple scattering model the origin of long range correlations is the fluctuations in the number of elementary scatterings \cite{50,127,128,130,131,132,133}.
Let us consider symmetric $F$ and $B$ intervals and having $N$ strings which decay into $\mu_1$ particles. Then, the slope $b$ can be split into short range (SR) and long range (LR) correlations \cite{130}
\begin{equation}
b=b^{SR}+b^{LR}=\frac{\delta_F \mu_1}{1+\delta_F \mu_1 [\omega_N+\Lambda(0)]}\Lambda(y_{FB})+\frac{\delta_F \mu_1}{1+\delta_F \mu_1 [\omega_N+\Lambda(0)]},
\end{equation}
where $\omega_N$ is given by
\begin{equation}
\omega_N=\frac{\langle N^2 \rangle-\langle N \rangle^2}{\langle N \rangle^2}.
\end{equation}
 $\Lambda(y_{FB})$ and $\delta_F$ are the correlation function of one string with rapidity separation $y_{FB}=y_F-y_B$ and the acceptance of the $F$ or $B$ rapidities, respectively. 
 We also take $\delta_F=\Delta y_{F}=\Delta y_B$.
For large rapidity window gap between the F and B intervals there are not long range correlations in a single string, then $\Lambda=0$ and $b$ becomes
\begin{equation}
b=\frac{1}{1+\frac{1}{\delta_F \mu_1 \omega_N}}.
\label{eq86}
\end{equation}
At low energy, there are not fluctuations in the number of strings, i. e., $\omega_N \to 0$, and according to Eq.~\eqref{eq86}, $b\to0$.
\begin{figure}
\centering
\includegraphics[scale=.2]{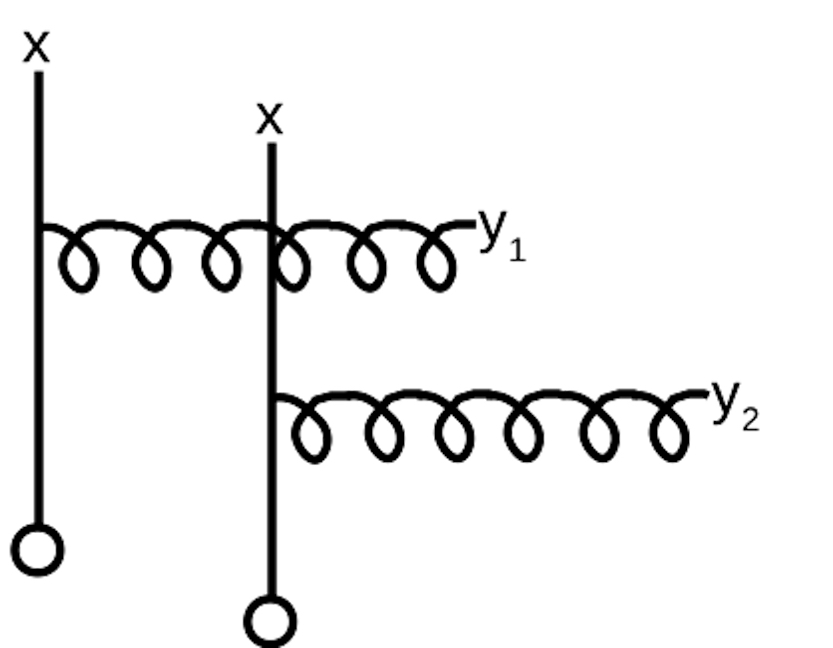}
\caption{The leading orden diagram which induces long range correlations in rapidity. The source of one nucleus is given by the $x$ and the other by the $o$. The produced gluon is denoted by the curled line.}

\label{figintercalar}
\end{figure}
As the energy or the centrality of the collisions increases $\omega$ increases as well as $b$. 
This behavior can be turned as a consequence of the formation of a large cluster of overlapping strings and consequently a decreasing of the number of independent color sources. 
Notice that if we fix the multiplicity, we eliminate many of the possible string fluctuations and therefore $b$ will be smaller.
In CGC the main contribution to long range correlations comes from the diagram of Fig.~\ref{figintercalar}, which only contributes to short range correlations in such a way that for a large rapidity gap between $F$ and $B$ intervals we have \cite{134,135,136}
\begin{equation}
b=\frac{1}{1+c\alpha_s^2},
\end{equation}
where $c$ is a constant independent on the energy and centrality degree. As the strong coupling constant, $\alpha_s$, decreases with energy and with centrality, $b$ increases. 
This behavior is very similar to the one described above for string percolation. 
The analysis of $F$-$B$ correlations has been extended not only to two rapidity separated windows but also to different azimuthal windows which help to separate short and long range correlations \cite{130}.
In this case, the coefficient $b$ is given by 
\begin{eqnarray}
b=b^{SR}+b^{LS}=\frac{\delta_F \mu_1}{1+\delta_F \mu_1 [\omega_N+\Lambda(0,0)]}\Lambda(y_{FB},\phi_{FB})+\nonumber\\
\frac{\delta_F \mu_1}{1+\delta_F \mu_1 [\omega_N+\Lambda(0,0)]},
\end{eqnarray}
where now $\delta_F$ is the product of the acceptance on rapidity and an azimuthal angle ($\delta_F=\Delta y_F\Delta\phi_F=\Delta y_B\Delta\phi_B$), $\Lambda(y_{FB},\phi_{FB})$ is the correlation function of the single string at rapidities and azimuthal angles separation, $y_{FB}=y_F-y_B$, and $\phi_{FB}=\phi_F-\phi_B$. 
\begin{figure}
\centering
\includegraphics[scale=.2]{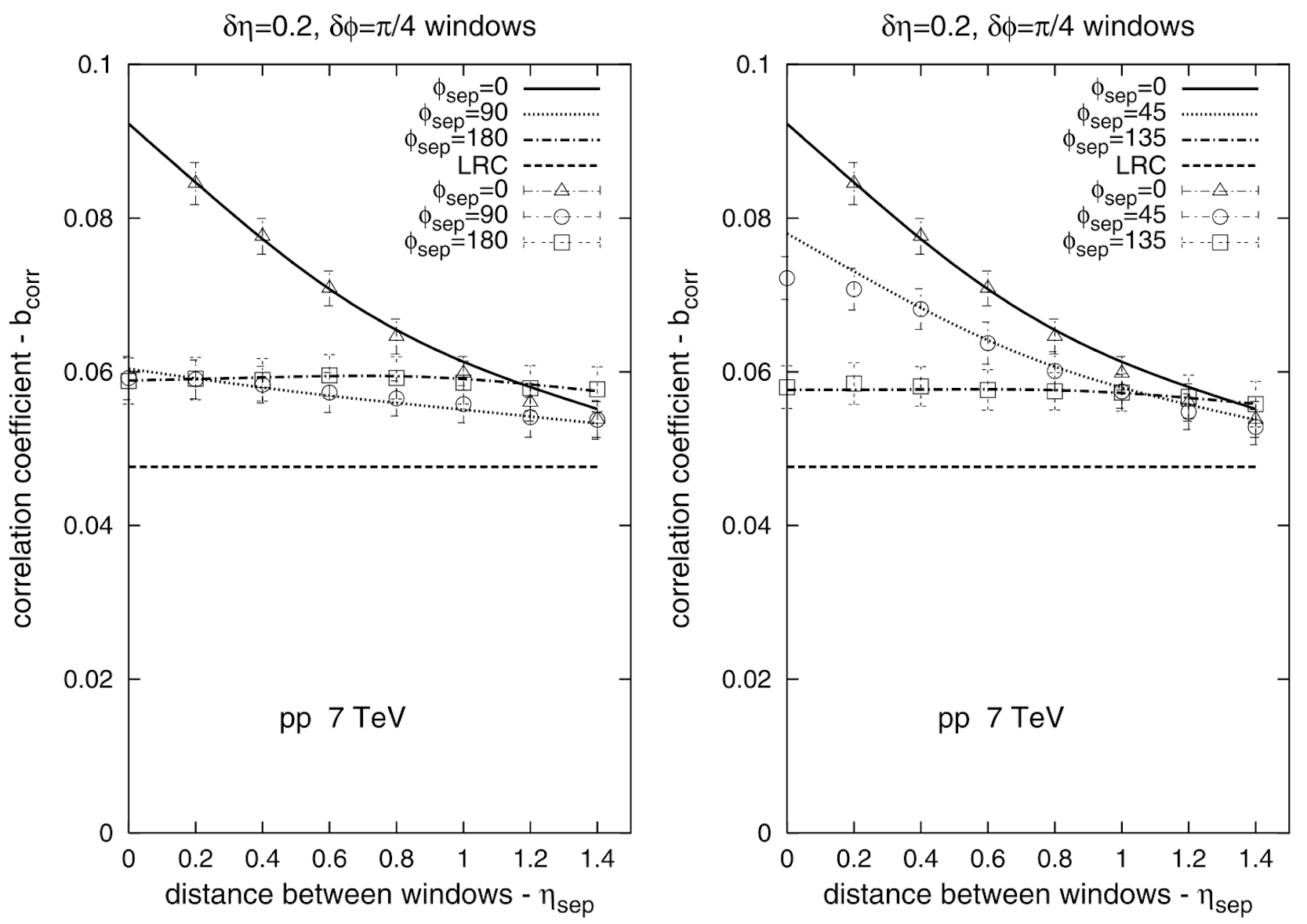}
\caption{The forward-backward ($F$-$B$) correlation coefficient in pp collisions at 7 TeV from reference \cite{130}.}

\label{fig20}
\end{figure}
In Fig.~\ref{fig20} it is shown the results of a separation azimuthal angle of 0º,45º and 135º for pp collisions together the ALICE data at 7 TeV. The agreement is good also at 0.9 and 2.76 TeV \cite{130}.
The $F$-$B$ correlations have been studied not only for multiplicities in the $F$-$B$ intervals but also for transverse momentum-multiplicity $(p_T -n)$ and transverse momentum correlations $(p_T -p)$. In the $(p_T - p_T )$ case, the asymptotic equation for the slope coefficient is
\begin{equation}
b=\frac{\omega_\rho \mu_F}{\omega_\rho \mu_F +16\gamma \sqrt{\rho}},
\end{equation}
where
\begin{eqnarray}
\omega_\rho=\frac{D_\rho}{\langle \rho \rangle}, & & \gamma=\frac{D_{p_T}}{\langle \rho \rangle^2},
\end{eqnarray}
$D_\rho$ and $D_{p_T}$ are the string density and transverse momentum dispersions respectively, $\mu_F$ is the multiplicity of one of the symmetric intervals, and $\gamma$ is a dimensionless coefficient which depends only of the form of the distribution. 
For a Tsallis shape distribution, $\gamma$ takes the value $(k-1)/2(k-4)$, which is related to the width of the distribution.
In the case of a thermal distribution its value is 1/2.
\begin{figure}
\centering
\includegraphics[scale=.1]{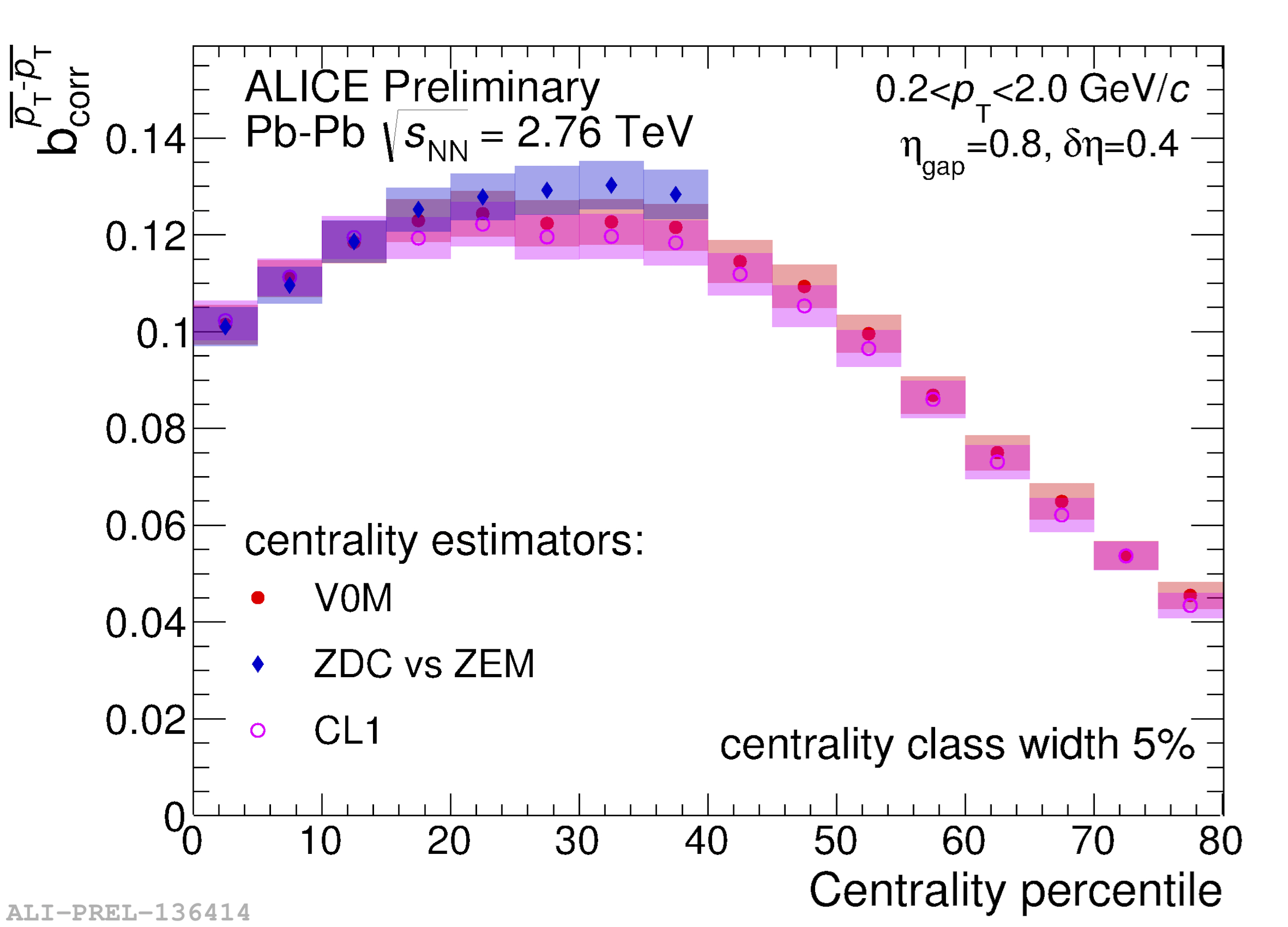}
\caption{Dependence of $b_{corr}^{\bar{p}_t-\bar{p}_t}$ on centrality clases 5\% width determined by the V0, ZDCvsZEM and CL1 estimators from ALICE \cite{fig21ref}.}

\label{fig21}
\end{figure}
In Fig.~\ref{fig21} is shown the ALICE preliminary data for Pb-Pb collisions at 2.76 TeV as a function of the centrality for a rapidity gap of 0.8 and a rapidity width of 0.4. 
The data are in qualitative agreement with the string percolation model \cite{138,139} which predicted a rise of $b$ with centrality up to around 30\% decreasing above this centrality value.

\subsection{Underlying event of high $p_T$ particles and KNO scaling}

The study of the underlying effect can be useful to understand the particle production mechanism. It has been shown that selecting events of determinate high $p_T$ particle and looking at the particles that are in the azimuthal range angles, say $\pi/3 < |\Delta \phi|<2\pi/3$, the associated multiplicity distribution satisfy approximately KNO \cite{142}.
Let us show that this is a rather general property which is satisfied by events of a determined class of scatterings, for instance diffractive and non diffractive or inelastic and elastic or soft and hard scatterings.
In a multiple scattering approach, there will be events that is sufficient to have one elementary scattering of being of this class to be the final result of this class. 
The non diffractive, the inelastic and the hard events satisfy this requirement.
It is said that these events are only shadowed by themselves and in fact the evaluation of the cross section for these selected events only appears the cross section of the elementary cross section of these events not the elementary cross section of all kind of events \cite{143}.
Concerning the associated multiplicity distribution to these events is shown that in terms of multiple scatterings, the original distribution and the new one are related for a factor $N$ which translate into a multiplicative factor $n$ in the multiplicity in such a way that \cite{106,107,108,110}
\begin{equation}
P_c(n)=nP(n)/\langle n \rangle.
\end{equation}
If we go on the process of the selection of high $p_T$ particles, we will have the chain
\begin{equation}
P(x)\to xP(x)/\langle x \rangle \to \dots  \to x^kP(x)/\langle x^k \rangle.
\end{equation}
In a similar way than the one in Sec.~\ref{mult}, the only stable distributions under these transformation are the generalized gamma function, being the gamma function the most simple of them (see Eq.~\eqref{eq55}). 
This function satisfies KNO scaling if k is independent of energy. 
We have seen above that $k$ increases with the energy for pp collisions in the studied range, as $1/k$ controls the width of $\langle n \rangle P(n)$, this distribution should be narrow as the energy increase as the experimental data show.

\subsection{Bose-Einstein correlations}

\begin{figure}
\centering
\includegraphics[scale=0.05]{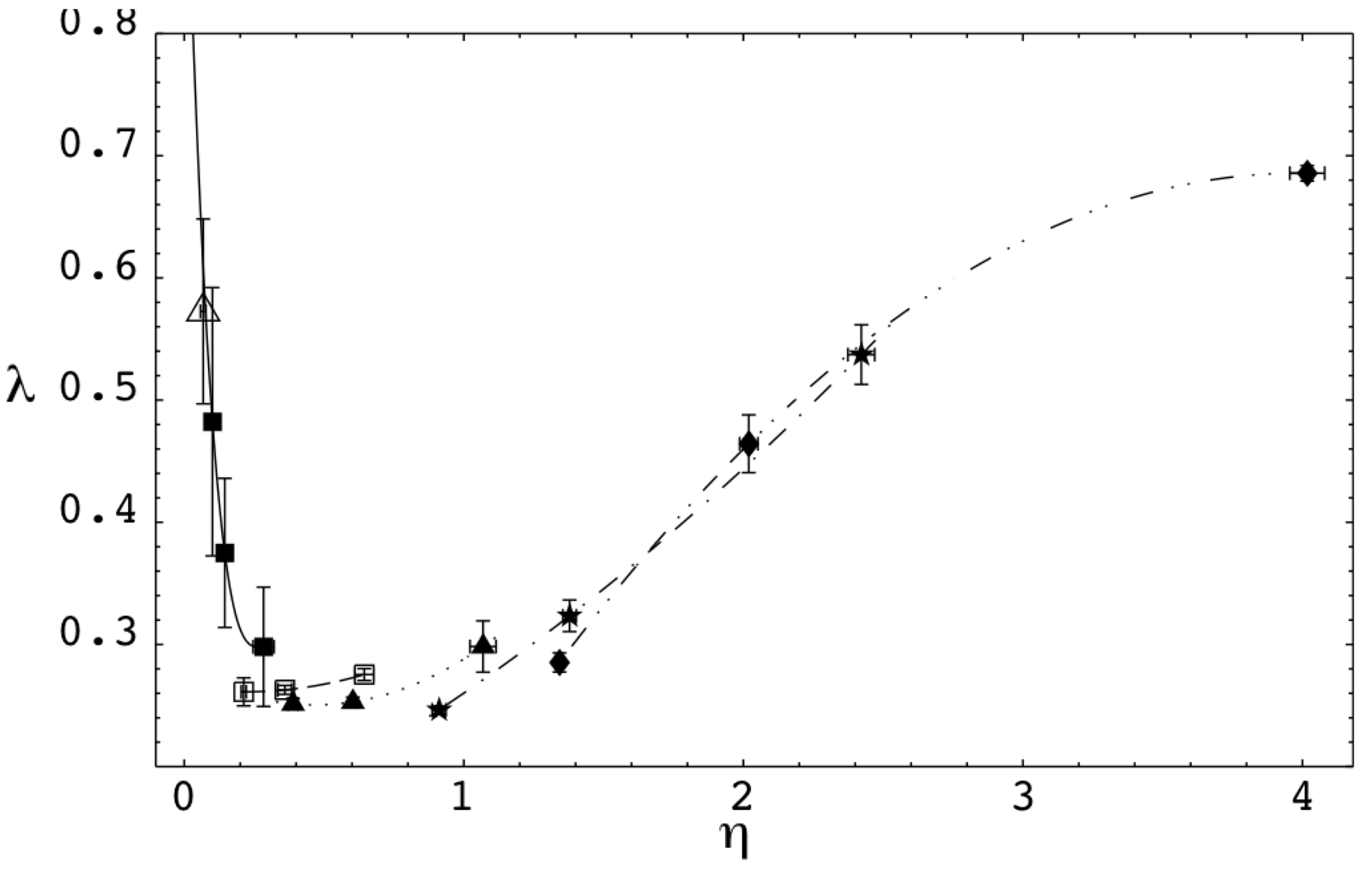}
\caption{Dependence of $\lambda$ on $\eta$ for different nucleus–nucleus collisions in the percolating strings framework taking into account the energy-momentum of the strings. Each point represents a specific type of nucleus–nucleus collisions. Correlations are calculated between identical pions for $y_{1cm} = y_{2cm} = 0.5$ and $m_{T1} = m_{T2} = 0.35$ GeV/$c^2$ \cite{154}.}

\label{fig22}
\end{figure}

The Bose-Einstein correlations (BEC) are very interesting in order to determinate the extension of the source of multi-particle production as well as to know the degree of coherence of the emitted particles.
The correlation strength is characterized by the parameter $\lambda$, which can also be interpreted as a measure of the chaotically of the degree of coherence of the collisions \cite{148,149,150}.
In this interpretation $\lambda=1$ means totally chaotic emission, whereas $\lambda=0$ means radiation in a coherent way.
This interpretation should be taken with caution, because in $e^+e^-$, $\lambda=1$ at energies where there are not production of more than two jets and higher energies, $\lambda$ decreases with increasing multiplicity.
These facts would apparent indicate a systematic increase of the coherence from $e^+e^-$ to pp collisions, which does not seem reasonable.
The experimental data on $\lambda$ have been in different kinematic conditions assuming different extrapolations, normalizations and corrections which makes difficult the comparison with models, however the ISR, SPS, RHIC and LHC data allow us to distinguish some trends.
First, for a not very large number of collisions the data of SPS with p and O as a projectiles show a decrease of $\lambda$ with multiplicity \cite{151,152,153}.
As the number of collisions increases no longer decreases even it increases reaching values of 0.6-0.7. 
At SPS energies the values of $\lambda$ are larger at forward than at central rapidity.
Notice that the particle multiplicity is larger at central than in forward rapidity. 
All these trends of data can be understood in the framework of percolation of strings \cite{154,155}.
The strings of the Lund type fragmentation according to totally chaotic sources, $\lambda=1$, and usually is assumed that there is not BEC from particles emitted from different strings \cite{156}.
Under this assumptions one can write
\begin{equation}
\lambda=n_s/n_T,
\end{equation}
where $n_s$  is the number of identical particles pairs produced from the same string and $n_T$ is the total number of identical pairs produced in the same kinetic range of $p_T$ and $y$.
The number of identical pairs produced by each cluster is
\begin{equation}
n_s=\frac{1}{2}\mu_1^2 \left\langle \sum_{n=1}^{N_s}  \frac{a_nnS_n}{S_1} \right\rangle,
\end{equation}
and the total number of pairs of identical particles produced is
\begin{equation}
n_T=\frac{1}{2}\mu_1^2 \left\langle \left( \sum_{n=1}^{N_s}  \frac{a_n\sqrt{n}S_n}{S_1} \right)^2 \right\rangle,
\end{equation}
where $a_n$ is the number of clusters with $n$ strings.
The numerical results of the Monte-Carlo simulation that includes energy conservation to different energies and collisions type shows the right change of the behavior and approximate to the scaling of $\lambda$ for string densities around $\rho \simeq 0.8-1$, which is in agreement with the experimental data, as is shown in Fig.~\ref{fig22} \cite{154}.

The three body BEC have been also studied in percolation \cite{155}, predicting the strength of the three particle BEC, which is in good overall agreement with data \cite{157}.

\subsection{$J/\Psi$ production dependence on the multiplicity}

\begin{figure}
\centering
\includegraphics[scale=0.15]{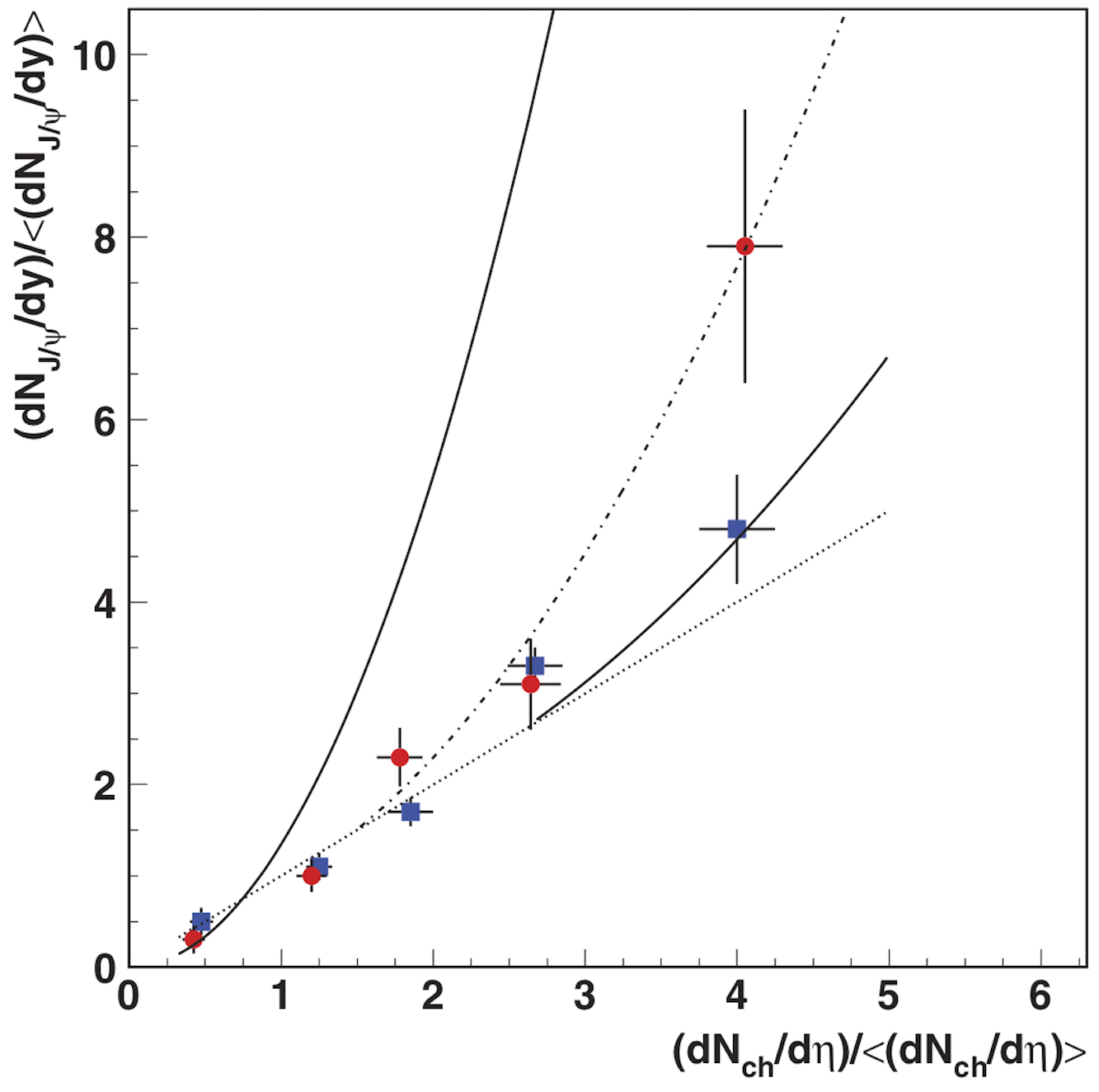}
\caption{Results for pp collisions in the central $|y| < 0.9$ rapidity range (dashed line) and forward $2.5 < y < 2.4$ (dotted line), together with the experimental data fro the central (circles) and forward (squares) rapidly regions from the ALICE Collaboration. The linear behavior (solid line) and the prediction for pPb collisions (dashed-dotted line) at 7 TeV are also plotted \cite{158}.}

\label{fig23}
\end{figure}

The ALICE collaboration have found a departure from linearity on the dependence of $J/\Psi$ production on the multiplicity at very high multiplicity. 
This departure is larger at central than a forward rapidity region. 
This behavior can be explained in the frameworks of string percolation \cite{158}.
In fact assuming that as in any hard process, the number of produced $J/\Psi$ is proportional to the number of elementary collisions, $N_s$, we have
\begin{equation}
\frac{n_{J/\Psi}}{\langle n_{J/\Psi} \rangle}=\frac{N_s}{\langle N_s \rangle}.
\end{equation}
From Eq.~\eqref{eq38} we can write
\begin{equation}
\frac{dN/dy}{\langle dN/dy \rangle}=\frac{N_s F(\rho)}{\langle N_s \rangle F(\langle \rho \rangle)},
\end{equation}
thus
\begin{equation}
\frac{dN/dy}{\langle dN/dy \rangle}=\left( \frac{n_{J/\Psi}}{\langle n_{J/\Psi} \rangle} \right)^{1/2} \left( \frac{1-\exp(-n_{J/\Psi}\langle \rho \rangle/\langle n_{J/\Psi} \rangle)}{1-\exp(-\langle \rho \rangle)} \right)^{1/2}.
\end{equation}
At low multiplicities, $N_s$ is small and the above equation give rise to a linear dependence
\begin{equation}
\frac{n_{J/\Psi}}{\langle n_{J/\Psi} \rangle}=\frac{dN/dy}{\langle dN/dy \rangle},
\end{equation}
therefore
\begin{equation}
\frac{n_{J/\Psi}}{\langle n_{J/\Psi} \rangle}=\langle \rho \rangle \left( \frac{dN/dy}{\langle dN/dy \rangle} \right)^2.
\end{equation}
Note that the linear behavior changes to quadratic at high multiplicities. 
In Fig. \ref{fig23} we show the results together with the experimental data \cite{33} as well as the results for the forward rapidity region together with the experimental data.
In the forward rapidity region we have less number of strings and as a consequence the departure from the linear behavior starts at higher multiplicity. 
In both cases central and forward rapidity region, a good agreement is obtained.
\begin{figure}
\centering
\includegraphics[scale=0.1]{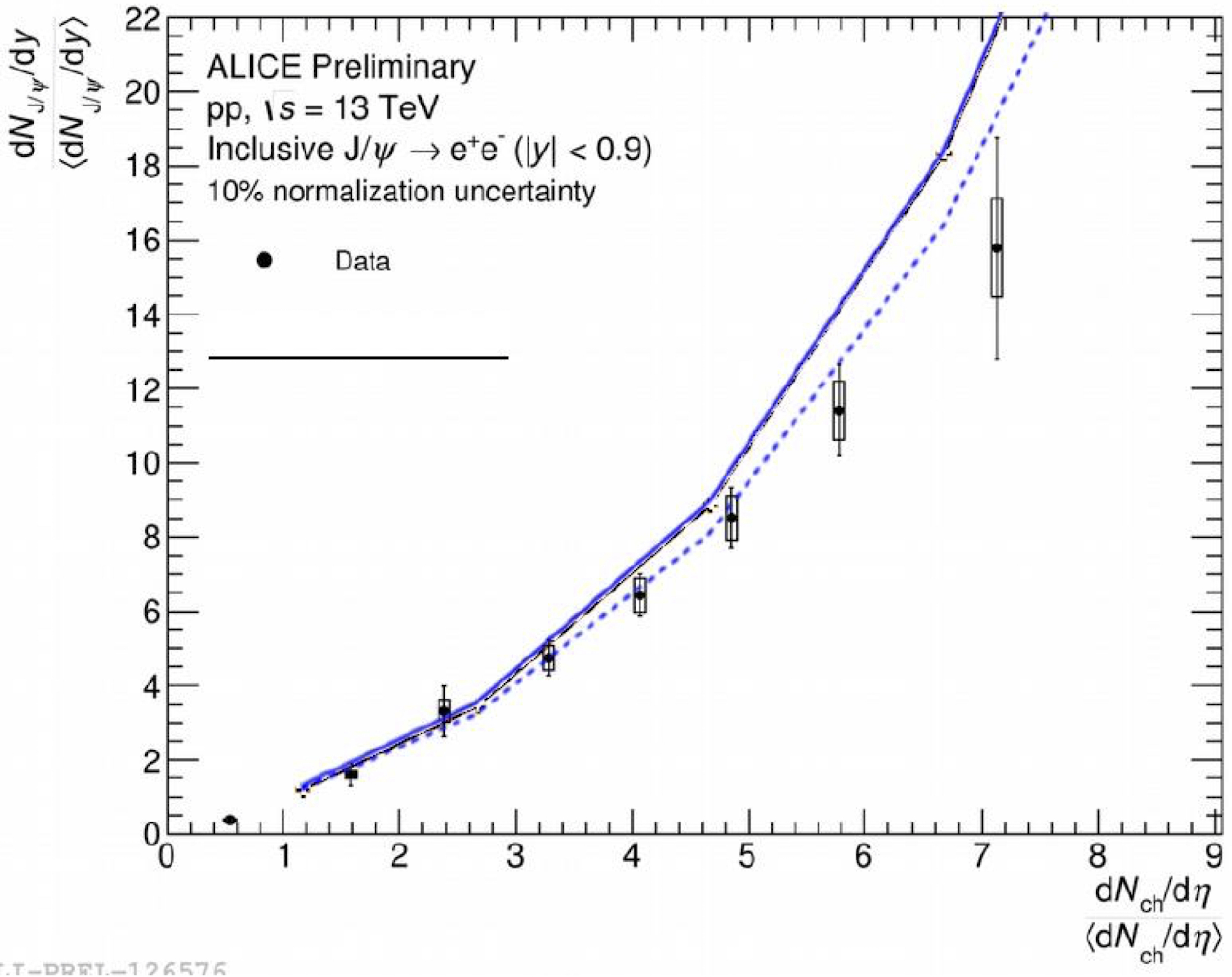}
\caption{Results without and with $J/\Psi$ suppression together the experimental data.}

\label{fig24}
\end{figure} 
Notice that only there are two assumptions, namely, the $J/\Psi$ is produced by a hard mechanism and the attenuation (saturation) of the increase of the multiplicity with $N_s$.
At low multiplicity behaves proportional to $N_s$, but at high multiplicities goes like $\sqrt{N_s}$.
The departure of the linear behavior is a consequence of this attenuation (saturation). 
At 14 TeV there is some possibility that $J/\Psi$ melts \cite{7} due to the high density reached.
In this case there is not nuclear suppression effects, then we assume that the suppression is proportional to the collision area covered by strings.
\begin{figure}
\centering
\includegraphics[scale=0.2]{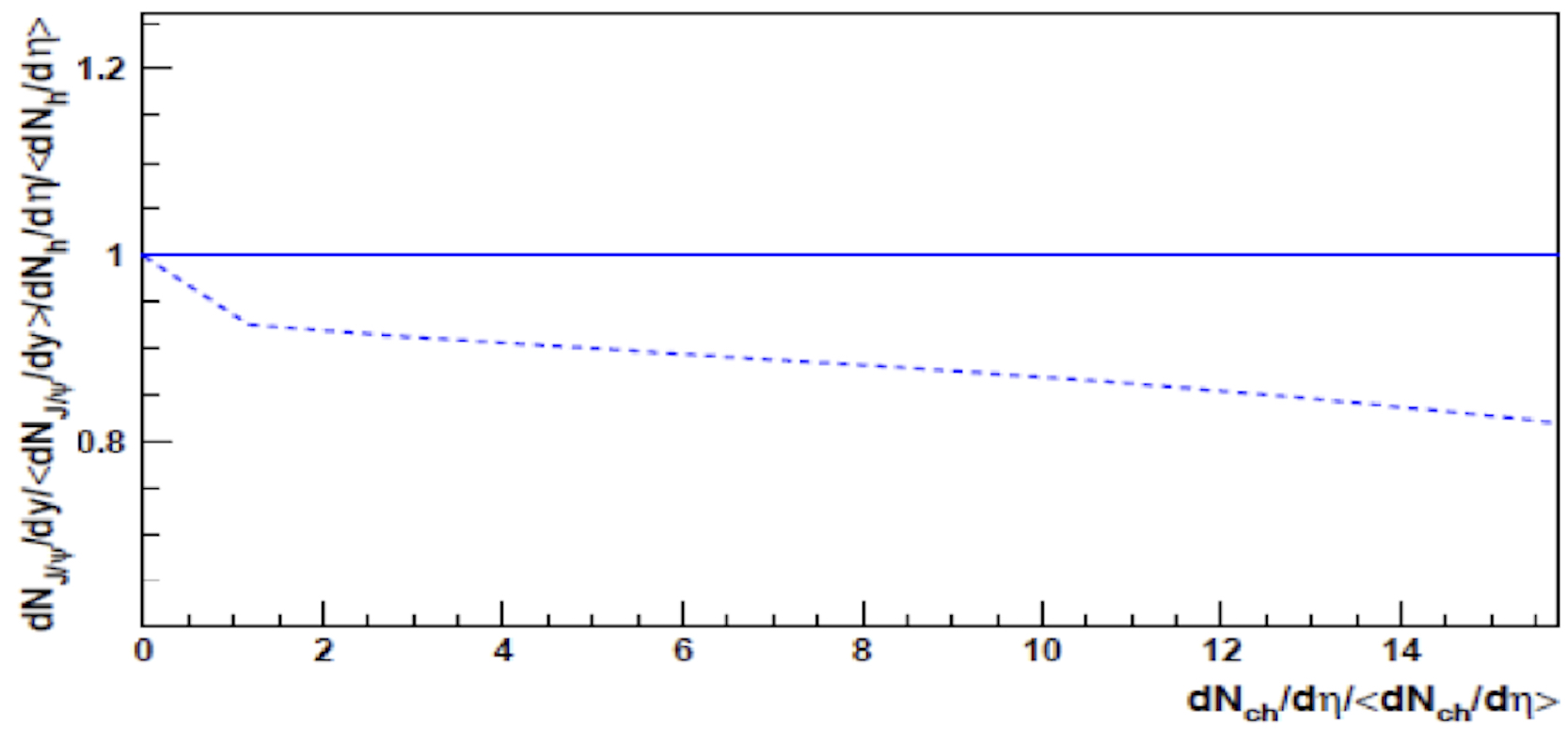}
\caption{Scale ratio of $J/\Psi$ production over the total charged particle production.}

\label{fig25}
\end{figure}
In Fig.~\ref{fig24} we show the results without and with $J/\Psi$ suppression together the experimental data. 
The $J/\Psi$ suppression could be clearly seen by looking at the dependence on the multiplicity of the ratio between the $J/\Psi$ production and events with a high $p_T$ particle(thus with a linear dependence on $N_s$ and consequently on the multiplicity). 
The result for this ratio is shown in Fig.~\ref{fig25}.

\subsection{Incoherent $J/\Psi$ photoproduction}

The incoherent photo production of $J/\Psi$ it has been studied experimentally \cite{162,163} and theoretically \cite{164}. 
The cross section of  $\mathrm{Pbp}\to\mathrm{Pb}J/\Psi X$ probes the fluctuations on the number of elementary scattering of the dipole $q\bar{q}$ (obtained from the virtual photon) on the partons of the proton via the reaction $\gamma (q^2) \mathrm{p} \to J/\Psi X$. 
The increase of these fluctuations gives rise to an increase of the cross section in agreement with data, but as the energy increases the number of elementary collisions increases and assuming that these collisions have a transverse size around 0.3fm, they start to overlap forming clusters of these hot spots. 
Above a critical point, the number of independent sources decreases and so the fluctuations and therefore the cross section. 
Above a critical percolation energy (around 500 GeV), the cross section starts to decrease. 
This prediction can be tested at LHC experiments.

\subsection{Strangeness enhancement}

\begin{figure}
\centering
\includegraphics[scale=.2]{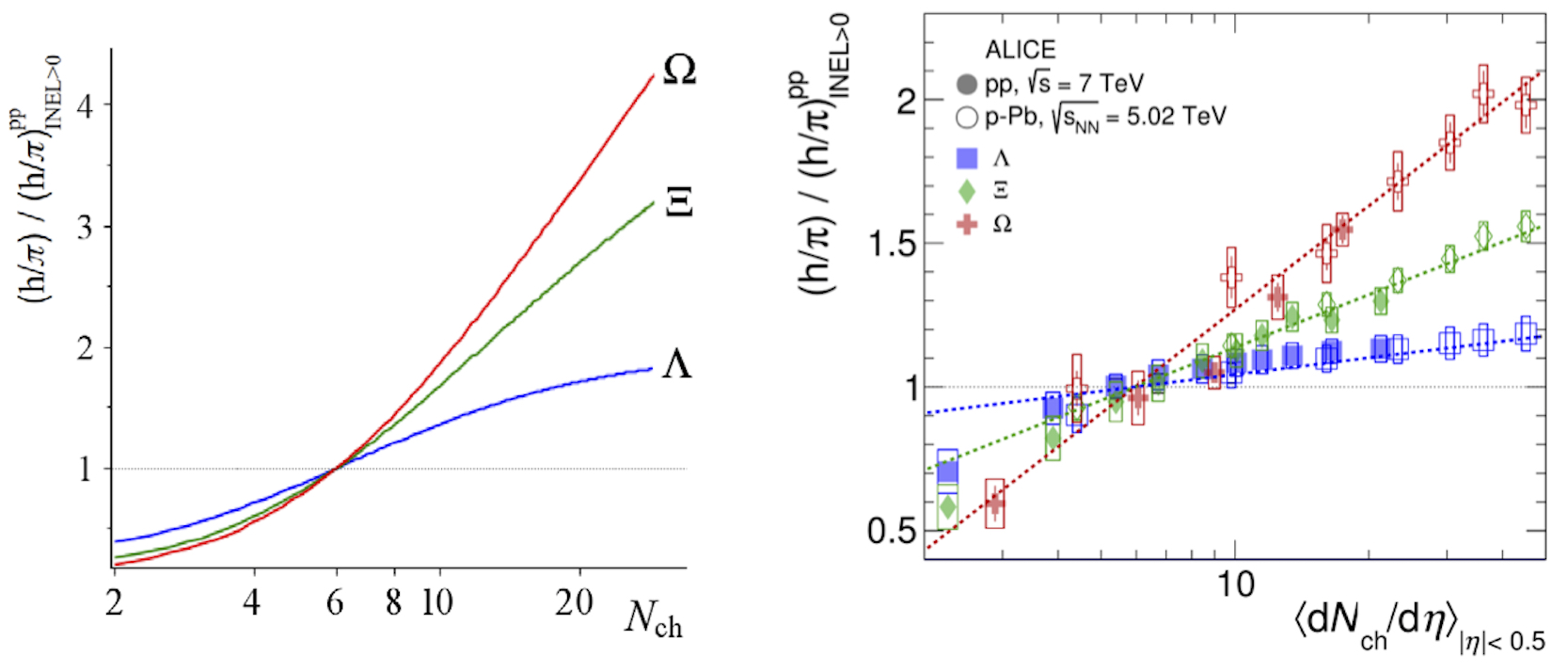}
\caption{Comparison of the multiplicity dependence of the relative yield of $\Omega$, $\Theta$ and $\Lambda$ baryons normalized to pion multiplicity for pp and pPb collisions for the model (left) and experimental data (right) \cite{167}.}

\label{fig27}
\end{figure}

The overlapping of the strings modify the strength of the color field and hence the string tension of the formed cluster. 
Due to this, the decay of these clusters produced naturally an enhancement of the strangeness \cite{167,165,166}. 
In addition to this effect, as the clusters have at their extremes complex flavor $F$ and $\bar{F}$ formed from the individual flavors of the single strings, the decay will produce more baryons and anti baryons than in the fragmentation of single strings. 
There is not any quantitative evaluation of this effect in the production of strange baryons. 
In the case of the strangeness enhancement with multiplicity seen in pp collisions \cite{168}, a simplified model of string percolation which taken into account only the different string tension of the cluster is able to describe qualitatively the data \cite{167}. In Fig.~\ref{fig27} are shown the model results (left) and the experimental data (right).

\section{Azimuthal dependence of the momentum distributions}

\subsection{Collective flow and ridge structure}

\begin{figure}
\centering
\includegraphics[scale=.15]{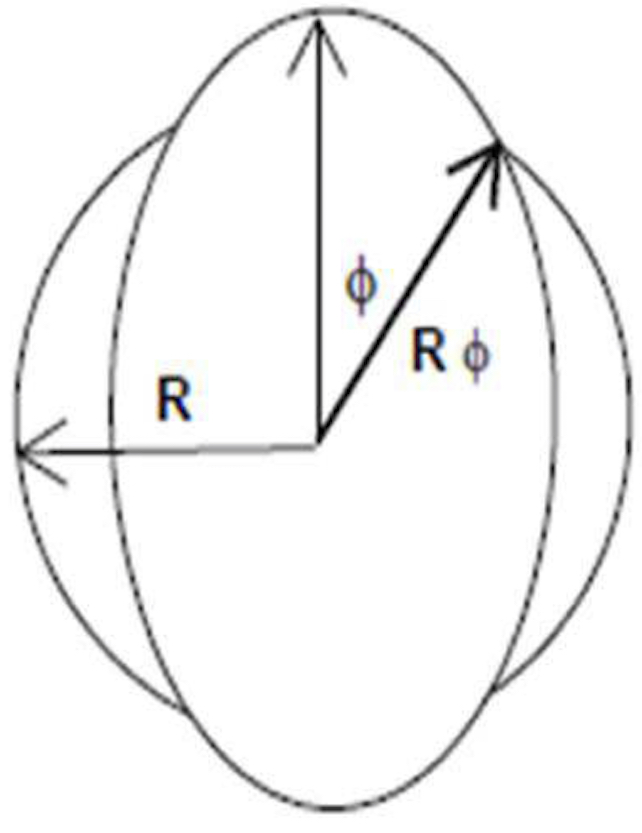}
\caption{Scheme of the azimuthal dependence modify by the escape probability of a parton on the nuclear overlap \cite{171}.}

\label{fig29}
\end{figure}

The clusters formed by the strings have an asymmetric form in the transverse plane and acquires dimensions comparable to the nuclear overlap. 
This azimuthal asymmetry is at the origin of the elliptic flow in string percolation. 
The partons emitted at some point inside the cluster have to pass through the strong color field before appearing in the surface.
The energy loss by the parton is proportional to the length and therefore the $p_T$ of the particle will depend on the direction of the emission as shown in Fig.~\ref{fig29}.
Monte Carlo simulation have been done taking into account this energy loss \cite{169}.
The results of this simulation for the different harmonics \cite{169,170} are in reasonable agreement with experimental data on the $p_T$ and centrality dependence. 
The azimuthal dependence in this way is very similar to evaluate the probability to escape a parton of the nuclear overlap from the initial clusters location.
One way of doing that is defining
\begin{eqnarray}
R_\phi =\frac{R_A\sin(\phi-\alpha)}{\sin\alpha},& & \alpha=\arcsin \left(\frac{b}{2R_A}\sin\phi \right),
\end{eqnarray}
where $b$ is the impact parameter. We also define $\rho_\phi=\rho(R/R_\phi)^2$, and substituting in the $p_T$ distribution, we obtain
\begin{equation}
f(F(\rho_\phi),p_T^2)\simeq f(F(\rho),p_T^2)\left[ 1+\frac{\partial \ln f(F(\rho),R^2)}{\partial R^2}(R_\phi^2-R^2) \right].
\end{equation}
Thus, the elliptic flow can be computed as follows
\begin{equation}
v_2(p_T^2)=\frac{2}{\pi}\int_0^{\pi/2} d\phi \cos(2\phi)\left[ 1+\frac{\partial \ln f(F(\rho),R^2)}{\partial R^2}(R_\phi^2-R^2) \right].
\label{eq105}
\end{equation}
Note that the latter is an analytical close expression for all energies, centralities, projectiles, and targets.

\begin{figure}
\centering
\includegraphics[scale=0.2]{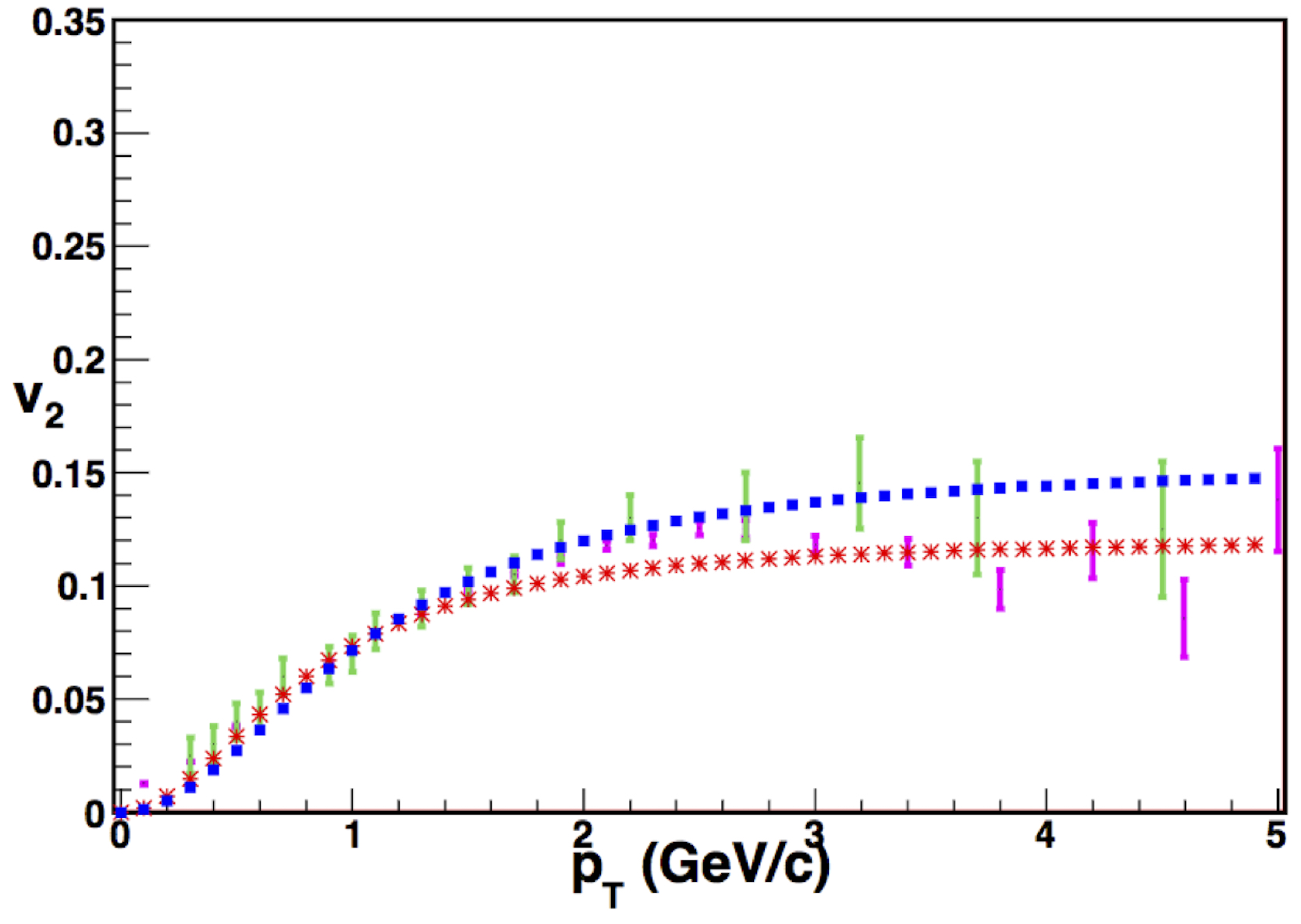}
\caption{Comparison between the prediction of percolation model (red stars and blue squares) and the experimental data (error-bars in green and pink) for $\sqrt{s}=200$ GeV and $\sqrt{s}=2.76$ TeV (centralities 10\%-20\%) \cite{172}.}

\label{fig30}
\end{figure}

The transverse momentum dependence of $v_2$ computed using Eq.~\eqref{eq105} for Pb-Pb at 2.76 TeV and Au-Au at 200 GeV for 10\%-20\% centrality is shown in Fig.~\ref{fig30} together the experimental data. 
A good agreement is also obtained at all centralities and rapidities \cite{171,172,173} as well as the hierarchy on $v_2$ of $\pi$, $k$ and p. 
The ridge structure was seen first at RHIC in Au-Au and later at LHC in Pb-Pb collisions. 
This structure has been also observed in pp and pA collisions at high multiplicity at LHC, as it was anticipated by string percolation \cite{174}.

\begin{figure}
\centering
\includegraphics[scale=0.2]{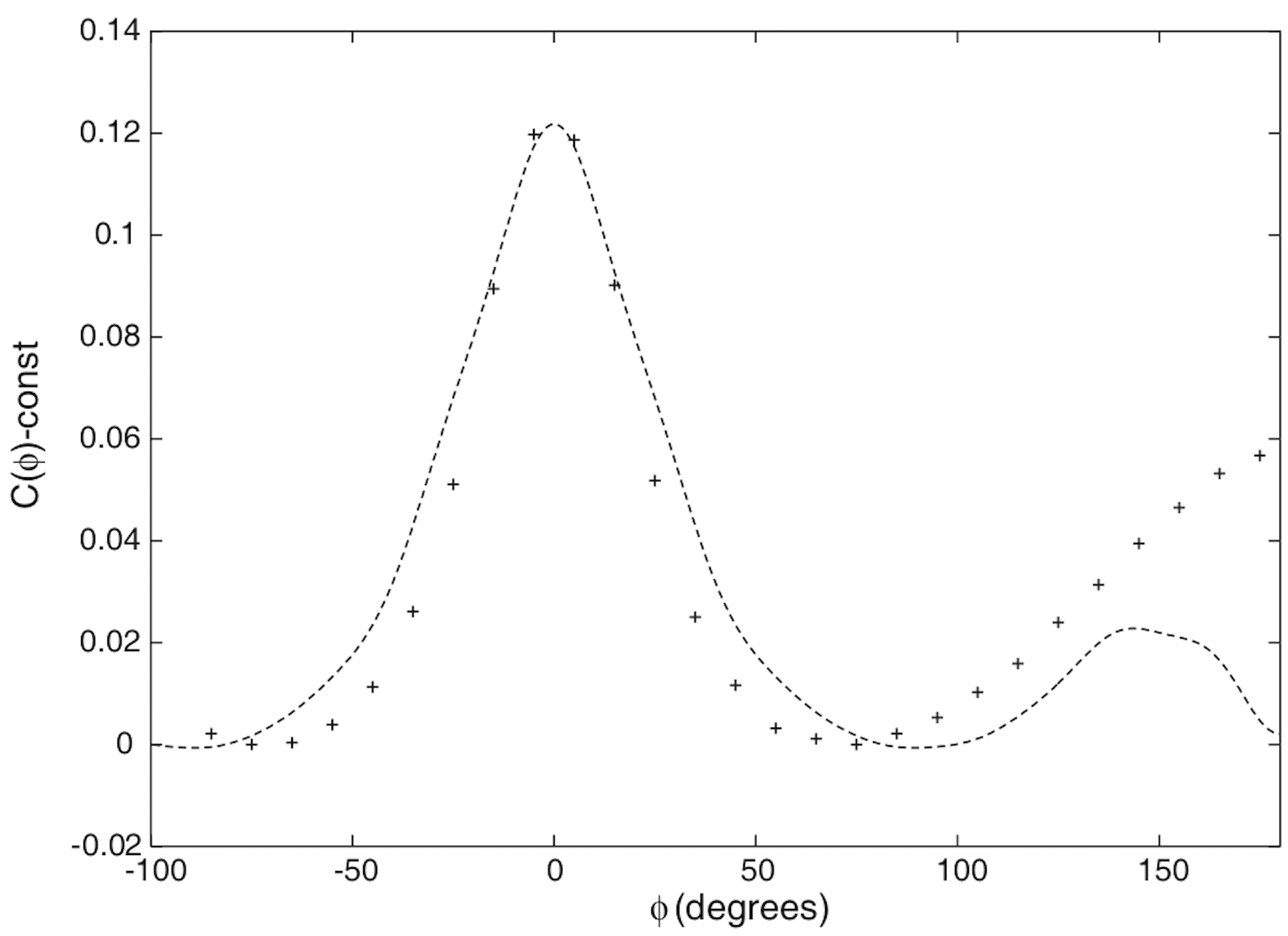}
\caption{Correlation coefficient $C(\phi)$ for pp collisions at 7 TeV with triple multiplicity \cite{175}.}

\label{fig31}
\end{figure}

In string percolation correlations can arise from the superposition of many events with different number and type of string. 
In this way, there appears long range correlations in rapidity.
However passing to the azimuthal dependence, if the emission of strings is isotropic, the correlations due to their distribution in different events will be also isotropic. 
Also in the central rapidity region, the inclusive cross section is approximately independent of rapidity. 
This generates a plateau in the $y-\phi$ distribution rather than a ridge, with only a peak at small y and $\phi$ due to short range correlations. 
This conclusion is also valid if one averages the inclusive cross sections over all events with the resulting loss of azimuthal angle dependence. 
So the ridge can only be obtained in an event by event basis. In this way we performed our evaluations \cite{173}.

\begin{figure}
\centering
\includegraphics[scale=0.2]{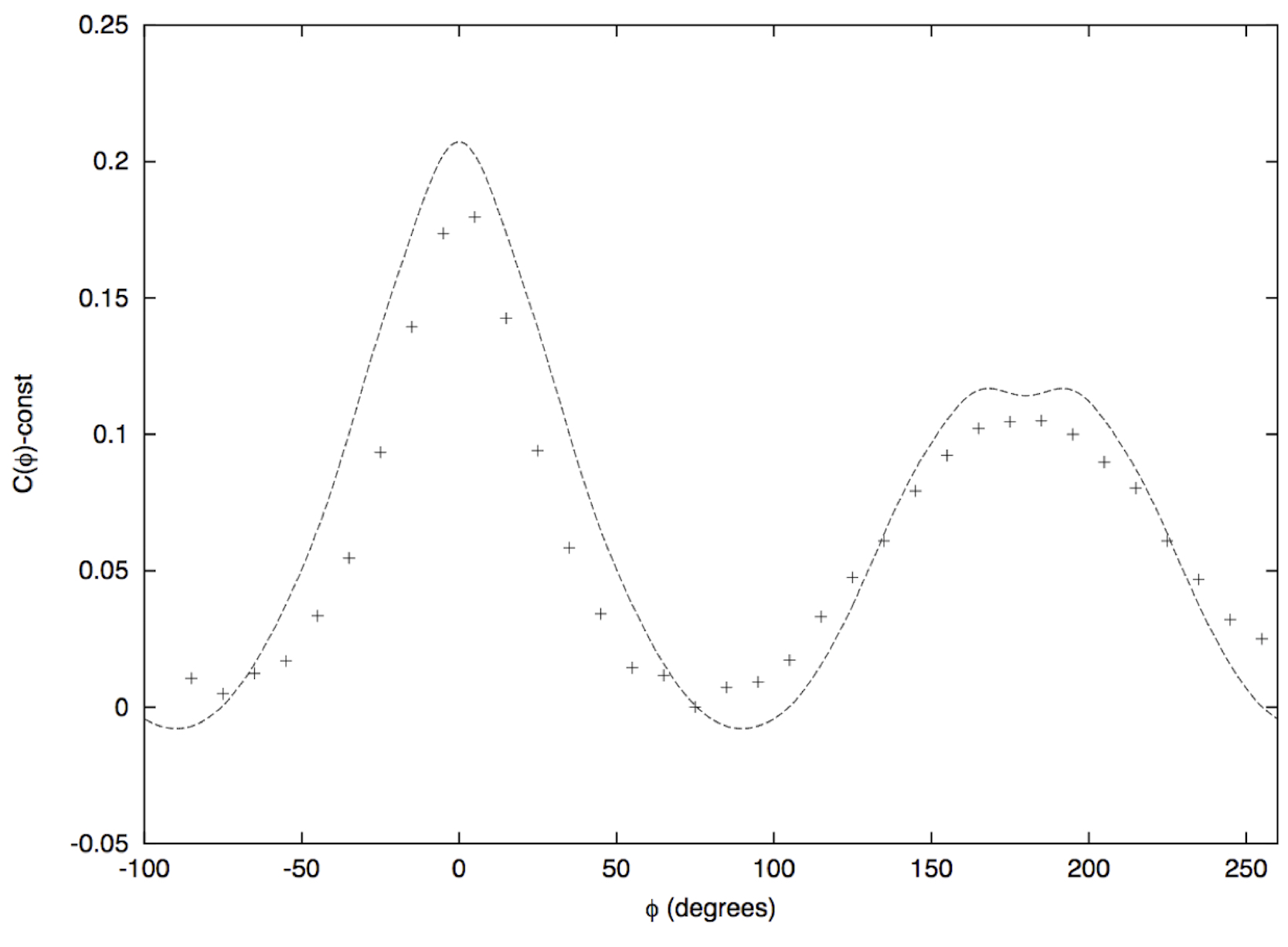}
\caption{Correlation coefficient $C(\phi)$ for p-Pb collisions at 5.02 TeV for central collisions compared to the data (ZYAM procedure) \cite{175}.}

\label{fig32}
\end{figure}

\begin{figure}
\centering
\includegraphics[scale=0.2]{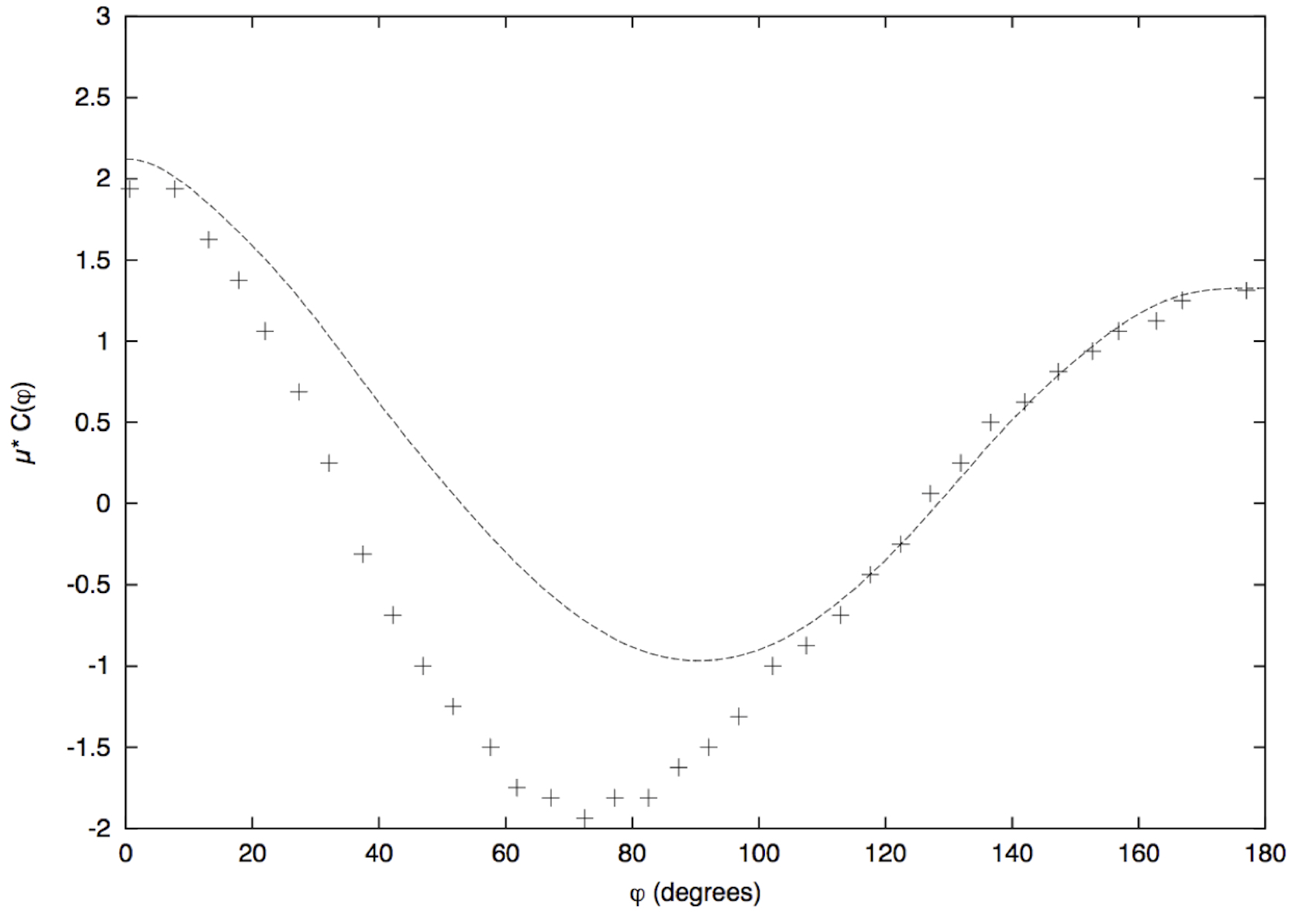}
\caption{Correlation coefficient $C(\phi)$ for Au-Au at 200 GeV for 10\% of the most central events against the experimental data \cite{175}.}

\label{fig33}
\end{figure}

In Fig.~\ref{fig31} we show the results \cite{175} for $C(\phi_{12})$ for event with triple multiplicity than minimum bias in pp collisions at 7 TeV compared to the experimental data. 
In Fig.~\ref{fig32} we show the results for central pPb at 5.02 TeV compared to the data and in Fig.~\ref{fig33} the results for Au-Au 0\%-10\% of centrality at 200 GeV and its comparison with experimental data \cite{175}. 
An overall agreement is obtained in spite of the approximations done in the computation.

In the case of pp collisions, to obtain the ridge structure we need to consider high multiplicity events (three times the minimum bias multiplicity shown in Fig.~\ref{fig33}). This is due to the fluctuations needed to have sizable long correlations which are only obtained for these events. These fluctuations are also crucial to describe the higher harmonics of the azimuthal distributions.

We can conclude that string percolation is able to describe the ridge structure seen in pp, pA and AA collisions. 
The ridge is obtained from the superposition of many events with different number and types of clusters of strings. 
There is not any essential difference between high multiplicity pp for pA collisions and AA collisions. 
The collective flow is obtained from the configuration of the initial state as clusters of overlapping clusters and the interaction of the produced partons with the color field of the clusters. 
This interaction could be interpreted as final state interaction but as far as the parton have these interactions before hadronization should be regarded as well as initial state interaction.
In the production of heavy particles, due to their short formation time, they can be formed before than the parton get out the surface collision area. This is certainly true for central heavy ion collisions. In this case, the energy loss by the parton would be smaller and thus the elliptic flow. As the elliptic flow for central collisions is small, the effect is difficult to be observed.

\subsection{Elliptic flow scaling and energy loss}

\begin{figure}
\centering
\includegraphics[scale=0.15]{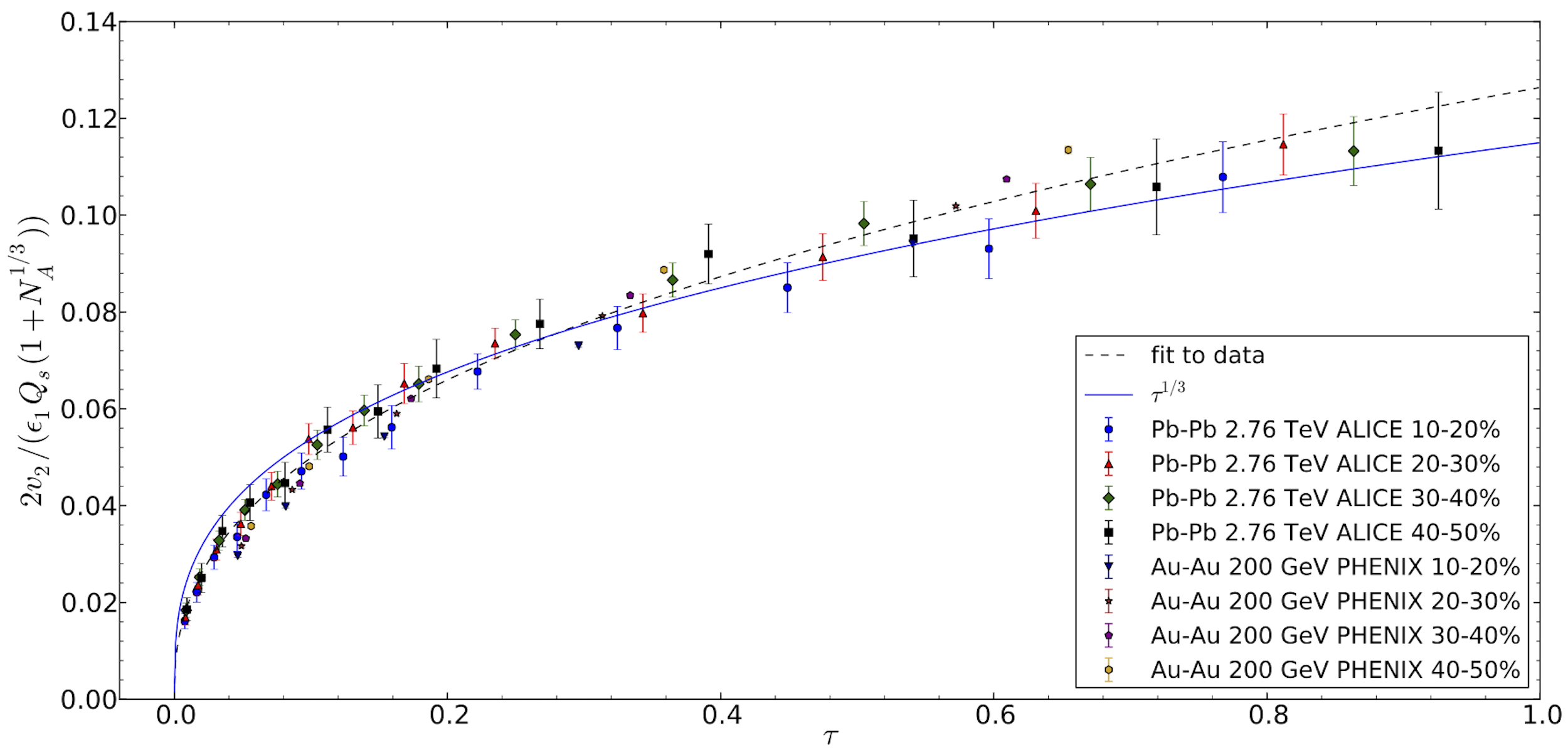}
\caption{$v_2$ scaled by $\epsilon_1 Q^A_s L$ for 10-20\%, 20-30\%, 30-40\% and 40-50\% Au-Au collisions at 200 GeV and Pb-Pb collisions at 2.76 TeV vs $\tau$. The dashed black line is a fit to data according to $a\tau^b$ with $a$=0.126$\pm$0.0076 and $b$=0.404$\pm$0.025, solid blue line corresponds to $\tau^{1/3}$ \cite{179}.}

\label{fig34}
\end{figure}

In Sec.~\ref{quenching}, we discussed the quenching of low $p_T$ partons. 
A parton emitted from the decay of a cluster with tension $t$ and momentum $p$ due to the energy loss in his way to get out the overlap collision area, obeys the distribution
\begin{equation}
P(p_T,\phi)=C\exp(-p_T/T)\exp(-8p_T^{2/3}T^{1/3}l(\phi)).
\end{equation}
Here the temperature, $T$, is proportional to the squared root of the string tension t. The departure from the thermal distribution is due to the quenching formula in Eq.~\eqref{eq42}.
The length $l(\phi)$ is the length of the path needed by the parton to get out. 
We will take proportional to the product of the eccentricity of the overlap area and $L$ which is the length independent of the eccentricity that we will take proportional to $(1-N_A^{1/3})/2$, the number of collisions of a parton with a nucleus. We define the eccentricity as
\begin{equation}
\epsilon=\frac{2}{\pi}\int_0^\pi d\phi \cos 2\phi \frac{R^2-R^2_\phi}{R^2}.
\end{equation}
We expect that the elliptic flow be proportional to the strength of the quenching, so
\begin{equation}
v_2\sim p_T^{2/3}T^{1/3}L\epsilon.
\label{eq108}
\end{equation}
Using the dependence of $Q_s$ on the energy and centrality \cite{173} and taking $T$ proportional to $Q_s$, we have that \cite{179}
\begin{equation}
\frac{v_2}{Q_s^A\epsilon L} \sim \left(\frac{p_T}{Q_s} \right)^{2/3}=\tau^{1/3},
\end{equation}
where we have choice the scaling variable $\tau=p_T^2/Q_s^2$.
In Fig.~\ref{fig34}, the experimental data of Phenix \cite{180} and ALICE \cite{181} at different centralities are shown versus the scaling function $\tau^{1/3}$. Also the best fit of the form $t^b$ is shown, giving a value of $b$=0.404, which is not very different from 1/3. Taking into account the crude approximations done in deriving the scaling formula Eq.~\eqref{eq108}, the result is very remarkable, confirming the quenching of partons inside the overlap surface of the colliding objects.

\section{Thermodynamics of string percolation}

The thermodynamics of the string percolation can be addressed by extracting the temperature from the transverse momentum distribution. We also can extract the suppression factor $F (\rho)$ and hence the local initial temperature as well as the Bjorken initial energy density $\epsilon$, which are given by \cite{116,118}
\begin{eqnarray}
T=\sqrt{\frac{\langle p_T^2 \rangle_1}{2F(\rho)}} ,& & \epsilon=\frac{3}{2}\frac{\langle m_T \rangle}{S\tau_p} \frac{dN}{dy},
\end{eqnarray}
where $S$ is the overlap area and $\tau_p$ is the production time, that we take $\tau_p=2.4 \hbar /\langle m_T \rangle$ \cite{183}.
\begin{figure}
\centering
\includegraphics[scale=.25]{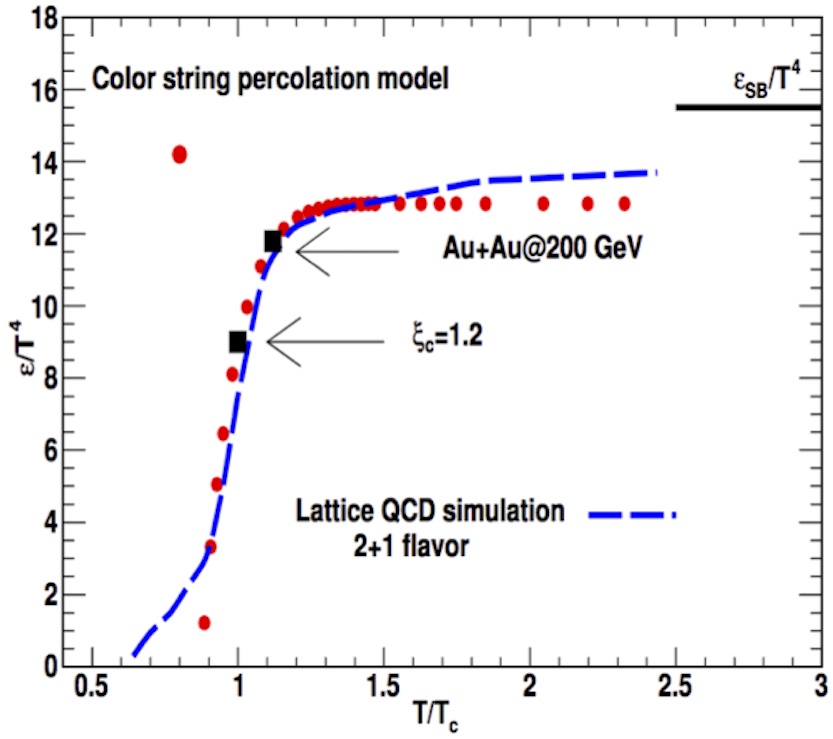}
\caption{Energy density $\epsilon/T^4$ vs $T/T_c$.}

\label{fig35}
\end{figure}
In Fig.~\ref{fig35}, we plot the obtained energy density over $T^4$ as a function of $T/T_c$ together the lattice result. 
Notice that $T$, which characterizes the percolation clusters, measures the initial temperature of the system, since the clusters cover most of the area of the collision this local temperature becomes a global temperature. 
In this way, the critical string density corresponds to the critical temperature. 
In relativistic kinetic theory the ratio between the shear viscosity and the entropy density is give by \cite{184}
\begin{equation}
\frac{\eta}{s}=\frac{T\lambda_{mfp}}{5},
\end{equation}
where the mean free path is $\lambda_{mfp}\sim1/n\sigma_{tot}$, being $n$ the number density of a free gas of quarks and gluons, and $\sigma_{tot}$ the transport cross section.
In string percolation the density numbers is the effective number of sources per unit of volume is $n=N_{sources}/5L$ \cite{54}, where $L$ is the longitudinal string length $\sim$1 fm. 
The effective number of sources is the area covered by strings $(1-\exp (-\rho))S$ divided by the area of one effective string $F(\rho)S_1$. 
On the other hand, the transport cross section is the area of the effective string. 
Collecting all these, we have
\begin{equation}
\frac{\eta}{s}=\frac{TL}{5(1-\exp(-\rho))}.
\end{equation}
In Fig.~\ref{fig36}, we show the behavior of $\lambda_{mfp}$, $T$ and $\lambda_{mfp}T$ as a function of the string density. 
\begin{figure}
\centering
\includegraphics[scale=.2]{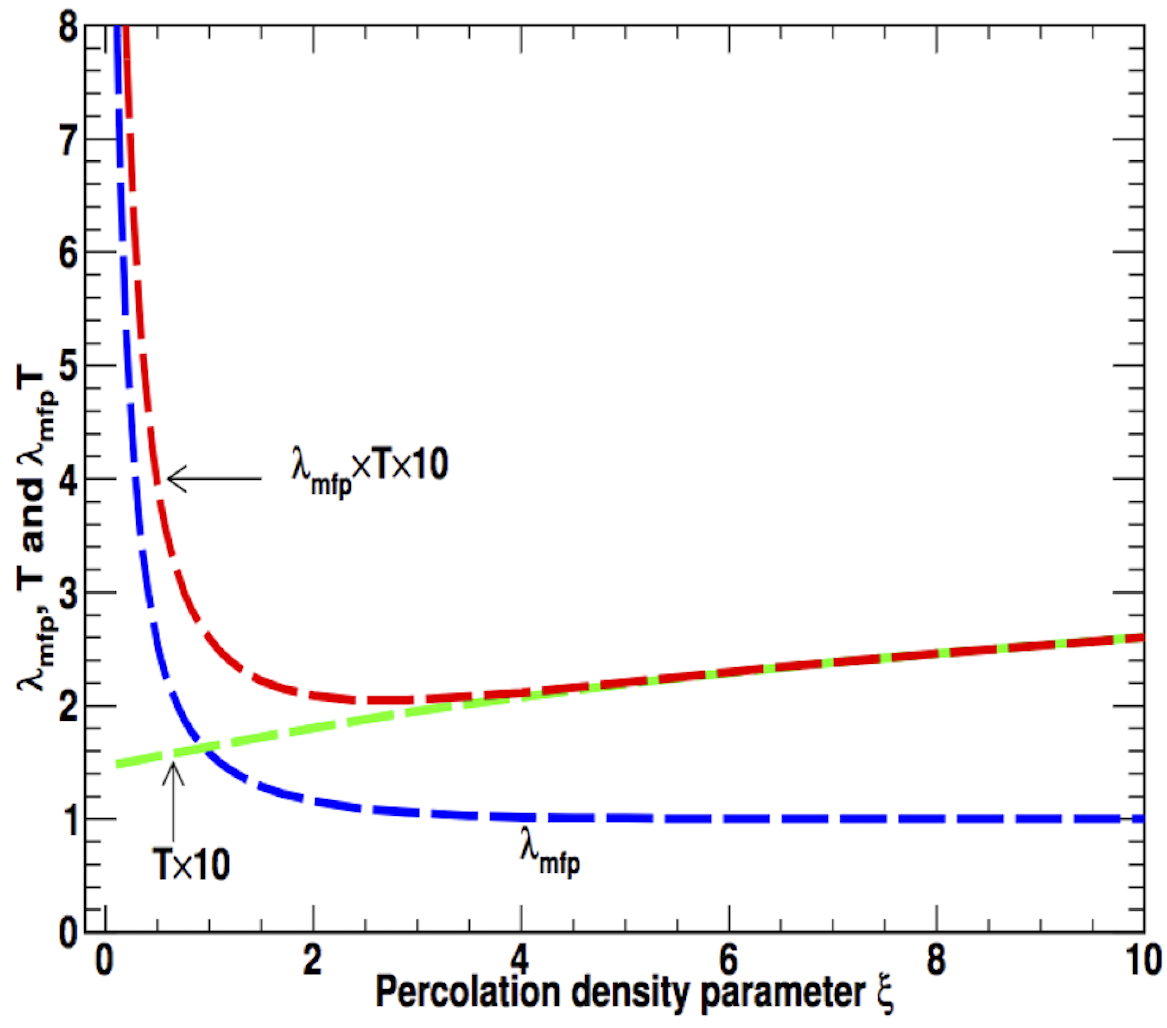}
\caption{$\lambda_{mfp}$, $T$ and $\lambda_{mfp}T$ as functions of the percolation density parameter $\rho$.}

\label{fig36}
\end{figure}
In Fig.~\ref{fig37}, we show the ratio $\eta/s$ as a function of the temperature. In the same figure are plotted evaluations in the case of weak QGP and string quark gluon plasma (sQGP) as well as AdS/CFT result \cite{18}. 
\begin{figure}
\centering
\includegraphics[scale=.2]{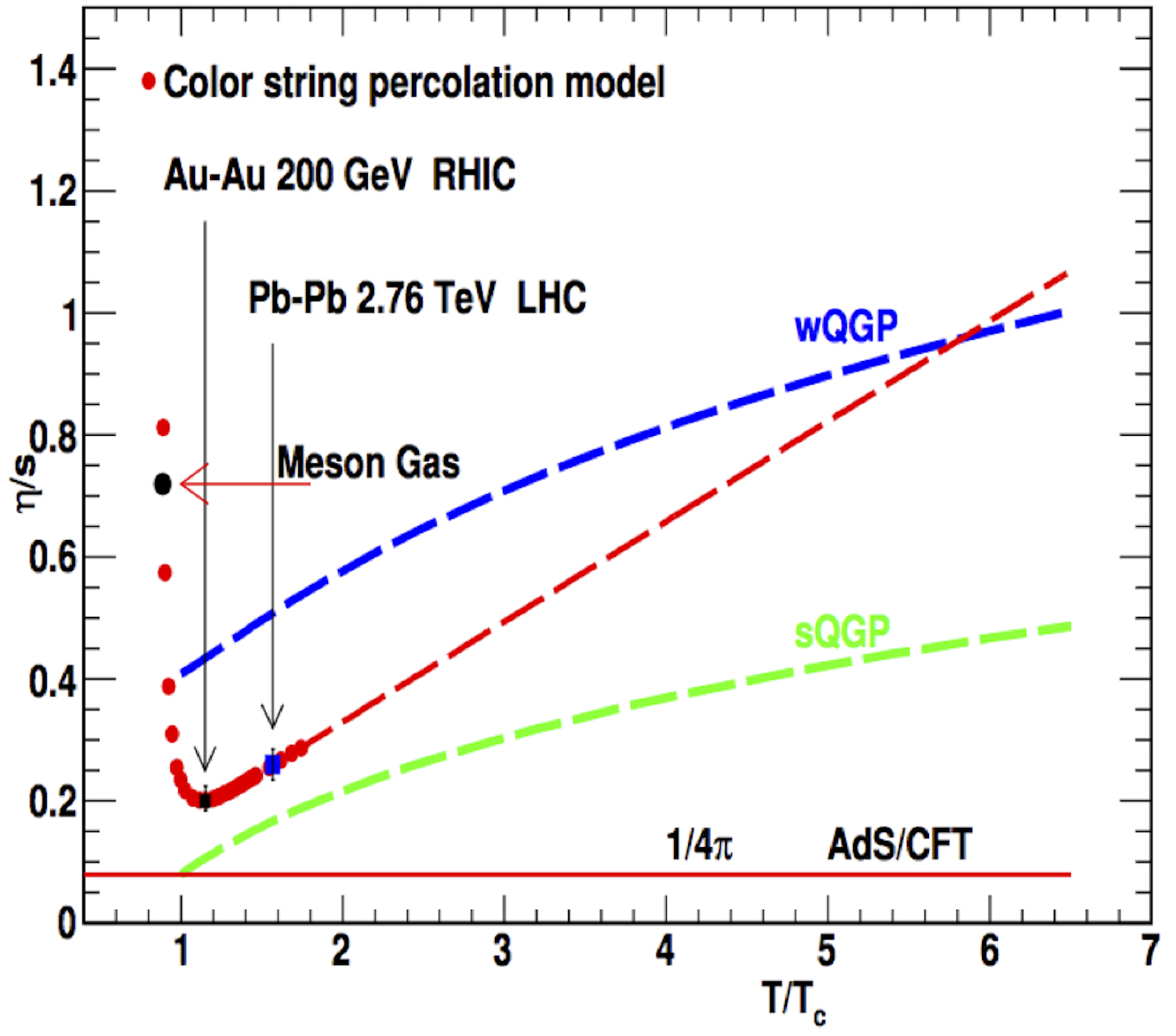}
\caption{Shear viscosity ratio $\eta/s$ vs $T/T_c$.}

\label{fig37}
\end{figure}
The arrows marks are the result of string percolation for Au-Au and Pb-Pb at RHIC and LHC energies. 
Below $\rho_c$ as the temperature becomes close to $T_c$, the string density increases and the area is filled rapidly and $\lambda_{mfp}$ and $\eta/s$ decrease sharply.
Above $Tc$, the area is not covered as fast and the relatively decreasing of $\lambda_{mfp}$ is compensated by the rising of temperature, resulting in a smooth increase of the $\eta/s$. 
The behavior of $\eta/s$ is governed by the fractional are covered by strings, what is not surprising because $\eta/s$ is the ability to transport momenta at large distances and that has to do with the density of voids in the matter. 
Notice that the values of the ratio for high $T$ values approach the weak coupling limit.

\begin{figure}
\centering
\includegraphics[scale=.2]{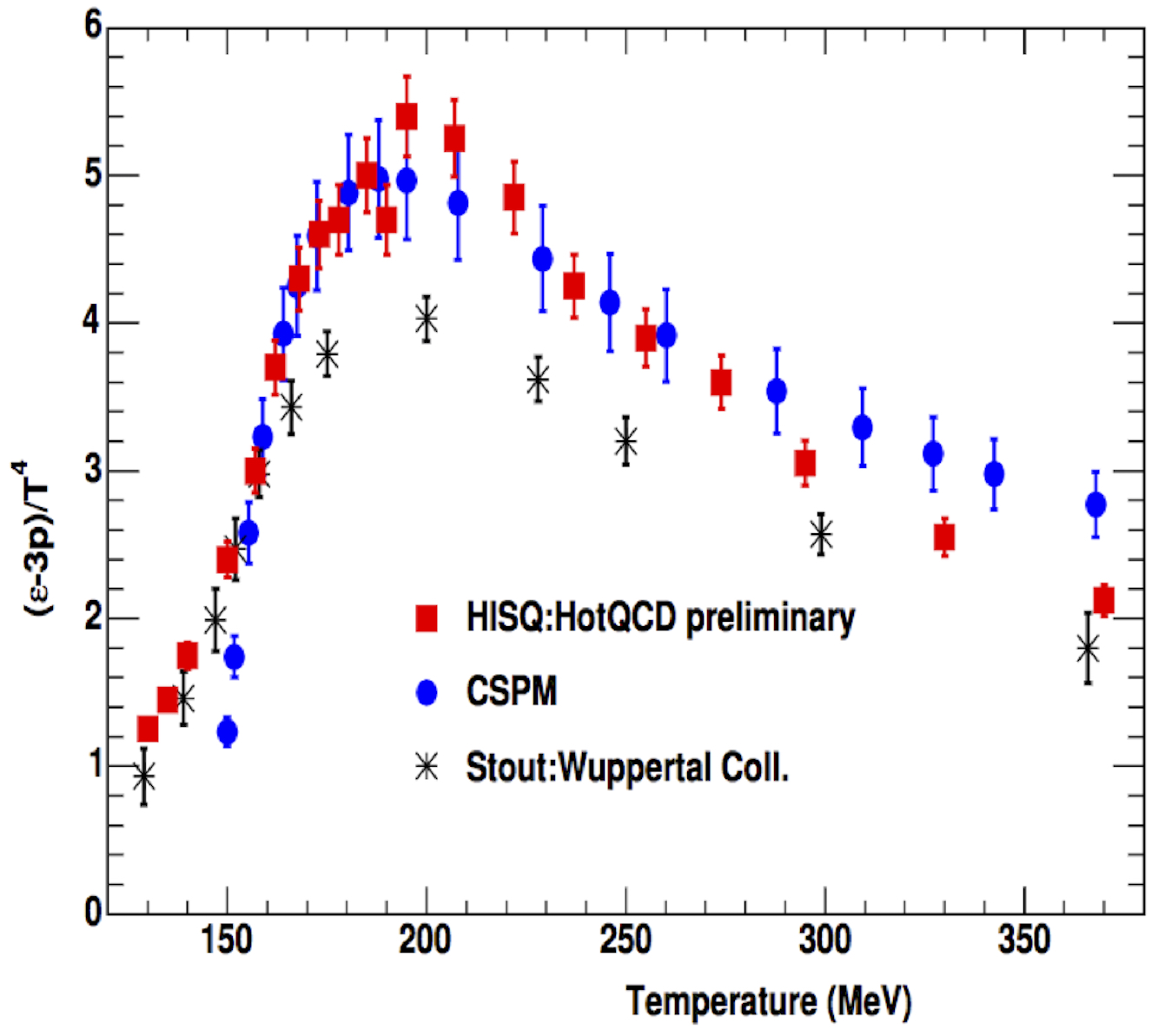}
\caption{Comparison between the trace anomaly of the energy momentum tensor and inverse of the $\eta/s$ ratio. Note that both variables have a maximum value at the same temperature point.}

\label{fig38}
\end{figure}

Moreover, the mean value of the trace of the energy momentum tensor $T^\mu_\mu=\epsilon-3P$ is a measure of the deviation of the conformal behavior and thus identifies the interaction still present in the medium. 
In a classical theory with massless quarks vanishes, but in any quantum field theory is not zero, because the scale needed to be renormalized, breaking the conformal symmetry. 
It is the well known trace anomaly.
We find that the reciprocal of $\eta/s$ is in quantitative agreement with the trace anomaly over a wide range of temperatures. 
The minimum corresponds to the maximum of $(\epsilon-3P )/T$ as it is seen in Fig.~\ref{fig38}.

\begin{figure}
\centering
\includegraphics[scale=.2]{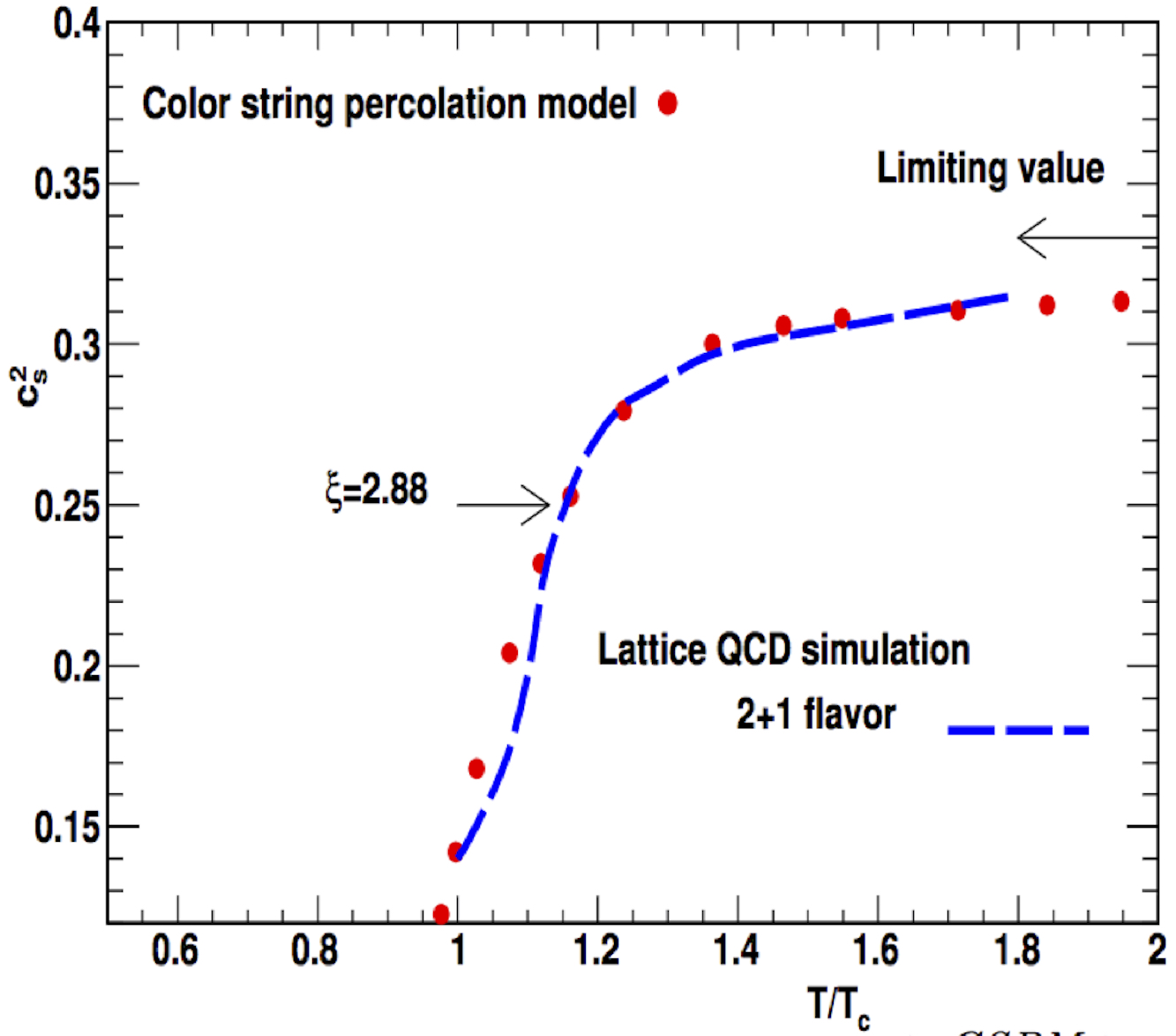}
\caption{Squared sound speed as a function of $T/T_c$.}

\label{fig40}
\end{figure}

On the other hand, it is possible to determine the speed of sound, $c_s$, by assuming the 1D Bjorken expansion, using the energy density, the initial temperature, and the trace anomaly given by the string percolation. Starting from the equations
\begin{eqnarray}
\frac{1}{T} \frac{dT}{d\tau}=-\frac{c_s^2}{\tau}, & & \frac{dT}{d\epsilon} \frac{d\epsilon}{d\tau}=\frac{T}{\tau},
\end{eqnarray}
where $\tau$ is the proper time and $c_s$ is the sound speed. Since $s=(\epsilon+P)/T$ and $\Delta=(\epsilon-3P)/T^4$, one gets
\begin{equation}
\frac{dT}{d\epsilon}=\frac{c_s^2}{s}.
\end{equation}
From the above equations, it is possible write $c_s$ in terms of $\rho$ in the following way
\begin{equation}
c_s^2=-\frac{1}{3}\left(\frac{\exp(-\rho)}{F(\rho)^2}-1 \right)+0.0191\left(\frac{\Delta}{3} \right)\left( \frac{\exp(-\rho)}{\rho F(\rho)^4}-\frac{1}{\rho F(\rho)^2} \right),
\end{equation}
where $F(\rho)$ is the scaling function in Eq.~\eqref{eq37}.
In Fig. ~\ref{fig40}, we show $c_s$ as a function of the temperature. 
It is observed a very good agreement with lattice calculations.

\begin{figure}
\centering
\includegraphics[scale=.2]{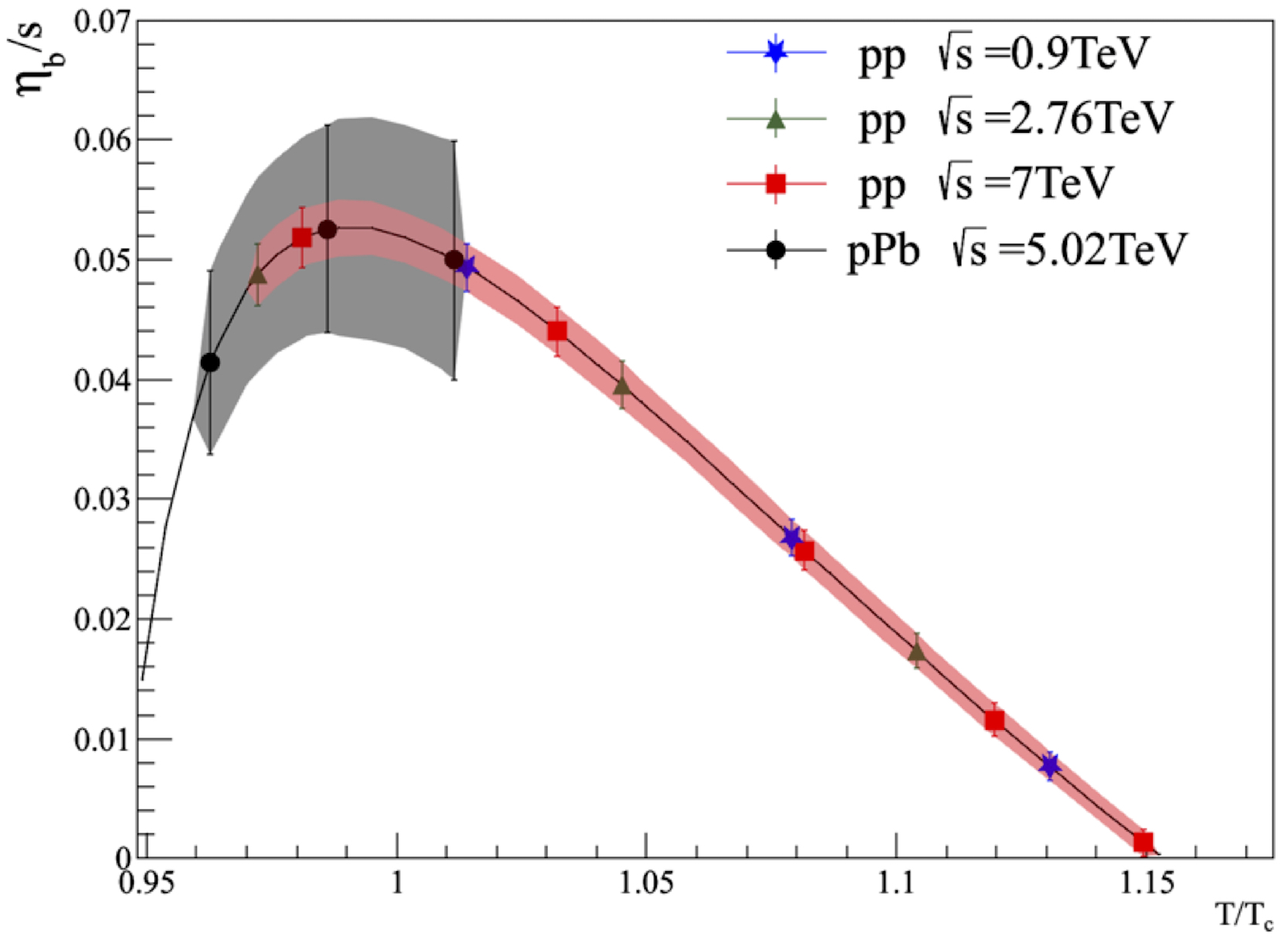}
\caption{Ratio between the bulk viscosity and the entropy density as a function of $T/T_c$.}

\label{fig39}
\end{figure}

Another interesting thermodynamic variable which can be determined is the bulk viscosity. Starting from \cite{bvis}
\begin{equation}
\frac{\eta_b}{\tau_\Pi}=\left(\frac{1}{3}-c_s^2 \right)(\epsilon+P)-\frac{\alpha}{p}(\epsilon-3P),
\end{equation}
where $\tau_\Pi$ is the corresponding relaxation time. Substituting the entropy density and the trace anomaly in the latter, we found that
\begin{equation}
\frac{\eta_b}{\tau_\Pi s}=\left(\frac{1}{3}-c_s^2 \right)T-\frac{\alpha}{9}\frac{\Delta T^4}{s}.
\end{equation}
Note that this last expression depends on the sound speed, trace anomaly and entropy density, which has already been computed in the string percolation context.
In Fig.~\ref{fig39}, we plot the bulk viscosity over the entropy density as a function of the temperature, which has a maximum close to $T_c$.

\section{Summary}

The string percolation describes successfully most of the experimental data in the soft region, namely, rapidity distributions, probability distributions of multiplicities and transverse momentum, strength of BE correlations as a function of multiplicities, forward-backward multiplicities as $p_T$ correlations, strangeness enhancement, elliptic flow and ridge structure. 

The string percolation although is not derived directly from QCD has a clean physical grounds and has the fundamental QCD feature. The non abelian character is reflected in the coherent sum of the color fields which gives rise to an enhancement of the mean $p_T$ and a suppression of the multiplicity. The confinement of the fields is reflected in the small transverse size of the strings as well as the transverse correlations length. The scaling observed in the transverse momentum distribution is a consequence of the invariance under the size of the clustersof strings. 

The collective behavior of the multiparticle production has its origin in the cluster configuration formed in the initial state of the collisions followed by the interactions between the produced partons with the color fields, given rise to energy loss. Due to that, the elliptic flow satisfies an universal scaling law valid for all centralities and energies. At low $p_T$, the thermal distribution of the $p_T$ distributions allows us to define temperature as a function of the string density, which can be regarded, at large cluster size, as the global temperature and compute the energy and entropy density, which presents a jump at the critical temperature corresponding to the percolation critical density. 
Using the kinetic transport theory, it is shown that the ratio between the shear viscosity over entropy density, which presents a minimum close to $T_c$ (and a maximum in the bulk viscosity). 
The inverse of this ratio is very close to the trace anomaly, including it smooth decreasing with temperature. 
The behavior of the speed of sound with temperature is also in agreement with lattice QCD calculations. 
It is remarkable that string percolation reproduce the behavior of the main thermodynamics magnitudes as functions of the temperature.

\section*{Acknowledgments}
C.P. was supported by the grant Maria de Maeztu Unit of Excellence MDM-20-0692 and FPA project 2017-83814-P of Ministerio de Ciencia e Inovación of Spain, FEDER and Xunta de Galicia.
I.B. thanks the CONACYT cátedras project 043 and the WS. Grant Loreal UNESCO CONACYT AMC.
J.E.R. acknowledges financial support from CONACYT (postdoctoral fellowship Grant no. 289198). We also thank Prof. M. A. Braun, J. Dias de Deus, R. Scharenberg, B. Srivastava for their collaboration in the work presented here.

\bibliographystyle{ieeetr}
\bibliography{bib.bib}

\begin{thebibliography}{100}

\bibitem{1}
T.~D. Lee and G.~C. Wick, ``Vacuum stability and vacuum excitation in a spin-0
  field theory,'' {\em Phys. Rev. D}, vol.~9, pp.~2291--2316, 1974.

\bibitem{2}
J.~C. Collins and M.~J. Perry, ``Superdense matter: Neutrons or asymptotically
  free quarks?,'' {\em Phys. Rev. Lett.}, vol.~34, pp.~1353--1356, 1975.

\bibitem{3}
N.~Cabibbo and G.~Parisi, ``Exponential hadronic spectrum and quark
  liberation,'' {\em Phys. Lett. B}, vol.~59, no.~1, pp.~67 -- 69, 1975.

\bibitem{4}
E.~Shuryak, ``Quark-gluon plasma and hadronic production of leptons, photons
  and psions,'' {\em Phys. Lett. B}, vol.~78, no.~1, pp.~150 -- 153, 1978.

\bibitem{5}
J.~Hofmann, H.~Stocker, W.~Scheid, and W.~Greiner, ``Bear mountain workshop,''
  {\em Z. Phys. A}, vol.~273, p.~359, 1975.

\bibitem{6}
J.~D. Bjorken, ``Highly relativistic nucleus-nucleus collisions: The central
  rapidity region,'' {\em Phys. Rev. D}, vol.~27, pp.~140--151, 1983.

\bibitem{7}
T.~Matsui and H.~Satz, ``{$J/\Psi$} suppression by quark-gluon plasma
  formation,'' {\em Phys. Lett. B}, vol.~178, no.~4, pp.~416 -- 422, 1986.

\bibitem{8}
J.~D. Bjorken, ``Energy loss of energetic partons in quark-gluon plasma:
  possible extinction of high {$p_t$} jets in hadron-hadron collisions,'' Tech.
  Rep. Pub-82/59-THY, Fermi-Lab, 1982.

\bibitem{9}
M.~Gyulassy and X.~nian Wang, ``Multiple collisions and induced gluon
  bremsstrahlung in {QCD},'' {\em Nucl. Phys. B}, vol.~420, no.~3, pp.~583 --
  614, 1994.

\bibitem{10}
G.~Baym, ``Confinement of quarks in nuclear matter,'' {\em Physica A}, vol.~96,
  no.~1, pp.~131 -- 135, 1979.

\bibitem{11}
T.~Celik, F.~Karsch, and H.~Satz, ``A percolation approach to strongly
  interacting matter,'' {\em Phys. Lett. B}, vol.~97, no.~1, p.~128, 1980.

\bibitem{12}
B.~Schwarzschild, ``Mapping the interstellar cloud we live in,'' {\em Physics
  Today}, vol.~53, no.~1, p.~17, 2000.

\bibitem{13}
M.~Gyulassy and L.~McLerran, ``New forms of {QCD} matter discovered at
  {RHIC},'' {\em Nucl. Phys. A}, vol.~750, no.~1, pp.~30 -- 63, 2005.
\newblock Quark-Gluon Plasma. New Discoveries at RHIC: Case for the Strongly
  Interacting Quark-Gluon Plasma. Contributions from the RBRC Workshop held May
  14-15, 2004.

\bibitem{14}
J.~{Adams et al. (STAR Collaboration)}, ``Experimental and theoretical
  challenges in the search for the quark--gluon plasma: The {STAR}
  collaboration's critical assessment of the evidence from {RHIC} collisions,''
  {\em Nucl. Phys. A}, vol.~757, no.~1, pp.~102 -- 183, 2005.

\bibitem{15}
K.~{Adcox et al. (PHENIX Collaboration)}, ``Formation of dense partonic matter
  in relativistic nucleus- nucleus collisions at {RHIC}: Experimental
  evaluation by the {PHENIX} collaboration,'' {\em Nucl. Phys. A}, vol.~757,
  p.~184, 2005.

\bibitem{16}
B.~{B. Back et al. (PHOBOS collaboration)}, ``The {PHOBOS} perspective on
  discoveries at {RHIC},'' {\em Nucl. Phys. A}, vol.~757, no.~1, pp.~28 -- 101,
  2005.

\bibitem{17}
I.~{Arsene et al. (BRAHMS Collaboration)}, ``Quark--gluon plasma and color
  glass condensate at {RHIC}? the perspective from the {BRAHMS} experiment,''
  {\em Nucl. Phys. A}, vol.~757, no.~1, pp.~1 -- 27, 2005.

\bibitem{18}
P.~K. Kovtun, D.~T. Son, and A.~O. Starinets, ``Viscosity in strongly
  interacting quantum field theories from black hole physics,'' {\em Phys. Rev.
  Lett.}, vol.~94, p.~111601, 2005.

\bibitem{19}
K.~{Aamodt et al. (ALICE Collaboration)}, ``Elliptic flow of charged particles
  in {Pb-Pb} collisions at $\sqrt{{s}_{NN}}=2.76$ {TeV},'' {\em Phys. Rev.
  Lett.}, vol.~105, p.~252302, 2010.

\bibitem{20}
S.~{Chatrchyan et al. (CMS Collaboration)}, ``Measurement of the elliptic
  anisotropy of charged particles produced in {Pb-Pb} collisions at
  $\sqrt{{s}_{NN}}=2.76$ {TeV},'' {\em Phys. Rev. C}, vol.~87, p.~014902, 2013.

\bibitem{21}
G.~{Aad et al. (ATLAS Collaboration)}, ``Measurement of the pseudorapidity and
  transverse momentum dependence of the elliptic flow of charged particles in
  lead--lead collisions at $\sqrt{{s}_{NN}}=2.76$ {TeV} with the {ATLAS}
  detector,'' {\em Physics Letters B}, vol.~707, no.~3, pp.~330 -- 348, 2012.

\bibitem{22}
K.~{Aamodt et al. (ALICE Collaboration)}, ``Higher harmonic anisotropic flow
  measurements of charged particles in {Pb-Pb} collisions at
  $\sqrt{{s}_{NN}}=2.76$ {TeV},'' {\em Phys. Rev. Lett.}, vol.~107, p.~032301,
  2011.

\bibitem{23}
S.~{Chatrchyan et al. (CMS Collaboration)}, ``Studies of azimuthal dihadron
  correlations in ultra-central {PbPb} collisions at {$\sqrt{s_{NN}}$}=2.76
  {TeV},'' {\em J. High Energ. Phys.}, vol.~2014, no.~2, p.~88, 2014.

\bibitem{24}
A.~{Adare et al. (PHENIX Collaboration)}, ``Dihadron azimuthal correlations in
  {Au+Au} collisions at $\sqrt{{s}_{\mathit{NN}}}=200$ {GeV},'' {\em Phys. Rev.
  C}, vol.~78, p.~014901, 2008.

\bibitem{25}
M.~{Aggarwal et al. (STAR Collaboration)}, ``Azimuthal di-hadron correlations
  in $d+$ {Au} and {Au} $+$ {Au} collisions at $\sqrt{{s}_{\mathit{NN}}}=200$
  {GeV} measured at the {STAR} detector,'' {\em Phys. Rev. C}, vol.~82,
  p.~024912, 2010.

\bibitem{26}
B.~{Abelev et al. (ALICE Collaboration)}, ``Long-range angular correlations on
  the near and away side in {p--Pb} collisions at {$\sqrt{s_{NN}}=5.02$} tev,''
  {\em Phys. Lett. B}, vol.~719, no.~1, pp.~29 -- 41, 2013.

\bibitem{27}
S.~{Chatrchyan et al. (CMS Collaboration)}, ``Observation of long-range,
  near-side angular correlations in {pPb} collisions at the {LHC},'' {\em Phys.
  Lett. B}, vol.~718, no.~3, pp.~795 -- 814, 2013.

\bibitem{28}
S.~{Chatrchyan et al. (CMS Collaboration)}, ``Multiplicity and transverse
  momentum dependence of two- and four-particle correlations in {pPb} and
  {PbPb} collisions,'' {\em Phys. Lett. B}, vol.~724, no.~4, pp.~213 -- 240,
  2013.

\bibitem{29}
V.~{Khachatryan et al. (CMS Collaboration)}, ``Observation of long-range,
  near-side angular correlations in proton-proton collisions at the {LHC},''
  {\em J. High Energ. Phys.}, vol.~2010, no.~9, p.~91, 2010.

\bibitem{30}
S.~{Chatrchyan et al. (CMS Collaboration)}, ``Observation of sequential
  $\ensuremath{\Upsilon}$ suppression in {PbPb} collisions,'' {\em Phys. Rev.
  Lett.}, vol.~109, p.~222301, 2012.

\bibitem{31}
B.~{Abelev et al. (ALICE Collaboration)}, ``Centrality, rapidity and transverse
  momentum dependence of {$J/\Psi$} suppression in {Pb--Pb} collisions at
  $\sqrt{s_{NN}}$=2.76 tev,'' {\em Phys. Lett. B}, vol.~734, pp.~314 -- 327,
  2014.

\bibitem{32}
G.~{Aad et al. (ATLAS Collaboration)}, ``Measurement of the centrality
  dependence of {$J/\Psi$} yields and observation of {Z} production in
  lead--lead collisions with the {ATLAS} detector at the {LHC},'' {\em Phys.
  Lett. B}, vol.~697, no.~4, pp.~294 -- 312, 2011.

\bibitem{33}
B.~{Abelev et al. (ALICE Collaboration)}, ``{$J/\Psi$} production as a function
  of charged particle multiplicity in pp collisions at $\sqrt{s}$ =7 {TeV},''
  {\em Phys. Lett. B}, vol.~712, no.~3, pp.~165 -- 175, 2012.

\bibitem{34}
Q.~{ Yang et al. (STAR Collaboration)}, ``{$J/ \Psi$} production in p+p at
  $\sqrt{s}$=500 {GeV} collisions {$\sqrt{s}=500\,{\rm{GeV}}$} collisions and
  {Au+Au} collisions at {$\sqrt{{s}_{NN}}=200$ GeV} at the {STAR} experiment,''
  {\em J. of Phys.: Conf. Ser.}, vol.~832, no.~1, p.~012026, 2017.

\bibitem{35}
S.~Weber, ``Measurement of {$J/\Psi$} production as a function of event
  multiplicity in pp collisions at $\sqrt{s}$=13 {TeV} with {ALICE},'' {\em
  Nucl. Phys. A}, vol.~967, pp.~333 -- 336, 2017.

\bibitem{36}
S.~{Chatrchyan et al. (CMS Collaboration)}, ``Study of high-{$p_T$} charged
  particle suppression in {PbPb} compared to pp collisions at
  {$\sqrt{s_{NN}}$=2.76 TeV},'' {\em Eur. Phys. J. C}, vol.~72, no.~3, p.~1945,
  2012.

\bibitem{37}
{STAR collaboration}, ``Global {$\Lambda$} hyperon polarization in nuclear
  collisions,'' {\em Nature}, vol.~548, no.~7665, pp.~62--65, 2017.

\bibitem{38}
L.~McLerran and R.~Venugopalan, ``Computing quark and gluon distribution
  functions for very large nuclei,'' {\em Phys. Rev. D}, vol.~49,
  pp.~2233--2241, 1994.

\bibitem{39}
L.~McLerran and R.~Venugopalan, ``Gluon distribution functions for very large
  nuclei at small transverse momentum,'' {\em Phys. Rev. D}, vol.~49,
  pp.~3352--3355, 1994.

\bibitem{40}
L.~McLerran and R.~Venugopalan, ``Green's function in the color field of a
  large nucleus,'' {\em Phys. Rev. D}, vol.~50, pp.~2225--2233, 1994.

\bibitem{41}
E.~Iancu, A.~Leonidov, and L.~McLerran, ``Nonlinear gluon evolution in the
  color glass condensate: I,'' {\em Nucl. Phys. A}, vol.~692, no.~3, pp.~583 --
  645, 2001.

\bibitem{42}
E.~G. Ferreiro, E.~Iancu, K.~Itakura, and L.~McLerran, ``Froissart bound from
  gluon saturation,'' {\em Nucl. Phys. A}, vol.~710, no.~3, pp.~373 -- 414,
  2002.

\bibitem{43}
L.~Gribov, E.~Levin, and M.~Ryskin, ``Semihard processes in {QCD},'' {\em Phys.
  Rep.}, vol.~100, no.~1, pp.~1 -- 150, 1983.

\bibitem{44}
D.~Kharzeev and M.~Nardi, ``Hadron production in nuclear collisions at {RHIC}
  and high-density {QCD},'' {\em Phys. Lett. B}, vol.~507, no.~1, pp.~121 --
  128, 2001.

\bibitem{45}
D.~Kharzeev and E.~Levin, ``Manifestations of high density {QCD} in the first
  {RHIC} data,'' {\em Phys. Lett. B}, vol.~523, no.~1, pp.~79 -- 87, 2001.

\bibitem{46}
T.~Lappi and L.~McLerran, ``Some features of the glasma,'' {\em Nucl. Phys. A},
  vol.~772, no.~3, pp.~200 -- 212, 2006.

\bibitem{47}
N.~Armesto, M.~A. Braun, E.~G. Ferreiro, and C.~Pajares, ``Percolation approach
  to quark-gluon plasma and {$J/\Psi$} suppression,'' {\em Phys. Rev. Lett.},
  vol.~77, pp.~3736--3738, 1996.

\bibitem{48}
M.~Nardi and H.~Satz, ``String clustering and {$J/\Psi$} suppression in nuclear
  collisions,'' {\em Phys. Lett. B}, vol.~442, no.~1, pp.~14 -- 19, 1998.

\bibitem{49}
M.~A. Braun, C.~Pajares, and J.~Ranft, ``Fusion of strings vs. percolation and
  the transition to the quark--gluon plasma,'' {\em Int. J. Mod. Phys. A},
  vol.~14, no.~17, pp.~2689--2704, 1999.

\bibitem{50}
M.~Braun and C.~Pajares, ``Implications of color-string percolation on
  multiplicities, correlations, and the transverse momentum,'' {\em Eur. Phys.
  J. C}, vol.~16, no.~2, pp.~349--359, 2000.

\bibitem{51}
M.~A. Braun and C.~Pajares, ``Transverse momentum distributions and their
  forward-backward correlations in the percolating color string approach,''
  {\em Phys. Rev. Lett.}, vol.~85, pp.~4864--4867, 2000.

\bibitem{Campostrini1984}
M.~Campostrini, A.~Di~Giacomo, and G.~Mussardo, ``Correlation length of the
  vacuum condensate in lattice gauge theories,'' {\em Zeitschrift f{\"u}r
  Physik C Particles and Fields}, vol.~25, pp.~173--177, Jun 1984.

\bibitem{DOSCH1988}
H.~Dosch and Y.~Simonov, ``The area law of the wilson loop and vacuum field
  correlators,'' {\em Physics Letters B}, vol.~205, no.~2, pp.~339 -- 344,
  1988.

\bibitem{52}
M.~B. Isichenko, ``Percolation, statistical topography, and transport in random
  media,'' {\em Rev. Mod. Phys.}, vol.~64, pp.~961--1043, 1992.

\bibitem{54}
J.~D. de~Deus and C.~Pajares, ``String percolation and the {Glasma},'' {\em
  Phys. Lett. B}, vol.~695, no.~1, pp.~211 -- 213, 2011.

\bibitem{55}
H.~Satz, ``Extreme states of matter in strong interaction physics,'' {\em
  Lecture Notes in Physics}, vol.~841, pp.~52--56, 2012.

\bibitem{56}
S.~{Borsanyi et al.}, ``Is there still any {$T_c$} mystery in lattice {QCD}?
  results with physical masses in the continuum limit {III},'' {\em J. High
  Energ. Phys.}, vol.~2010, no.~9, p.~73, 2010.

\bibitem{57}
P.~Petreczky, ``Lattice {QCD} at non-zero temperature,'' {\em J. Phys. G Nucl.
  Partic.}, vol.~39, no.~9, p.~093002, 2012.

\bibitem{perc2}
E.~T. Gawlinski and H.~E. Stanley, ``Continuum percolation in two dimensions:
  Monte carlo tests of scaling and universality for non-interacting discs,''
  {\em J. of Phys. A}, vol.~14, no.~8, p.~L291, 1981.

\bibitem{perc3}
J.~Li and S.-L. Zhang, ``Finite-size scaling in stick percolation,'' {\em Phys.
  Rev. E}, vol.~80, p.~040104, 2009.

\bibitem{perc4}
J.~A. Quintanilla and R.~M. Ziff, ``Asymmetry in the percolation thresholds of
  fully penetrable disks with two different radii,'' {\em Phys. Rev. E},
  vol.~76, p.~051115, 2007.

\bibitem{perc1}
S.~Mertens and C.~Moore, ``Continuum percolation thresholds in two
  dimensions,'' {\em Phys. Rev. E}, vol.~86, p.~061109, 2012.

\bibitem{61}
C.~Andr\'es, A.~Moscoso, and C.~Pajares, ``Onset of the ridge structure in
  {$AA,pA$, and $pp$} collisions,'' {\em Phys. Rev. C}, vol.~90, p.~054902,
  2014.

\bibitem{62}
J.~E. Ram{\'\i}rez, {A. Fern{\'a}ndez T{\'e}llez}, and I.~Bautista, ``String
  percolation threshold for elliptically bounded systems,'' {\em Physica A},
  vol.~488, pp.~8 -- 15, 2017.

\bibitem{63}
C.~Gattringer, ``Coherent center domains in {SU(3)} gluodynamics and their
  percolation at {$T_c$},'' {\em Phys. Lett. B}, vol.~690, no.~2, pp.~179 --
  182, 2010.

\bibitem{67}
M.~Braun, J.~D. de~Deus, A.~Hirsch, C.~Pajares, R.~Scharenberg, and
  B.~Srivastava, ``De-confinement and clustering of color sources in nuclear
  collisions,'' {\em Phys. Rep.}, vol.~599, pp.~1 -- 50, 2015.

\bibitem{68}
A.~Capella, U.~Sukhatme, C.-I. Tan, and J.~T.~T. Van, ``Dual parton model,''
  {\em Phys. Rep.}, vol.~236, no.~4, pp.~225 -- 329, 1994.

\bibitem{69}
A.~Capella, U.~Sukhatme, C.-I. Tan, and J.~T.~T. Van, ``Jets in small-{$p_T$}
  hadronic collisions, universality of quark fragmentation, and rising rapidity
  plateaus,'' {\em Phys. Lett. B}, vol.~81, no.~1, pp.~68 -- 74, 1979.

\bibitem{70}
K.~Werner, ``Strings, pomerons and the {VENUS} model of hadronic interactions
  at ultrarelativistic energies,'' {\em Phys. Rep.}, vol.~232, no.~2, pp.~87 --
  299, 1993.

\bibitem{71}
A.~Kaidalov and K.~Ter-Martirosyan, ``Pomeron as quark-gluon strings and
  multiple hadron production at sps-collider energies,'' {\em Phys. Lett. B},
  vol.~117, no.~3, pp.~247 -- 251, 1982.

\bibitem{72}
K.~Werner, T.~Hirano, I.~Karpenko, T.~Pierog, S.~Porteboeuf, M.~Bleicher, and
  S.~Haussler, ``Gribov-regge theory, partons, remnants, strings -- and the
  epos model for hadronic interactions,'' {\em Nucl. Phys. B Proc. Suppl.},
  vol.~196, pp.~36 -- 43, 2009.

\bibitem{74}
V.~Abramovsky, V.~Gribov, and O.~Kancheli, ``Character of inclusive spectra and
  fluctuations produced in inelastic processes by multipomeron exchange,'' {\em
  Sov. J. Nucl. Phys.}, vol.~18, p.~308, 1974.

\bibitem{75}
A.~Capella, C.~Pajares, and A.~Ramallo, ``High energy nucleus-nucleus
  collisions in the dual parton model,'' {\em Nucl. Phys. B}, vol.~241, no.~1,
  pp.~75 -- 98, 1984.

\bibitem{76}
A.~Bialas, M.~Bleszy{\'n}ski, and W.~Czy{\.z}, ``Multiplicity distributions in
  nucleus-nucleus collisions at high energies,'' {\em Nucl. Phys. B}, vol.~111,
  no.~3, pp.~461 -- 476, 1976.

\bibitem{77}
K.~Boreskov and A.~Kaidalov, ``Nucleus-nucleus scattering in the glauber
  approach,'' {\em Sov. J. Nucl. Phys.}, vol.~48, p.~367, 1988.

\bibitem{78}
T.~Biro, H.~Nielsen, and J.~Knoll, ``Colour rope model for extreme relativistic
  heavy ion collisions,'' {\em Nucl. Phys. B}, vol.~245, pp.~449 -- 468, 1984.

\bibitem{79}
M.~Braun and C.~Pajares, ``Particle production in nuclear collisions and string
  interactions,'' {\em Phys. Lett. B}, vol.~287, no.~1, pp.~154 -- 158, 1992.

\bibitem{80}
N.~Amelin, M.~Braun, and C.~Pajares, ``Multiple production in the monte carlo
  string fusion model,'' {\em Phys. Lett. B}, vol.~306, no.~3, pp.~312 -- 318,
  1993.

\bibitem{81}
N.~S. Amelin, M.~A. Braun, and C.~Pajares, ``String fusion and particle
  production at high energies: Monte-carlo string fusion model,'' {\em Z. Phys.
  C}, vol.~63, no.~3, pp.~507--516, 1994.

\bibitem{82}
A.~Bialas and W.~Czyz, ``Conversion of color field into qq matter in the
  central region of high-energy heavy ion collisions,'' {\em Nucl. Phys. B},
  vol.~267, no.~1, pp.~242 -- 252, 1986.

\bibitem{83}
R.~Baier, Y.~Dokshitzer, A.~Mueller, S.~Peign{\'e}, and D.~Schiff, ``Radiative
  energy loss of high energy quarks and gluons in a finite-volume quark-gluon
  plasma,'' {\em Nucl. Phys. B}, vol.~483, no.~1, pp.~291 -- 320, 1997.

\bibitem{84}
R.~Baier, Y.~Dokshitzer, A.~Mueller, S.~Peign{\'e}, and D.~Schiff, ``Radiative
  energy loss and {$p_T$}-broadening of high energy partons in nuclei,'' {\em
  Nucl. Phys. B}, vol.~484, no.~1, pp.~265 -- 282, 1997.

\bibitem{85}
A.~I. Nikishov and V.~I. Ritus, ``Interaction of electrons and photons with a
  very strong electromagnetic field,'' {\em Sov. Phys. Uspekhi}, vol.~13,
  no.~2, p.~303, 1970.

\bibitem{86}
A.~Mikhailov, ``Nonlinear waves in {AdS/CFT} correspondence,'' {\em arXiv
  preprint hep-th/0305196}, 2003.

\bibitem{87}
I.~Bautista, J.~G. Milhano, C.~Pajares, and J.~D. de~Deus, ``Multiplicity in pp
  and {AA} collisions: the same power law from energy--momentum constraints in
  string production,'' {\em Phys. Lett. B}, vol.~715, no.~1, pp.~230 -- 233,
  2012.

\bibitem{88}
C.~{Albajar et al. (UA1 Collaboration)}, ``A study of the general
  characteristics of proton-antiproton collisions at $\sqrt{s}$=0.2 to 0.9
  tev,'' {\em Nucl. Phys. B}, vol.~335, no.~2, pp.~261 -- 287, 1990.

\bibitem{89}
V.~{Khachatryan et al. (CMS Collaboration)}, ``Transverse-momentum and
  pseudorapidity distributions of charged hadrons in $pp$ collisions at
  $\sqrt{s}=7\mathrm{TeV}$,'' {\em Phys. Rev. Lett.}, vol.~105, p.~022002,
  2010.

\bibitem{90}
B.~{Alver et al.}, ``Charged-particle multiplicity and pseudorapidity
  distributions measured with the phobos detector in $\text{Au}+\text{Au}$,
  $\text{Cu}+\text{Cu}$, $\mathbf{d}+\text{Au}$, and $\mathbf{p}+\mathbf{p}$
  collisions at ultrarelativistic energies,'' {\em Phys. Rev. C}, vol.~83,
  p.~024913, 2011.

\bibitem{91}
K.~{Aamodt et al. (ALICE Collaboration)}, ``Centrality dependence of the
  charged-particle multiplicity density at midrapidity in pb-pb collisions at
  $\sqrt{{s}_{\mathrm{NN}}}=2.76\mathrm{TeV}$,'' {\em Phys. Rev. Lett.},
  vol.~106, p.~032301, 2011.

\bibitem{92}
J.~D. de~Deus and J.~Milhano, ``Energy conservation and scaling violations in
  particle production,'' {\em Phys. Lett. B}, vol.~662, no.~2, pp.~129 -- 131,
  2008.

\bibitem{93}
P.~Brogueira, {Deus, J. Dias de}, and C.~Pajares, ``Limiting fragmentation in
  heavy-ion collisions and percolation of strings,'' {\em Phys. Rev. C},
  vol.~75, p.~054908, 2007.

\bibitem{94}
I.~Bautista, C.~Pajares, and J.~D. de~Deus, ``Evolution of particle density in
  high-energy pp collisions,'' {\em Nucl. Phys. A}, vol.~882, pp.~44 -- 48,
  2012.

\bibitem{95}
I.~Bautista, C.~Pajares, J.~G. Milhano, and J.~Dias~de Deus, ``Rapidity
  dependence of particle densities in $pp$ and {$AA$} collisions,'' {\em Phys.
  Rev. C}, vol.~86, p.~034909, 2012.

\bibitem{96}
G.~{Antchev et al. (TOTEM Collaboration)}, ``Measurement of the forward
  charged-particle pseudorapidity density in pp collisions at $\sqrt{s}$ = 7
  {TeV} with the {TOTEM} experiment,'' {\em Eur. Phys. Lett.}, vol.~98, no.~3,
  p.~31002, 2012.

\bibitem{97}
B.~{Alver et al. (PHOBOS Collaboration)}, ``System size, energy, and centrality
  dependence of pseudorapidity distributions of charged particles in
  relativistic heavy-ion collisions,'' {\em Phys. Rev. Lett.}, vol.~102,
  p.~142301, 2009.

\bibitem{98}
B.~{B. Back et al. (PHOBOS collaboration)}, ``Significance of the fragmentation
  region in ultrarelativistic heavy-ion collisions,'' {\em Phys. Rev. Lett.},
  vol.~91, p.~052303, 2003.

\bibitem{99}
S.~{Chatrchyan et al. (CMS Collaboration)}, ``Dependence on pseudorapidity and
  on centrality of charged hadron production in {PbPb} collisions at
  {$\sqrt{s_{NN}}=2.76$ TeV},'' {\em J. High Energ. Phys.}, vol.~2011, no.~8,
  p.~141, 2011.

\bibitem{100}
C.~Merino, C.~Pajares, and Y.~M. Shabelski, ``Production of secondaries in
  high-energy {d+Au} collisions,'' {\em Eur. Phys. J. C}, vol.~59, no.~3,
  p.~691, 2008.

\bibitem{101}
J.~Dias~de Deus, E.~G. Ferreiro, C.~Pajares, and R.~Ugoccioni, ``Universality
  of the transverse momentum distributions in the framework of percolation of
  strings,'' {\em Eur. Phys. J. C}, vol.~40, no.~2, pp.~229--241, 2005.

\bibitem{102}
C.~Pajares, ``String and parton percolation,'' {\em Eur. Phys. J. C}, vol.~43,
  no.~1, pp.~9--14, 2005.

\bibitem{103}
J.~D. de~Deus and R.~Ugoccioni, ``Large {$P_T$} distributions at {RHIC} and
  percolation of strings,'' {\em Eur. Phys. J. C}, vol.~43, no.~1-4,
  pp.~249--253, 2005.

\bibitem{105}
G.~Jona-Lasinio, ``The renormalization group: A probabilistic view,'' {\em
  Nuovo Cimento B}, vol.~26, no.~1, pp.~99--119, 1975.

\bibitem{106}
J.~D. de~Deus, C.~Pajares, and C.~Salgado, ``Moment analysis, multiplicity
  distributions and correlations in high energy processes: nucleus-nucleus
  collisions,'' {\em Phys. Lett. B}, vol.~407, no.~3, pp.~335 -- 340, 1997.

\bibitem{107}
J.~D. de~Deus, C.~Pajares, and C.~Salgado, ``Production associated to rare
  events in high energy hadron-hadron collisions,'' {\em Phys. Lett. B},
  vol.~408, no.~1, pp.~417 -- 421, 1997.

\bibitem{108}
J.~D. de~Deus, C.~Pajares, and C.~Salgado, ``Multiplicity and transverse energy
  distributions associated to rare events in nucleus-nucleus collisions,'' {\em
  Phys. Lett. B}, vol.~409, no.~1, pp.~474 -- 478, 1997.

\bibitem{109}
J.~D. de~Deus and C.~Pajares, ``Rare event triggers in hadronic and nuclear
  collisions,'' {\em Phys. Lett. B}, vol.~442, no.~1, pp.~395 -- 397, 1998.

\bibitem{110}
M.~Braun and C.~Pajares, ``Self-similarity of multiplicity distributions and
  the kno scaling,'' {\em Phys. Lett. B}, vol.~444, no.~3, pp.~435 -- 441,
  1998.

\bibitem{115}
I.~Bautista and C.~Pajares, ``Strong color fields and heavy flavor
  production,'' {\em Phys. Rev. C}, vol.~82, p.~034912, 2010.

\bibitem{138p}
I.~{Arsene et al. (BRAHMS Collaboration)}, ``Evolution of the nuclear
  modification factors with rapidity and centrality in $d+\mathrm{A}\mathrm{u}$
  collisions at $\sqrt{N^{S}N}=200\text{ }\text{
  }\mathrm{G}\mathrm{e}\mathrm{V}$,'' {\em Phys. Rev. Lett.}, vol.~93,
  p.~242303, 2004.

\bibitem{140p}
J.~D. de~Deus, E.~Ferreiro, C.~Pajares, and R.~Ugoccioni, ``Schwinger model and
  string percolation in hadron--hadron and heavy ion collisions,'' {\em Phys.
  Lett. B}, vol.~581, no.~3, pp.~156 -- 160, 2004.

\bibitem{ex1}
L.~Cunqueiro, J.~Dias~de Deus, E.~Ferreiro, and C.~Pajares, ``Universal
  behavior of transverse momentum distributions of baryons and mesons in the
  framework of percolation of strings,'' {\em Eur. Phys. J. C}, vol.~53, no.~4,
  pp.~585--589, 2008.

\bibitem{118}
J.~D. de~Deus and C.~Pajares, ``Percolation of color sources and critical
  temperature,'' {\em Phys. Lett. B}, vol.~642, no.~5, pp.~455 -- 458, 2006.

\bibitem{116}
A.~Bialas, ``Fluctuations of the string tension and transverse mass
  distribution,'' {\em Phys. Lett. B}, vol.~466, no.~2, pp.~301 -- 304, 1999.

\bibitem{119}
P.~Castorina, D.~Kharzeev, and H.~Satz, ``Thermal hadronization and
  {Hawking-Unruh} radiation in {QCD},'' {\em Eur. Phys. J. C}, vol.~52, no.~1,
  p.~187, 2007.

\bibitem{121}
R.~C. Hwa and C.~B. Yang, ``Centrality scaling of the ${p}_{T}$ distribution of
  pions,'' {\em Phys. Rev. Lett.}, vol.~90, p.~212301, 2003.

\bibitem{122}
W.~C. Zhang and C.~B. Yang, ``Scaling behaviour of charged hadron $p_{T}$
  distribution in pp and $\bar{p}\bar{p}$ collisions,'' {\em J. Phys. G},
  vol.~41, no.~10, p.~105006, 2014.

\bibitem{120}
L.~Zhu, H.~Zheng, and C.~Yang, ``Scaling behavior of transverse kinetic energy
  distributions in {Au + Au} collisions at {$\sqrt{s_{NN}}$=200} {GeV},'' {\em
  Nucl. Phys. A}, vol.~802, no.~1, pp.~122 -- 130, 2008.

\bibitem{125}
E.~G. Ferreiro, F.~del Moral, and C.~Pajares, ``Transverse momentum
  fluctuations and percolation of strings,'' {\em Phys. Rev. C}, vol.~69,
  p.~034901, 2004.

\bibitem{127}
A.~Capella and A.~Krzywicki, ``Unitarity corrections to short-range order:
  Long-range rapidity correlations,'' {\em Phys. Rev. D}, vol.~18,
  pp.~4120--4133, 1978.

\bibitem{128}
T.~Chou and C.~N. Yang, ``A unified physical picture: Narrow poisson-like
  distribution for $e^+e^-$ two-jet events and wide approximate {KNO}
  distribution for hadron-hadron collisions,'' {\em Phys. Lett. B}, vol.~167,
  no.~4, pp.~453 -- 456, 1986.

\bibitem{130}
V.~Vechernin, ``Forward--backward correlations between multiplicities in
  windows separated in azimuth and rapidity,'' {\em Nucl. Phys. A}, vol.~939,
  pp.~21 -- 45, 2015.

\bibitem{131}
M.~Braun, C.~Pajares, and V.~Vechernin, ``On the forward--backward correlations
  in a two-stage scenario,'' {\em Phys. Lett. B}, vol.~493, no.~1, pp.~54 --
  64, 2000.

\bibitem{132}
N.~S. Amelin, N.~Armesto, M.~A. Braun, E.~G. Ferreiro, and C.~Pajares, ``Long
  and short range correlations: A signature of string fusion,'' {\em Phys. Rev.
  Lett.}, vol.~73, pp.~2813--2816, 1994.

\bibitem{133}
P.~Brogueira, J.~D. de~Deus, and C.~Pajares, ``Long range forward--backward
  rapidity correlations in proton--proton collisions at lhc,'' {\em Phys. Lett.
  B}, vol.~675, no.~3, pp.~308 -- 311, 2009.

\bibitem{134}
N.~Armesto, L.~McLerran, and C.~Pajares, ``Long range forward--backward
  correlations and the color glass condensate,'' {\em Nucl. Phys. A}, vol.~781,
  no.~1, pp.~201 -- 208, 2007.

\bibitem{135}
A.~Dumitru, F.~Gelis, L.~McLerran, and R.~Venugopalan, ``Glasma flux tubes and
  the near side ridge phenomenon at {RHIC},'' {\em Nucl. Phys. A}, vol.~810,
  no.~1, pp.~91 -- 108, 2008.

\bibitem{136}
F.~G. Gelis, T.~Lappi, and R.~Venugopalan, ``High energy factorization in
  nucleus-nucleus collisions. ii. multigluon correlations,'' {\em Phys. Rev.
  D}, vol.~78, p.~054020, 2008.

\bibitem{138}
M.~A. Braun, R.~S. Kolevatov, C.~Pajares, and V.~V. Vechernin, ``Correlations
  between multiplicities and average transverse momentum in the percolating
  color strings approach,'' {\em Eur. Phys. J. C}, vol.~32, no.~4,
  pp.~535--546, 2004.

\bibitem{139}
V.~V. Vechernin, ``Asymptotic behavior of the correlation coefficients of
  transverse momenta in the model with string fusion,'' {\em Theo. Math.
  Phys.}, vol.~190, no.~2, pp.~251--267, 2017.

\bibitem{142}
A.~Ortiz and L.~V. Palomo, ``Universality of the underlying event in $pp$
  collisions,'' {\em Phys. Rev. D}, vol.~96, p.~114019, 2017.

\bibitem{143}
R.~Blankenbecler, A.~Capella, J.~T. Van, C.~Pajares, and A.~Ramallo, ``Unusual
  shadowing effects in particle production off nuclei,'' {\em Phys. Lett. B},
  vol.~107, no.~1, pp.~106 -- 110, 1981.

\bibitem{148}
W.~Kittel and E.~De~Wolf, {\em Soft Multihadron Dynamics}.
\newblock World Scientific, 2005.

\bibitem{149}
B.~Andersson and W.~Hofmann, ``{Bose-Einstein} correlations and color
  strings,'' {\em Phys. Lett. B}, vol.~169, no.~4, pp.~364 -- 368, 1986.

\bibitem{150}
B.~Andersson and M.~Ringn{\'e}r, ``{Bose-Einstein} correlations in the {Lund}
  model,'' {\em Nucl. Phys. B}, vol.~513, no.~3, pp.~627 -- 644, 1998.

\bibitem{151}
T.~{C. Awes et al. (WA80 Collaboration)}, ``{Bose-Einstein} correlations of
  soft pions in ultrarelativistic nucleus-nucleus collisions,'' {\em Z. Phys.
  C}, vol.~69, no.~1, pp.~67--76, 1995.

\bibitem{152}
C.~{Albajar et al. (UA1 Collaboration)}, ``{Bose-Einstein} correlations in pp
  interactions at $\sqrt{s}$=0.2 to 0.9 {TeV},'' {\em Phys. Lett. B}, vol.~226,
  no.~3, pp.~410 -- 416, 1989.

\bibitem{153}
I.~{G. Bearden et al. (NA44 Collaboration)}, ``High energy {Pb+Pb} collisions
  viewed by pion interferometry,'' {\em Phys. Rev. C}, vol.~58, pp.~1656--1665,
  1998.

\bibitem{154}
M.~Braun, F.~del Moral, and C.~Pajares, ``Chaotic sources and percolation of
  strings in heavy ion collisions,'' {\em Eur. Phys. J. C}, vol.~21, no.~3,
  pp.~557--562, 2001.

\bibitem{155}
M.~Braun, F.~del Moral, and C.~Pajares, ``Genuine three-body {Bose--Einstein}
  correlations and percolation of strings,'' {\em Phys. Lett. B}, vol.~551,
  no.~3, pp.~291 -- 295, 2003.

\bibitem{156}
M.~Biyajima, N.~Suzuki, G.~Wilk, and Z.~W{\l}odarczyk, ``Totally chaotic
  poissonian-like sources in multiparticle production processes?,'' {\em Phys.
  Lett. B}, vol.~386, no.~1, pp.~297 -- 303, 1996.

\bibitem{157}
M.~{Aggarwal et al. (WA98 Collaboration)}, ``Three-pion interferometry results
  from central {$Pb+Pb$} collisions at {158A GeV/c},'' {\em Phys. Rev. Lett.},
  vol.~85, pp.~2895--2899, 2000.

\bibitem{158}
E.~G. Ferreiro and C.~Pajares, ``High multiplicity $pp$ events and {$J/\psi$}
  production at energies available at the {CERN Large Hadron Collider},'' {\em
  Phys. Rev. C}, vol.~86, p.~034903, 2012.

\bibitem{162}
C.~{Alexa et al. (H1 Collaboration)}, ``Elastic and proton-dissociative
  photoproduction of j/$\psi$ mesons at hera,'' {\em Eur. Phys. J. C}, vol.~73,
  no.~6, p.~2466, 2013.

\bibitem{163}
B.~{Abelev et al. (ALICE Collaboration)}, ``Exclusive {$J/\Psi$}
  photoproduction off protons in ultraperipheral $p$-{Pb} collisions at
  {$\sqrt{{s}_{\text{NN}}}=5.02$ TeV},'' {\em Phys. Rev. Lett.}, vol.~113,
  p.~232504, 2014.

\bibitem{164}
J.~Cepila, J.~Contreras, and J.~T. Takaki, ``Energy dependence of dissociative
  {$J/\Psi$} photoproduction as a signature of gluon saturation at the {LHC},''
  {\em Phys. Lett. B}, vol.~766, pp.~186 -- 191, 2017.

\bibitem{167}
G.~Feofilov, V.~Kovalenko, and A.~Puchkov, ``Correlation between heavy flavour
  production and multiplicity in pp and {p-Pb} collisions at high energy in the
  multi-pomeron exchange model,'' {\em EPJ Web Conf.}, vol.~171, p.~18003,
  2018.

\bibitem{165}
M.~Braun and C.~Pajares, ``A probabilistic model of interacting strings,'' {\em
  Nucl. Phys. B}, vol.~390, no.~2, pp.~542 -- 558, 1993.

\bibitem{166}
N.~Armesto, M.~Braun, E.~Ferreiro, and C.~Pajares, ``Strangeness enhancement
  and string fusion in nucleus-nucleus collisions,'' {\em Phys. Lett. B},
  vol.~344, no.~1, pp.~301 -- 307, 1995.

\bibitem{168}
J.~{Adam et al. (ALICE Collaboration)}, ``Enhanced production of multi-strange
  hadrons in high-multiplicity proton-proton collisions,'' {\em Nature},
  vol.~13, no.~6, pp.~535--539, 2017.

\bibitem{169}
M.~A. Braun and C.~Pajares, ``Elliptic flow from colour strings,'' {\em Eur.
  Phys. J. C}, vol.~71, no.~2, p.~1558, 2011.

\bibitem{170}
M.~Braun, C.~Pajares, and V.~Vechernin, ``Anisotropic flows from colour
  strings: Monte carlo simulations,'' {\em Nucl. Phys. A}, vol.~906, pp.~14 --
  27, 2013.

\bibitem{171}
I.~Bautista, L.~Cunqueiro, J.~D. de~Deus, and C.~Pajares, ``Particle production
  azimuthal asymmetries in a clustering of color sources model,'' {\em J. Phys.
  G}, vol.~37, no.~1, p.~015103, 2010.

\bibitem{172}
I.~Bautista, J.~Dias~de Deus, and C.~Pajares, ``Elliptic flow at {RHIC} and
  {LHC} in the string percolation approach,'' {\em Eur. Phys. J. C}, vol.~72,
  no.~6, p.~2038, 2012.

\bibitem{173}
I.~Bautista, J.~D. de~Deus, and C.~Pajares, ``Elliptic flow: Pseudorapidity and
  number of participants dependence,'' {\em Phys. Lett. B}, vol.~693, no.~3,
  pp.~362 -- 365, 2010.

\bibitem{174}
L.~Cunqueiro, J.~Dias~de Deus, and C.~Pajares, ``Nuclear-like effects in
  proton--proton collisions at high energy,'' {\em Eur. Phys. J. C}, vol.~65,
  no.~3, pp.~423--426, 2010.

\bibitem{175}
M.~A. Braun, C.~Pajares, and V.~V. Vechernin, ``Ridge from strings,'' {\em Eur.
  Phys. J. A}, vol.~51, no.~4, p.~44, 2015.

\bibitem{179}
C.~Andr{\'e}s, M.~Braun, and C.~Pajares, ``Energy loss as the origin of a
  universal scaling law of the elliptic flow,'' {\em Eur. Phys. J. A}, vol.~53,
  no.~3, p.~41, 2017.

\bibitem{180}
A.~{Adler et al. (PHENIX Collaboration)}, ``Identified charged particle spectra
  and yields in {Au-Au} collisions at {$\sqrt{s_{NN}}=200$} {GeV},'' {\em Phys.
  Rev. C}, vol.~69, p.~034909, 2004.

\bibitem{181}
J.~{Adam et al. (ALICE Collaboration)}, ``Measurement of jet quenching with
  semi-inclusive hadron-jet distributions in central {Pb-Pb} collisions at
  {$\sqrt{s_{NN}}=2.76$ TeV},'' {\em J. High Energ. Phys.}, vol.~2015, no.~9,
  p.~170, 2015.

\bibitem{183}
R.~Scharenberg, B.~Srivastava, and A.~Hirsch, ``Percolation of color sources
  and the equation of state of {QGP} in central {Au-Au} collisions at
  {$\sqrt{s_{NN}}$}=200 {GeV},'' {\em Eur. Phys. J. C}, vol.~71, no.~1,
  p.~1510, 2011.

\bibitem{184}
T.~Hirano and M.~Gyulassy, ``Perfect fluidity of the quark--gluon plasma core
  as seen through its dissipative hadronic corona,'' {\em Nucl. Phys. A},
  vol.~769, pp.~71 -- 94, 2006.

\bibitem{bvis}
X.-G. Huang and T.~Koide, ``Shear viscosity, bulk viscosity, and relaxation
  times of causal dissipative relativistic fluid-dynamics at finite temperature
  and chemical potential,'' {\em Nucl. Phys. A}, vol.~889, pp.~73 -- 92, 2012.

\bibitem{53}
A.~{Andronic et al.}, ``Hadron production in ultra-relativistic nuclear
  collisions: Quarkyonic matter and a triple point in the phase diagram of
  qcd,'' {\em Nucl. Phys. A}, vol.~837, no.~1, pp.~65 -- 86, 2010.

\bibitem{64}
J.~Danzer, C.~Gattringer, S.~Borsanyi, and Z.~Fodor, ``Center clusters and
  their percolation properties in lattice {QCD},'' {\em PoS Lattice2010}, 2010.

\bibitem{fig21ref}
I.~{Altsybeev et al., (ALICE Collaboration)}, ``Forward--backward correlations
  between mean transverse momenta in {Pb--Pb} collisions with {ALICE},'' {\em
  KnE Energy}, vol.~3, no.~1, pp.~304--312, 2018.

\end{thebibliography}

\end{document}